\RequirePackage{fix-cm} 
\documentclass[fleqn,usenatbib]{mnras}

\usepackage{amsmath}
\usepackage{amssymb}
\usepackage{amsfonts}
\usepackage{textcomp}
\usepackage{newtxtext}
\usepackage{bm}
\usepackage{lmodern}

\usepackage[T1]{fontenc}

\DeclareRobustCommand{\VAN}[3]{#2}
\let\VANthebibliography\thebibliography
\def\thebibliography{\DeclareRobustCommand{\VAN}[3]{##3}\VANthebibliography}

\usepackage{graphicx}	
\usepackage{longtable}
\usepackage{multirow}
\usepackage{booktabs}
\usepackage{amsmath}	
\usepackage{pdflscape} 
\usepackage{orcidlink}

\newcommand{\kmsmpc} {$\rm {km~s^{-1}}~Mpc^{-1}$}
\newcommand{\ii}{~\textsc{ii}}
\newcommand{\iii}{~\textsc{iii}}
\newcommand{\iv}{~\textsc{iv}}

\newcommand{\hii}{H\,{\sc ii }\rm}

\newcommand{\Lx}{L_{\mathrm{X}}}

\newcommand{\aox}{\alpha_{\mathrm{ox}}}
\newcommand{\Ne}{$N_{\rm e}$}
\newcommand{\Te}{$T_{\rm e}$}

\newcommand{\halpha}{\mathrm{H}\alpha}

\newcommand{\hgamma}{\mathrm{H}\gamma}

\newcommand{\oiii}{\mathrm{[O\,\textsc{iii}]}}
\newcommand{\oii}{\mathrm{[O\,\textsc{ii}]}}

\newcommand{\nii}{\mathrm{[N\,\textsc{ii}]}}
\newcommand{\sii}{\mathrm{[S\,\textsc{ii}]}}

\newcommand{\ntwo}{\mathrm{N_2}}
\newcommand{\otnt}{\mathrm{O_3 N_2}}
\newcommand{\ns}{\mathrm{N_2 S_2}}
\newcommand{\no}{\mathrm{N_2 O_2}}
\newcommand{\otot}{\mathrm{O_3 O_2}}

\defcitealias{2026MNRAS.548ag560A}{Paper I}
\defcitealias{1998AJ....115..909S}{SB98}
\DeclareRobustCommand{\SBcalib}[1]{\hyperlink{cite.1998AJ....115..909S}{SB98\textcolor{blue}{#1}}}

\defcitealias{2017MNRAS.467.1507C}{C17}  
\defcitealias{2020MNRAS.492.5675C}{C20}
\defcitealias{2021MNRAS.507..466D}{D21}

\title[Identifying AGNs from X-ray detections---II]{Identifying AGNs from X-ray detections---II: Metallicity calibrations for the \texorpdfstring{\boldmath $\no$}{N_2 O_2} and \texorpdfstring{\boldmath $\ns$}{N_2 S_2} diagnostics}

\author[M.~Armah et al.]{Mark Armah$^{{\orcidlink{0000-0002-9746-3938}}1}$\thanks{E-mail: mrkrmh@gmail.com (MA)},
O. L. Dors$^{\orcidlink{0000-0003-4782-1570}1}$, 
Rog\'erio Riffel$^{\orcidlink{0000-0002-1321-1320}2}$,
M.~V. Cardaci$^{\orcidlink{0000-0002-8856-602X}3,4}$, G.~F. H\"agele$^{\orcidlink{0000-0002-9011-8517}3,4}$,
\newauthor{Rogemar A. Riffel$^{\orcidlink{0000-0003-0483-3723}5}$ and
J. M. V\'ilchez$^{\orcidlink{0000-0001-7299-8373}6}$}
\\
$^{1}$Universidade do Vale do Para\'iba, Instituto de Pesquisa e Desenvolvimento, Av. Shishima Hifumi, 2911, CEP: 12244-000, São José dos Campos, SP, Brazil\\
$^2$Departamento de Astronomia, Instituto de F\'isica, Universidade Federal do Rio Grande do Sul, CP 15051, 91501-970, Porto Alegre, RS, Brazil \\
$^3$ Facultad de Ciencias Astron\'omicas y Geof\'{\i}sicas, Universidad Nacional de La Plata, Paseo del Bosque s/n, 1900 La Plata, Argentina.\\
$^4$ Instituto de Astrofísica de La Plata (CONICET-UNLP), Avenida Centenario (Paseo del Bosque) S/N, B1900FWA, La Plata, Argentina\\
$^5$Departamento de F\'isica, CCNE, Universidade Federal de Santa Maria, 97105-900, Santa Maria, RS, Brazil\\
$^{6}$Instituto de Astrof\'isica de Andaluc\'ia, CSIC, Apartado de correos 3004, E-18080 Granada, Spain
}

\date{Accepted XXX. Received YYY; in original form ZZZ}

\pubyear{2026}

\begin{document}
\label{firstpage}
\pagerange{\pageref{firstpage}--\pageref{lastpage}}
\maketitle

\begin{abstract}
Emission-line ratios from ions with nearly identical ionization potentials offer a robust solution to the degeneracies inherent to traditional active galactic nuclei (AGNs) metallicity diagnostics. We introduce new semi-empirical metallicity calibrations for the $\no$ and $\ns$ diagnostics, explicitly designed to isolate the chemical abundance from the incident radiation field. By coupling an extensive grid of {\sc cloudy} photoionization simulations directly to the intrinsic 2–10 keV X-ray luminosity ($\Lx$) and benchmarking against Seyfert~2 nuclei from the Burst Alert Telescope AGN Spectroscopic Survey (BASS), we establish robust relations valid across the metallicity regime of $ 8.0 \lesssim 12+\log({\rm O/H}) \lesssim 9.1$ ($ 0.2 \lesssim Z/Z_{\odot} \lesssim 2.6$). The $\no$ and $\ns$ indices trace co-spatial emitting volumes within the narrow-line regions (NLRs), enabling this framework to resolve the significant $\Lx$-driven systematic biases we previously identified in the standard $\ntwo$ and $\otnt$ indices.
While the $\no$ ratio proves to be virtually independent of nebular structural variations, the $\ns$ index exhibits a subtle, yet discernible, electron density (\Ne) susceptibility due to the low critical density  ($N_{\rm c}$) of the  $\sii\lambda\lambda6716,6731$ doublet. 
Nevertheless, both diagnostics yield highly precise metallicity constraints with tight root-mean-square residual dispersions of $\sim 0.081$ dex for $\no$ and $\sim 0.121$ dex for $\ns$. 
We propose the $\no$ and $\ns$ calibrations as highly optimized, unbiased metallicity tracers for AGNs.
\end{abstract}





\section{Introduction}
\label{sec:introduction}
The metallicity of the interstellar medium (ISM) is a key parameter for tracing the chemical enrichment of galaxies and reconstructing their global evolutionary histories. The chemical abundances of heavy elements (e.g. O, N, C) record the time-integrated impact of star formation, stellar evolution, and gas (in/out)flows over cosmic time, thereby providing stringent constraints on the mechanisms that govern galaxy evolution. Within this context, active galactic nuclei (AGNs) constitute particularly powerful laboratories for investigating galactic chemical enrichment, owing to their high luminosities and characteristic, strong emission-line spectra \citep[e.g.][]{1999ARA&A..37..487H, 2019A&ARv..27....3M, 2023MNRAS.524.5640R,2023MNRAS.520.1687A, 2025MNRAS.542.3181D}.

The most fundamental approach for deriving gas-phase metallicities and elemental abundances in the ISM is the ``direct'' electron-temperature ($T_{\rm e}$) method.  Originating from early efforts to determine nebular composition \citep{1936Natur.138..503P, 1939LicOB..19....1B, 1942ApJ....95..356W} and first applied to gaseous nebulae \citep{1954ApJ...120..401A, 1959ApJ...130...45A}, this approach was eventually adapted for \hii\ regions \citep{1969BOTT....5....3P} and later extended to the extreme environments of AGNs. Following its pioneering application to the radio galaxy Cygnus~A \citep{1975ApJ...197..535O}, the $T_{\rm e}$-method has progressively driven AGN abundance studies, from foundational samples of Seyfert galaxies \citep[e.g.][]{1978ApJ...223...56K,1992A&A...266..117A} and dwarf hosts \citep[e.g.][]{2008ApJ...687..133I} to modern, large-scale systematic abundance frameworks \citep[e.g.][and references therein]{2015MNRAS.453.4102D, 2020MNRAS.492..468D,2020MNRAS.496.3209D,2020MNRAS.496.2191F} and empirical calibrations \citep[e.g.][]{2021MNRAS.507..466D}.
 This technique requires detections of auroral lines (e.g. $\oiii\lambda4363$) to constrain $T_{\rm e}$, however, these transitions are intrinsically faint (typically $\sim 100\times$ weaker than H$\beta$) and are therefore generally accessible only in low-metallicity and/or highly ionized \hii\ regions and AGNs (e.g. \citealt{1998AJ....116.2805V, 2020MNRAS.496.2191F, 2020MNRAS.496.3209D}). In AGNs, measuring auroral features is further complicated by line blending, most notably between $\oiii\lambda4363$ and $\hgamma$ because the permitted Balmer lines often exhibit much larger full widths at half-maximum (FWHM; larger by a factor of $\sim 10$) than in canonical \hii\-region spectra, forming complex profiles that blend with adjacent forbidden transitions \citep[e.g.][]{1981ApJ...249..462O}.
A further limitation is that the $T_{\rm e}$-method tends to yield oxygen abundances that are lower by $\sim 0.2$ dex than values inferred from photoionization models or recombination lines \citep[e.g.][]{2015ApJ...798...99B, 2020MNRAS.496.3209D}. This offset, commonly termed the abundance discrepancy factor \citep[ADF; e.g.][]{2004MNRAS.355..229E, 2007ApJ...670..457G, 2017PASP..129h2001P}, is often attributed to temperature inhomogeneities \citep[e.g.][]{1967ApJ...150..825P, 2021MNRAS.506L..11R}, shock heating \citep[e.g.][]{2002ApJ...572..753D, 2013ApJ...774..100K, 2021MNRAS.501.1370D}, and/or uncertainties in ionization correction factors \citep[ICFs; e.g.][]{2006A&A...448..955I, 2017PASP..129h2001P,2019MNRAS.489.2652P}. Consequently, the combination of weak auroral-lines and residual theoretical systematics often motivates the use of strong-line calibrations (e.g. \citealt{1979A&A....78..200A}; \citealt{1979MNRAS.189...95P}; \citealt[][hereafter \citetalias{1998AJ....115..909S}]{1998AJ....115..909S}; \citealt{2016Ap&SS.361...61D};  \citealt{2019MNRAS.485..367K}; \citealt[][hereafter \citetalias{2020MNRAS.492.5675C}]{2020MNRAS.492.5675C}; \citealt[][hereafter \citetalias{2021MNRAS.507..466D}]{2021MNRAS.507..466D}, and references therein), which exploit the brightest collisionally excited transitions.

The application of these strong-line diagnostics has been significantly expanded by the advent of powerful multiplexed spectroscopic facilities and integral field spectrographs (IFS). Large-scale surveys and instruments such as the Sloan Digital Sky Survey \citep[\href{https://www.sdss.org/}{SDSS};][]{2000AJ....120.1579Y}, including its Mapping Nearby Galaxies at Apache Point Observatory (MaNGA) survey \citep{2015ApJ...798....7B}, the Multi Unit Spectroscopic Explorer (MUSE) on the Very Large Telescope (VLT)  \citep{2010SPIE.7735E..08B}, the Gemini Multi-Object Spectrograph (GMOS) on the Gemini Telescopes \citep{2004PASP..116..425H}, the Dark Energy Spectroscopic Instrument (DESI) \citep{2022AJ....164..207D}, and the \textit{James Webb Space Telescope} (\textit{JWST}) \citep{2023PASP..135d8002R,2025arXiv251001033C,2025arXiv251001034S} now provide vast catalogs of high-quality spectra. 

Although these spectroscopic facilities enable highly reliable AGN classification via classical diagnostic diagrams \citep[e.g.][]{1981PASP...93....5B,1987ApJS...63..295V,2001ApJ...556..121K,2006MNRAS.372..961K}, deriving metallicities and elemental abundances from these spectra remains a challenge mainly for distant objects (e.g. \citealt{2024MNRAS.535..881J, 2025MNRAS.540.1608D, 2026MNRAS.545f2107S, 2025ApJ...994L..29Z, 2026ApJ...998....5Z, 2026ApJ...998..175M, 2026ApJ...998..141Z}).  Fundamentally, the quest for a universal metallicity scale in AGNs has historically been hindered by the profound degeneracy between chemical abundance, electron density ($N_{\rm e}$), and the local ionization state, parameterized by the dimensionless ionization parameter ($U$). This structural sensitivity drives significant systematic biases that severely impact strong-line metallicity derivations across both star-forming environments \citep[e.g.][]{2002ApJS..142...35K, 2006A&A...459...85N, 2016ApJ...816...23S} and AGNs \citep[e.g.][]{2004ApJS..153....9G,2006A&A...447..863N, 2014MNRAS.443.1291D, 2024ApJ...977..187Z}.

To overcome this limitation, in the first paper of this series (\citealp[][]{2026MNRAS.548ag560A}, hereafter \citetalias{2026MNRAS.548ag560A}), we demonstrated that replacing the unobservable $U$ with the intrinsic 2-10 keV X-ray luminosity ($\Lx$) reveals substantial systematic biases within standard metallicity diagnostics, specifically the $\ntwo = \log(\nii\lambda6584/\halpha)$ and 
$\otnt = \log(\oiii\lambda5007/\nii\lambda6584)$. These classic diagnostics are affected by the $\Lx$-driven expansion of the partially ionized zone (PIZ), a structural consequence of X-ray photoionization that alters nebular emission lines in Seyfert galaxies (e.g. \citealt{1974ApJ...191..309S}).
To establish a truly ionization-invariant metallicity scale, this second paper investigates the use of the $\no = \log(\nii\lambda6584/\oii\lambda3727)$ and $\ns = \log(\nii\lambda6584/\sii(\lambda6716+\lambda6731))$ indices as
metallicity indicators for the narrow-line regions (NLRs) of AGNs. Because these ratios utilize transitions from ions with virtually identical first ionization potentials (N$^+$: 14.53 eV; O$^+$: 13.62 eV; S$^+$: 10.36 eV), they trace co-spatial emitting volumes within the NLR. Theoretically, taking the ratio of these lines symmetrically cancels out structural variations driven by the incident radiation field, isolating the underlying elemental abundance ratio  \citep[e.g.][]{2002ApJS..142...35K, 2019MNRAS.489.2652P}.

 The $\no$ index is advantageous as it uses lines from ions with similar ionization potentials, reducing its sensitivity to the ionization parameter, and it avoids the bi-value issues common in some calibrations \citep[e.g.][]{2017MNRAS.467.1507C, 2019MNRAS.489.2652P}. Additionally, when plotted against an ionization-sensitive ratio such as $\otot = \log(\oiii\lambda5007/\oii\lambda 3727$), the $\no$ index forms a powerful diagnostic diagram capable of breaking the degeneracy found in  metallicity estimates from strong-line methods  (e.g. \citealt{2008ApJ...681.1183K}). However, $\no$ requires broad spectral coverage and accurate reddening corrections because the $\oii\lambda\lambda3726,3729$ doublet can be challenging to measure in high redshift or specific surveys \citep[e.g.][]{2011MNRAS.412.1145P, 2019A&ARv..27....3M,2021AJ....161...52L}.  For example, in powerful IFS such as the MUSE, the typical wavelength coverage of 4750--9350 \AA\ means that the $\oii$ line is shifted out of the blue end of the spectral range for sources at $z \lesssim 0.3$ \citep{2010SPIE.7735E..08B}.  Also, the DEEP2 Galaxy Redshift Survey, while designed to push to higher redshifts, still faced challenges in consistently measuring these lines for all targets, especially at the faint end of the AGN population \citep[e.g.][]{2013ApJS..208....5N}.

Similarly, the $\ns$ index is a robust diagnostic in star-forming \citep[e.g.][]{2016Ap&SS.361...61D} and non-star-forming environments \citep[e.g.][]{2017MNRAS.466.3217Z,2019MNRAS.485..367K}. The proximity of the ionization potentials of N$^{+}$ and S$^{+}$ (14.5 eV and 10.4 eV, respectively; \citealt{1986ApJ...301..727D}) renders $\ns$ comparatively insensitive to variations in the hardness and intensity of the ionizing field, mitigating the substantial scatter typically driven by $\Lx$ in other indicators \citep[e.g.][]{2008ApJ...681.1183K, 2016Ap&SS.361...61D}. Furthermore, the flux ratio of the $\nii$ and $\sii$ lines provides a robust proxy for the $\rm N^{+}/S^{+}$ ionic abundance ratio, as their similar atomic properties largely cancel out dependencies on local physical conditions \citep[][]{2006agna.book.....O}. This relationship is nearly independent of \Te\, enabling metallicity estimates without recourse to weak auroral lines \citep[see, e.g.][]{2009MNRAS.398..949P,2013ApJS..207...21N}.

In a comprehensive study, \citet{2024ApJ...977..187Z} built an extensive grid of photoionization models using the \textsc{mappings} code \citep{Sutherland_2018} to propose emission-line intensity ratios as metallicity indicators in the NLRs. This theoretical approach successfully covers the low-metallicity regime ($Z \: \lesssim \: 0.2 \times \rm Z_{\odot}$) and a wide range of ionization degrees, which are inaccessible to (semi-)empirical calibrations \citep[e.g.][]{2017MNRAS.467.1507C, 2020MNRAS.492.5675C, 2021MNRAS.507..466D} due to sample selections from diagnostic diagrams \citep[e.g.][]{1981PASP...93....5B, 1987ApJS...63..295V} that exclude objects with $Z \: \lesssim \: 0.2 \times \rm Z_{\odot}$ and/or low ionization parameters (e.g. \citealt{2006MNRAS.371.1559G, 2016MNRAS.456.3354F, 2024MNRAS.527.8193D}). Although limited by the completeness of the available observational samples (see \citealt{2020MNRAS.492.5675C}), (semi-)empirical calibrations remain highly advantageous over purely theoretical calibrations because nebular parameters (e.g. metallicity and ionization degree) are better constrained by observational data.

In this study, we propose two semi-empirical calibrations for the $\no$ and $\ns$ indices. Adopting the theoretical framework from \citetalias{2026MNRAS.548ag560A}, we leverage {\sc cloudy} models alongside Seyfert 2 observations from the Burst Alert Telescope AGN Spectroscopic Survey Data Release 2 \citep[BASS DR2;][]{2017ApJ...850...74K, 2017MNRAS.464.1466O, 2022ApJS..261....1K,2022ApJS..261....4O}. This paper is structured as follows: \S~\ref{method} outlines the methodology and sample; \S~\ref{results} presents the diagnostic diagrams; \S~\ref{smetal} details the calibrations; \S~\ref{discussion} discusses the physical drivers and compares our results with the literature; and \S~\ref{conclusions} summarizes our conclusions. 

Throughout this paper, we adopt a solar oxygen abundance of $\log(\text{O/H})_\odot = -3.31$ \citep{2021A&A...653A.141A} and a standard cosmology model with $\Omega_{\rm m}= 0.315\pm0.007$ and $H_{0} = 67.4\pm 0.5$ \kmsmpc\, \citep{2021A&A...652C...4P}.

\section{Methodology}
\label{method} 

\subsection{Photoionization models}
We adopt a theoretical framework consistent with the methodology developed in \citetalias{2026MNRAS.548ag560A}, utilizing the {\sc cloudy} spectral synthesis code \citep[C23.01;][]{2013RMxAA..49..137F,2017RMxAA..53..385F, 2023RNAAS...7..246G} to simulate the physical conditions in the NLRs of the AGNs. The cornerstone of our modelling strategy is the substitution of the conventional $U$ with the 2–10 keV $\Lx$. This directly connects the photoionization grids to an observable quantity, thereby alleviating the classical degeneracy between the incident ionizing flux and the geometric distance of the irradiated gas clouds. This formalism facilitates a more reliable calibration against empirical data from the X-ray-selected sample.

The model grid systematically varies the following principal parameters, consistent with the parameter space explored in \citetalias{2026MNRAS.548ag560A}:
\begin{itemize}
\item X-ray luminosity: $\log \Lx = 38, 40, 42, 44, 46, 48$.
\item Gas-phase metallicity: $Z/Z_{\odot} = 0.2, 0.5, 0.75, 1.0, 2.6$.
\item Electron density: $10^2\lesssim N_{\rm e}\left[{\rm cm}^{-3}\right]\lesssim 10^4$.
\item Spectral index: $\aox = -2.0, -1.7, -1.4, -1.1, -0.8$.
\end{itemize}

To explicitly contextualize our framework with standard photoionization models in the literature, we note that our explored X-ray luminosity range of $10^{38}$ to $10^{48}$ erg s$^{-1}$ corresponds to an ionization parameter range spanning approximately $\log U = -3.5$ to $-1.0$. This effectively encompasses the typical physical conditions observed within the high-ionization zones of NLRs \citep{2006A&A...447..863N, 2009A&A...503..721M,2020MNRAS.492.5675C, 2025A&A...696A.229P}.
The spectral energy distribution (SED) of the ionizing source is characterized by the power-law index $\alpha_{\rm ox}$, defined as the slope between monochromatic luminosities at $2500\,\text{\AA~(UV)}$ and $2\,\text{keV~ (X-ray)}$ \citep{1979ApJ...234L...9T, 1981ApJ...245..357Z}. While direct observations of AGN often derive values softer than $-1.2$ \citep[e.g.][]{1981ApJ...245..357Z,2011ApJ...726...20M,2011ApJS..196....2S}, standard photoionization frameworks consistently find that an index of $\alpha_{\rm ox} \approx -1.4$ is required to successfully reproduce observed optical emission-line intensities \citep[e.g.][]{2006MNRAS.371.1559G,2006A&A...458..405G, 2017MNRAS.468L.113D}. Consequently, we adopt this as our baseline while exploring a grid of spectral indices spanning $\aox \in [-2.0, -0.8]$, which encompasses the range utilized in established NLR abundance studies \citep[e.g.][]{2004ApJS..153....9G, 2004ApJS..153...75G}.

We also explore the impact of the characteristic peak temperature of the Big Blue Bump ($T_{\rm BB}$) component of the AGN ionizing continuum, varying it from $1.0\times10^{4}$ K to $3.0\times10^{5}$ K. Our fiducial models use $T_{\rm BB} = 1.0 \times 10^5$ K. For a comprehensive description of the AGN ionizing continuum, elemental abundance scaling (based on relations from \citealt{2022MNRAS.514.5506D} and \citealt{2024MNRAS.534.3040D}), and other fixed nebular parameters (such as the dust-free assumption and plane-parallel geometry) are detailed in \S~\textcolor{blue}{3} of \citetalias{2026MNRAS.548ag560A}.

\begin{figure*}
    \centering
\includegraphics[width=1.0\textwidth]{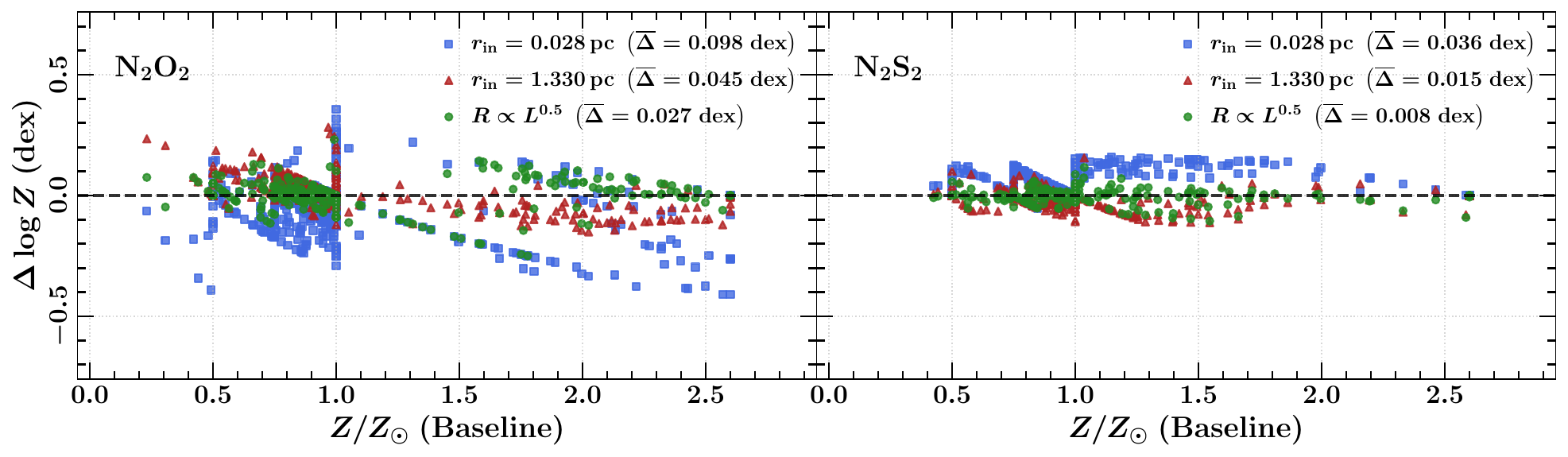}
\vspace{-10pt} 
\caption{Logarithmic residuals of derived gas-phase metallicity ($\Delta \log Z$ in dex) resulting from variations in the inner radius ($r_{\mathrm{in}}$) of the photoionized cloud for the $\no$ (left panel) and $\ns$  (right panel) diagnostics. The residuals are plotted as a function of the baseline metallicity derivation ($r_{\mathrm{in}} = 0.30$ pc). 
To evaluate the extreme spatial boundaries of the GRAVITY sample, the derived metallicities were analytically recalculated assuming fixed constant radii of $r_{\mathrm{in}} = 0.028$ pc (blue squares) and $r_{\mathrm{in}} = 1.330$ pc (red triangles).
The green circles represent the scenario where $r_{\mathrm{in}}$ is continuously coupled to the AGN luminosity ($R \propto L^{0.5}$), effectively evaluating the grid at a constant ionization parameter ($U$) corresponding to a pivot luminosity of $\Lx = 10^{43.5}$ erg s$^{-1}$. The dashed horizontal line indicates zero residual. The mean absolute deviations $\left(\overline{\Delta}\right)$ provided in the legend demonstrate that incorporating the $R \propto L^{0.5}$ correlation significantly reduces systematic bias across the sample compared to assuming unscaled extreme spatial boundaries.
}
\label{fig_1}
\end{figure*}

To rigorously evaluate the spatial boundaries of the simulated clouds, we adapt the geometric scaling analysis introduced in \citetalias{2026MNRAS.548ag560A}.When the extreme, fixed spatial boundaries of the \citet{2024A&A...690A..76G} sample are adopted without any luminosity-dependent scaling, the inferred metallicities exhibit an exceptionally stable baseline relative to conventional optical diagnostics. As shown in Figure~\ref{fig_1}, the maximum absolute deviations for $\no$ range from $-0.410$ to $+0.358$ dex, whereas the $\ns$ index shows even tighter constraints, with deviations restricted to the interval $-0.112$ to $+0.160$ dex.

At the extreme inner boundary ($r_{\mathrm{in}} = 0.028$ pc), these unscaled limits introduce only a modest mean (median) systematic bias of $\sim0.098$ (0.076) dex for $\no$ and merely $0.036$ (0.000) dex for $\ns$. At the outer boundary ($r_{\mathrm{in}} = 1.330$ pc), the corresponding mean (median) offsets decrease further to $\sim0.045$ (0.035) dex and $\sim0.015$ (0.000) dex, respectively. Despite this inherent robustness, the residual systematic dispersion is further minimized when the baseline radius (i.e. $r_{\rm in} = 0.30$ pc) is self-consistently coupled to the ionizing radiation field via the $R \propto L^{0.5}$ relation. Mapping the sample onto our pivot coordinate effectively eliminates the explicit geometric dependence, thereby reducing the average systematic offsets to negligible levels. Under this continuous spatial scaling, the final mean (median) absolute differences are only $\sim0.027$ (0.008) dex for $\no$ and an almost negligible $\sim0.008$ (0.000) dex for $\ns$.

\subsection{Observational data}
\label{bass_data}

Our empirical analysis uses the same BASS DR2 subset \citep[][]{2022ApJS..261....1K,2022ApJS..261....4O} defined in \citetalias{2026MNRAS.548ag560A}. This hard X-ray–selected (14–195 keV) parent sample is largely unaffected by the obscuration biases common in optical surveys. We define a Seyfert~2 sample, incorporating intermediate Sy~1.8 and 1.9 objects, whose weak broad Balmer components render them spectrally analogous to true Type~2 sources \citep[e.g.][]{1978ApJ...223...56K,1981ApJ...249..462O} by requiring  $\mathrm{FWHM} < 1000$ km s$^{-1}$ for the Balmer recombination lines and applying standard diagnostic diagrams \citep{1981PASP...93....5B,2001ApJ...556..121K,2011MNRAS.413.1687C}. This yields 617 Seyfert 2s, from which we select sub-samples for the $\no$ ($N_{\rm L} = 426$) and $\ns$ ($N_{\rm L} = 417$) diagnostics sources with reliable measurements of all lines used in the \S~\ref{results} diagnostic diagrams. See \S~\textcolor{blue}{2} of \citetalias{2026MNRAS.548ag560A} for full data details, including electron density distributions.

To align our empirical flux measurements with the theoretical photoionization grids, we de-reddened the data prior to calculating the diagnostic ratios. Following the analytical framework presented by \citet{2023MNRAS.520.1687A}, the intrinsic emission-line intensities were recovered from the attenuated observables via the standard formalism $F^{\rm obs}_{\lambda} = F_{\lambda} \times 10^{-0.4A_{\lambda}}$, where $A_{\lambda}$ defines the line-of-sight extinction profile. This dereddening procedure utilized the canonical \citet{1989ApJ...345..245C} Galactic extinction curve, adopting a total-to-selective extinction value of $R_V = A_V/E(B-V) = 3.1$ along with a fiducial unattenuated Balmer decrement of $\rm H\alpha/H\beta = 2.86$. Such an assumption is predicated on standard Case B recombination in nebular environments ($T \approx 10^4$ K, $N_{\rm e} \sim 10^2$ cm$^{-3}$; \citealt{2006agna.book.....O}) and is consistent with our sample properties (see \citetalias{2026MNRAS.548ag560A} for details). While an AGN-enhanced intrinsic Balmer decrement of $\rm H\alpha/H\beta = 3.1$ is consistent with previous studies from photoionization models
 \citep[e.g.][]{1982PhDT.........4H,1983ApJ...264..105F,1984A&A...131..159P} and observational determinations \citep[e.g.][]{1984ApL....24...43G,1984PASP...96..393G}, particularly to account for collisional excitation effects within the extended PIZ \citep[e.g.][]{1983ApJ...264..105F,1983ApJ...269L..37H}, we maintain $\rm H\alpha/H\beta = 2.86$ for consistency with our baseline photoionization grids. 

\section{Results}
\label{results}
\begin{figure*}
    \centering
    \includegraphics[width=0.94\textwidth]{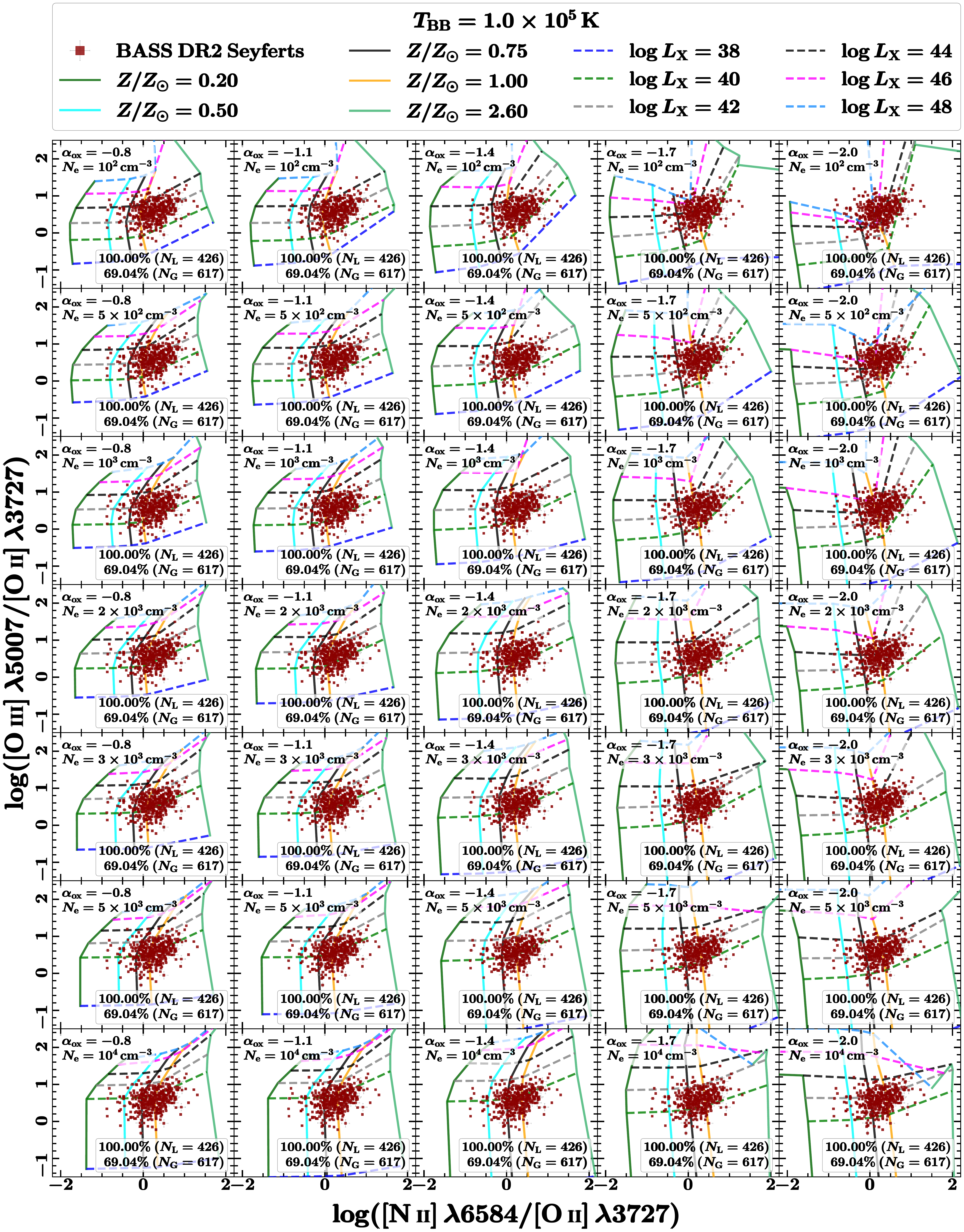}
    \caption{Diagnostic diagram of $\log(\oiii\lambda5007/\oii\lambda 3727$) versus $\log(\nii\lambda6584/\oii\lambda 3727$) from the models and the observational sample of Seyfert 2s. Each panel represents a grid of photoionization models stratified by specific values of the ionizing spectral index ($\aox$) and logarithm of electron density (\Ne\ [$\mathrm{cm}^{-3}$]), as indicated in the top-left corner of each subplot. The rows correspond to $\aox$ values of $-2.0, -1.7, -1.4, -1.1,$ and $-0.8$ (from bottom to top), while columns correspond to \Ne~values from $10^{2}$ to $10^{4}\,\mathrm{cm}^{-3}$ (top to bottom). All models assume a blackbody temperature $T_{\mathrm{BB}} = 100\,000 \, \mathrm{K}$  for the ionizing continuum. Within each panel, solid lines represent models of constant metallicity ($Z/Z_{\odot}=0.20, 0.50, 0.75, 1.00,$ and $2.60$; see legend), varying with $\log \Lx$. Dashed lines represent models of constant X-ray luminosity ($\log \Lx = 38, 40, 42, 44, 46, 48$; see legend), varying with metallicity. The red filled squares show the observational data from the BASS DR2. Note that for harder ionizing continua ($\aox \gtrsim -1.4$), the grid vectors are nearly orthogonal, facilitating robust calibration, whereas soft continua ($\aox \lesssim -1.7$) lead to severe degeneracies where the metallicity and luminosity vectors collapse onto each other.
    }  
    \label{fig_2}
\end{figure*}

\begin{figure*}
    \centering
    \includegraphics[width=1.0\textwidth]{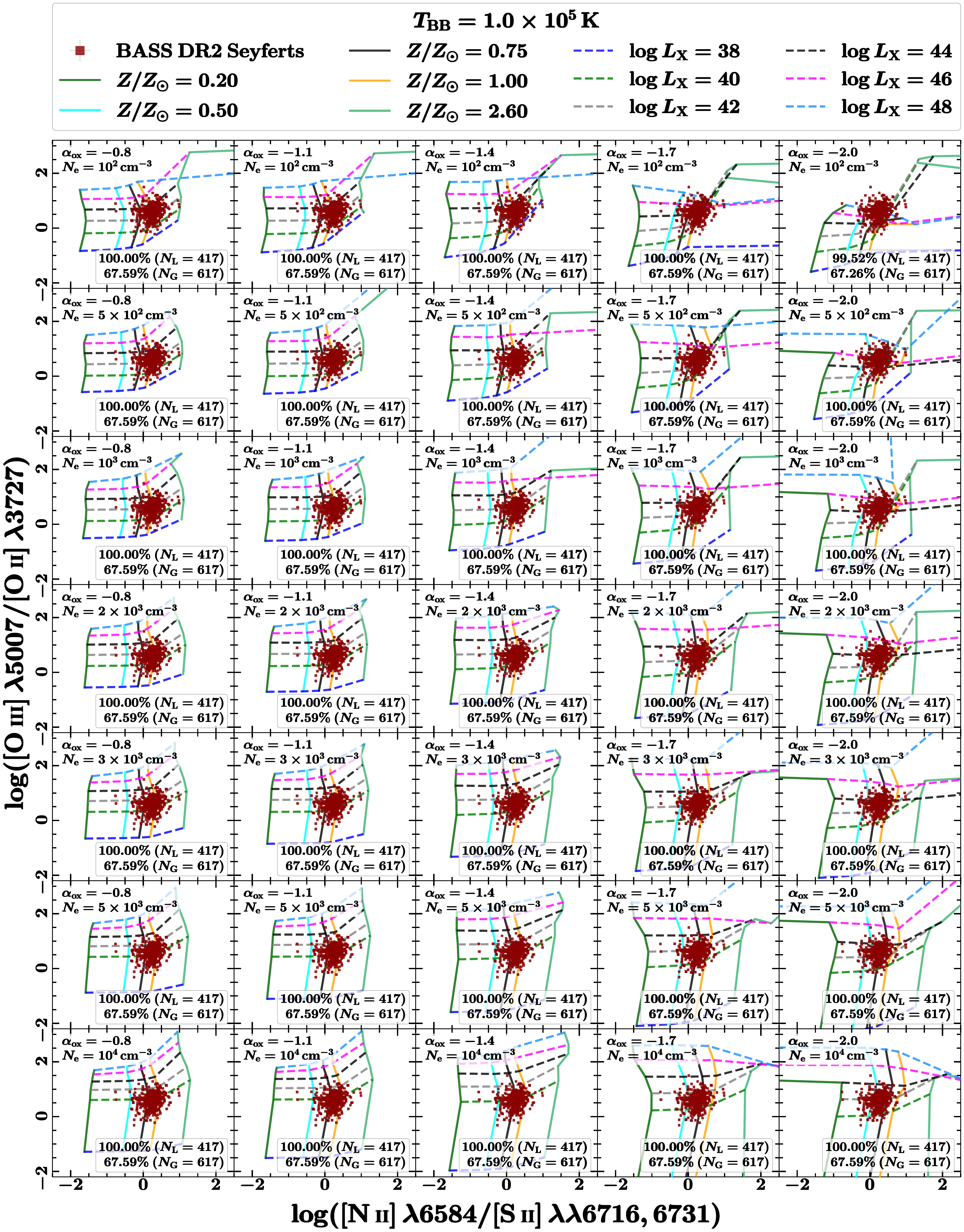}
    \caption{Same as Figure~\ref{fig_2}, but for the diagnostic diagram of $\log(\oiii\lambda5007/\oii\lambda 3727$) versus $\log(\nii\lambda6584/\sii\lambda\lambda 6716,6731$). 
    }
    \label{fig_3}
\end{figure*}

We extend our investigation of the physical conditions in the NLRs of AGNs  by employing optical diagnostic diagrams that utilize the $\no$ and $\ns$ indices. We compare the predictions from our extensive grid of photoionization models with the observational data from the BASS DR2 sample. In Figures~\ref{fig_2} and \ref{fig_3}, each panel shows a grid of photoionization models defined by a specific combination of ionizing spectral index ($\aox$) and electron density (\Ne). Within each grid, the dependence of the emission-line ratios (LRs) on gas-phase metallicity and X-ray luminosity is represented by solid and dashed lines, respectively.

\subsection{The {\texorpdfstring{\boldmath $\no$}{N2O2}} diagnostic diagram}
\label{nii_oii}
In Figure~\ref{fig_2}, we present the diagnostic diagram comparing $\otot$ versus $\no$. This diagram is particularly effective because the horizontal axis, the $\no$ index, is defined as a ratio of emission lines arising from ions with very similar ionization potentials, specifically N$^{+}$ (14.53 eV) and O$^{+}$ (13.62 eV). Owing to this similarity, the index exhibits only a weak dependence on the ionization state of the gas, thereby providing a robust metallicity diagnostic over a broad metallicity range \citep{2000ApJ...542..224D, 2002ApJS..142...35K}. The vertical axis, $\otot$, remains a sensitive tracer of the ionization parameter.

The structure of the photoionization model grids is dependent on the assumed ionizing spectral index, $\aox$, which fundamentally dictates the physical extent of the partially ionized zone \citep[e.g.][]{2004ApJS..153....9G}. For harder continua (e.g. $\aox = -0.8$, left column), the grid vectors for metallicity (solid lines) and X-ray luminosity (dashed lines) are nearly orthogonal. This orthogonality is ideal for calibration, as it allows a clear disentangling of the two parameters; the $\no$ index is almost purely a function of metallicity, while the $\otot$ ratio is a near-pure tracer of $\Lx$. Conversely, as the ionizing continuum softens (i.e. $\aox$ approaches $-2.0$, see the right column), the model grids become highly compressed and degenerate. The vectors for metallicity and $\Lx$ are no longer independent and become nearly parallel, sloping from the lower-left to the upper-right. This “overlapping” of the grid lines introduces a significant degeneracy, making it impossible to uniquely determine a metallicity or luminosity for an object observed in this part of the parameter space.  

Therefore, the observational data from the BASS DR2 sample (red squares) are well-described by models across the full range of electron densities explored (from $N_{\rm e} = 10^2$ to $10^4 \text{ cm}^{-3}$). Specifically, the bulk of the sample is clearly bounded between the $Z/Z_{\odot} = 0.75$ and $Z/Z_{\odot} = 2.60$ tracks. This confirms that the vast majority of the data points cluster firmly in the near-solar to super-solar regime ($Z/Z_{\odot} \gtrsim 1.0$), with a conspicuous scarcity of sources at sub-solar metallicities.
In alignment with the structural grid behaviours mapped in \citetalias{2026MNRAS.548ag560A}, the choice of the $\aox$ heavily dictates the diagnostic efficacy of the indices. While models adopting hard-to-intermediate continua ($\aox \in [-1.4, -0.8]$) generate well-defined, orthogonal grid vectors that sufficiently span the locus of the BASS DR2 observations, softer SEDs ($\aox \lesssim -1.7$) induce structural collapse. This strong degeneracy in the soft-continuum regime prevents stable metallicity derivations; therefore, to ensure the utmost precision, the final semi-empirical calibrations are strictly constrained to models featuring harder spectral indices.

\subsection{The \texorpdfstring{\boldmath $\ns$}{N2S2} diagnostic diagram}
In Figure~\ref{fig_3}, we present the diagnostic diagram of $\otot$ versus $\ns$. The $\ns$\ index is a robust tracer of the $\rm N/S$ abundance ratio. As nitrogen abundance increases with metallicity due to secondary nucleosynthetic production, while sulphur remains a primary $\alpha$-element, the $\rm N/S$ ratio is expected to correlate strongly with the overall gas-phase metallicity \citep{2016Ap&SS.361...61D}.

We find a consistent result with the $\no$ diagnostic: the observational data from the BASS DR2 sample (red squares) are well-described by models across the full range of electron densities explored (from $N_{\rm e} = 10^2$ to $10^4 \text{ cm}^{-3}$). Similarly, the bulk of the sample is bounded between the $Z/Z_{\odot} = 0.75$ and $Z/Z_{\odot} = 2.60$ tracks, confirming that the sample is dominated by near-solar to super-solar metallicities ($Z/Z_{\odot} \gtrsim 1.0$). The choice of ionizing continuum, $\aox$, again has a significant effect on the structure of the grid.  Models with harder continua (e.g. $\aox \gtrsim -1.4$) produce well-defined grids that are sufficiently spread to span the locus of the BASS data. In stark contrast, models with the softest ionizing continua (e.g. $\aox \lesssim -1.7$) exhibit a severe compression of the grid lines and profound structural degeneracies, particularly at $\log([\mathrm{N}\,\textsc{ii}]/[\mathrm{S}\,\textsc{ii}]) > 0$. 
In this soft-continuum regime, the iso-metallicity and iso-luminosity tracks become highly entangled, inherently precluding any reliable extraction of metallicities. Consequently, to maintain the structural integrity of our semi-empirical relations, all subsequent $\no$ calibrations in \S~\ref{smetal} are strictly confined to the stable parameter space defined by the harder continuum models ($\aox \in [-1.4, -0.8]$).

\section{Towards a new metallicity calibration for AGNs}
\label{smetal}
We derive the final metallicity calibrations for the $\no$\ and $\ns$\ indices using the quantitative analysis approach detailed in \citetalias{2026MNRAS.548ag560A}. This process involves interpolating the position of each observed BASS DR2 galaxy within our two-dimensional model grids (Figures~\ref{fig_2} and \ref{fig_3}) to estimate its metallicity. We subsequently fit a continuous exponential function to the resulting set of ($x_{\text{LR}}$, $Z_{\text{est}}$) data points, where $x_{\rm LR}$ represents the given logarithmic emission-line ratio and $Z_{\rm est}$ is the interpolated metallicity estimate, to derive a one-dimensional calibration.

As established in \citetalias{2026MNRAS.548ag560A}, and confirmed in our analysis of Figures~\ref{fig_2} and \ref{fig_3}, the models with the softest ionizing continua ($\aox = -2.0, -1.7$) introduce significant degeneracies due to grid compression that prevent a robust interpolation of metallicity. Consequently, only the models with harder ionizing continua ($\aox = [-1.4, -1.1, -0.8]$) were used to derive the final calibrations presented in Figure~\ref{fig_4}. This selection ensures the stability and reliability of the derived relations.

Figure~\ref{fig_4} maps the parameter dependencies of the $\no$\ and $\ns$\ diagnostics across our model grids. Overall, both calibrations demonstrate a high degree of stability. Most notably, they are remarkably robust against the severe $\Lx$-driven degeneracies that biased the $\ntwo$ and $\otnt$ indices in \citetalias{2026MNRAS.548ag560A}, as evidenced by the tight clustering of model points around the global fits. However, while generally stable, the models do exhibit specific, localized sensitivities to secondary parameters, such as the $\aox$ and \Ne, particularly at extreme metallicities. A comprehensive, parameter-by-parameter analysis of these specific sensitivities for each index is provided in \S~\ref{n202_index} and \S~\ref{n2s2_index} below.

Figure~\ref{fig_5} quantifies the high precision of the $\no$\ and $\ns$\  indices by presenting the global calibrations derived from the complete model grid. As shown, the interpolated observational data (grey squares) tightly track the global best-fit exponential curves (solid black lines) across the entire metallicity range. The coefficients for the global fits are provided in the bottom row of Table~\ref{table_1}. Note that the apparent horizontal steps in the parameter contours are numerical artifacts resulting from the 2D interpolation over our discrete model grid, and do not represent physical limitations or abrupt transitions in the ionizing continuum. To rigorously evaluate this stability, the inset panels show the residual distributions ($\log(Z/Z_{\odot})_{\rm grid} - \log(Z/Z_{\odot})_{\rm fit}$) for each diagnostic. A comparison with the $\ntwo$ and $\otnt$ diagnostics reveals a marked reduction in intrinsic scatter. We derive root-mean-square (RMS) dispersions of just $0.081$~dex for $\no$\ and $0.121$~dex for $\ns$. This factor-of-two improvement relative to the standard optical diagnostics discussed in \citetalias{2026MNRAS.548ag560A} (which yielded dispersions of $\approx 0.22$ dex for $\ntwo$ and $\approx 0.20$ dex for $\otnt$) underscores the advantage of employing line ratios that originate from similar ionization zones, thereby minimizing sensitivity to the hardness of the AGN power-law.

For the $\no$\ and $\ns$\  diagnostics, we adopt an exponential parameterization. The general functional form is given by
\begin{equation}
    (Z/ {\rm Z_{\odot}}) = \sum_{n=0}^{N} c_n \mathcal{F}_n(x)
    \label{exp_calibration}
\end{equation}
where $N=1$, $x$ is the logarithmic line ratio, and the basis function $\mathcal{F}_n(x)$ is defined as:
\begin{equation}
    \mathcal{F}_n(x) = \delta_{n,0} + \delta_{n,1} \cdot c_n^{x-1}
\end{equation}
This formulation recovers the exponential relation, thus, $Z/Z_{\odot} = c_1^x + c_0$, where $c_1$ represents the base and $c_0$ represents the offset.

\subsection{Calibration of the \texorpdfstring{\boldmath $\no$}{N2O2} index}
\label{n202_index}

\begin{figure*}
    \centering
\includegraphics[width=1.0\textwidth]{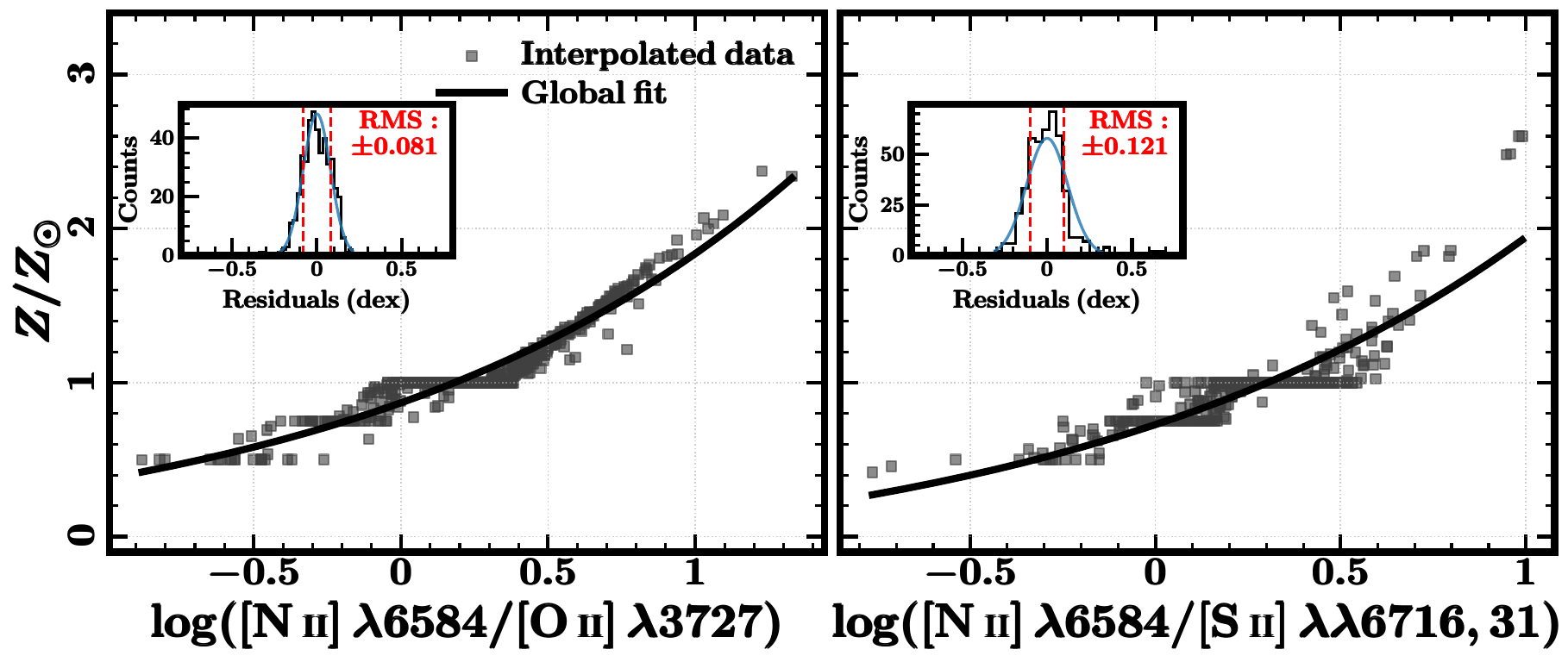}
\vspace{-12pt} 
\caption{Global metallicity calibrations for the $\no$\ (left) and $\ns$\  (right) indices. The grey squares represent the interpolated metallicity estimates ($Z_{\rm est}$) for the observational sample derived from the full 2D photoionization model grid. The solid black curves trace the global best-fit exponential functions (Equation~\ref{eqn_rms}). The inset panels show the residual distributions ($\log(Z/Z_{\odot})_{\rm grid} - \log(Z/Z_{\odot})_{\rm fit}$), with the vertical dashed red lines indicating the root-mean-square (RMS) dispersion. The solid blue curve in the inset panels represents a standard Gaussian distribution fit to the residuals. 
The coefficients for the global fits are provided in the bottom row of Table~\ref{table_1}.
}
\label{fig_5}
\end{figure*}

In the left panel of Figure~\ref{fig_5}, we present the calibration relating the estimated metallicity $Z/Z_{\odot}$ to the $\no = \log(\nii\lambda6584/\oii\lambda3727)$ index. 
The global relation, derived from all stable model points and represented by the dashed black line, is described by the following relation:
 \begin{eqnarray}
       \begin{array}{l@{}l@{}l}
(Z/{\rm Z_{\odot}})\, & = \,  &  c_1^x + c_0,   \\
       \end{array}
\label{cal_1}
\end{eqnarray}
where $ x = \no $.
A strong positive correlation is evident, establishing the $\no$ index as a reliable metallicity tracer for AGNs. This relation is driven by the secondary production of nitrogen, which increases the $\rm N/O$ ratio at high metallicity. 
To quantify the statistical precision of our semi-empirical calibrations, we calculate the RMS dispersion of the residuals. The RMS dispersion is quantified by the following relation:
\begin{equation}
\mathrm{RMS} = \sqrt{ \frac{1}{N} \sum_{i=1}^{N} \left[ \log(Z/Z_{\odot})_{{\rm grid}, i} - \log(Z/Z_{\odot})_{{\rm fit}, i} \right]^2 }
\label{eqn_rms}
\end{equation} 
where $N$ is the total number of observational data points, $\log(Z/Z_{\odot})_{{\rm grid}, i}$ is the metallicity derived via 2D interpolation of the photoionization grid, and $\log(Z/Z_{\odot})_{{\rm fit}, i}$ is the predicted metallicity evaluated from our 1D global best-fit calibration curves. As quantified by Equation~\ref{eqn_rms} and shown in Figure~\ref{fig_5}, this derived calibration is characterized by a remarkably tight RMS dispersion of $\sim 0.081$ dex over the index range $-0.88 \lesssim \no \lesssim 1.33$. This indicates that the $\no$ index serves as a highly precise tracer of metallicity across the modeled AGN parameter space. However, while the exponential trend is highly stable, we note that the interpolated data points at the extreme low-metallicity end ($\no \lesssim -0.5$) are less densely populated, consistent with the overwhelming concentration of the observational sample in the near-solar to super-solar metallicity regime (see Figure~\ref{fig_2} and \ref{fig_3}).

\begin{figure*}
    \centering
\includegraphics[width=1.0\textwidth]{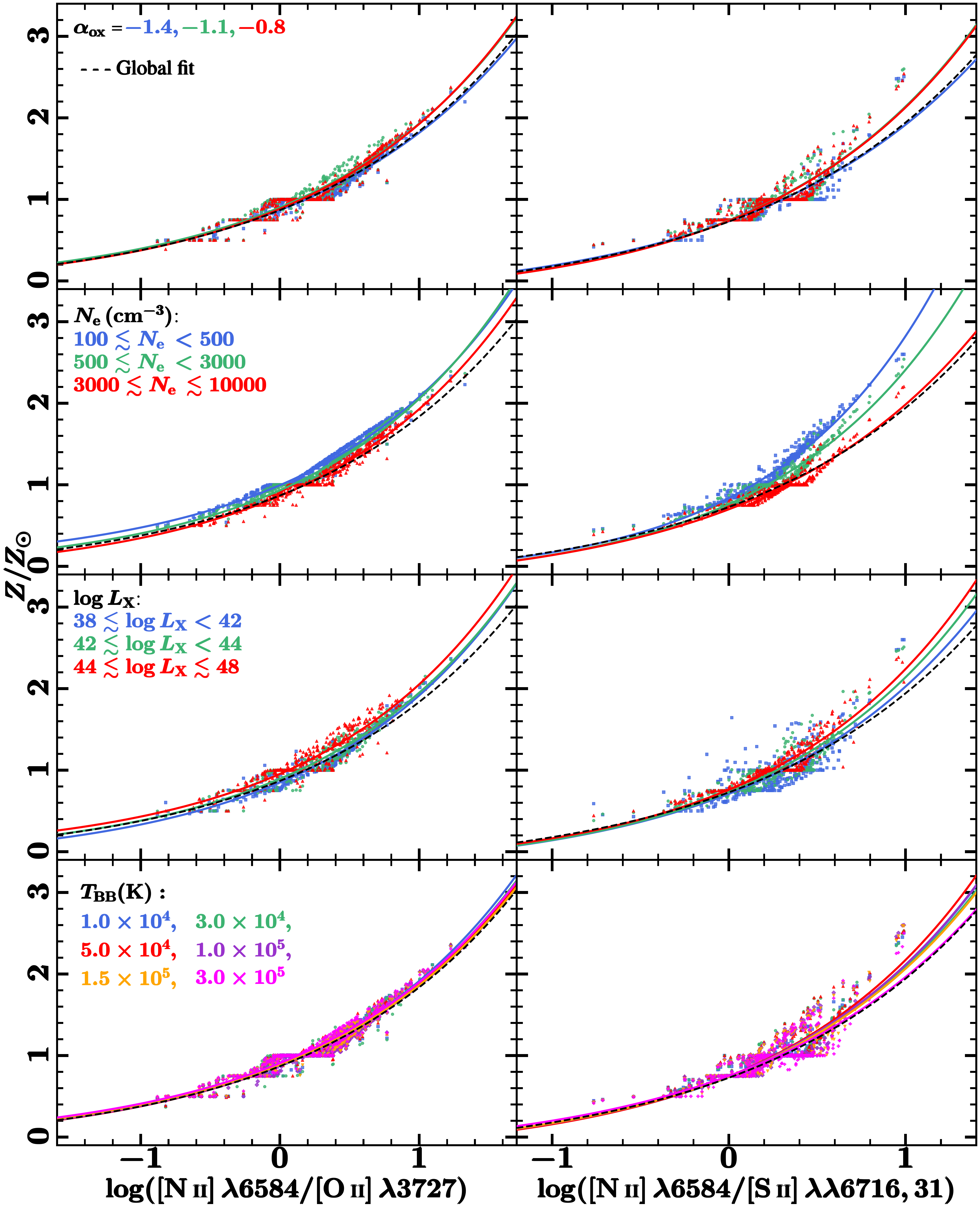}
\vspace{-10pt} 
    \caption{Metallicity calibrations for the $\no$\ and $\ns$\  strong-line indices. \textit{Left column:} Metallicity ($Z/Z_{\odot}$) as a function of the $\log(\nii\lambda6584/\oii\lambda3727)$ ratio ($\no$). \textit{Right column:} Metallicity as a function of the $\log(\nii\lambda6584/\sii\lambda\lambda6716,6731)$ ratio ($\ns$). In both columns, each panel explores the effect of a different model parameter on the estimated metallicity. 
The colored markers represent the interpolated metallicity estimate ($Z_{\rm est}$) obtained for each observational data point from a subset of the model grid, while the lines are the corresponding  exponential fits (see Equation~\ref{exp_calibration}).
The dashed black line in each panel represents the global fit using all model data. From top to bottom, the rows are stratified by the $\aox$, the \Ne, $\log \Lx$, and the $T_{\rm BB}$. 
The coefficients for each fit are presented in Table~\ref{table_1}.}
    \label{fig_4}
\end{figure*}

A detailed inspection of the parameter dependencies in Figure~\ref{fig_4} (left column) reveals the specific sensitivities of the $\no$ diagnostic.  
Regarding the X-ray luminosity ($\Lx$) for the $\no$ index, this dependence is not considered dominant because it lacks significant systematic vertical offsets throughout the entire calibration range ($-0.88 \lesssim \no \lesssim 1.33$); instead, much like the divergences driven by other model parameters (e.g. $\aox$, \Ne, $T_{\rm BB}$), it becomes prominent primarily in the high-metallicity regime ($\no > 0.5$). We observe that the vertical separation between calibration curves for distinct luminosity ranges is regime-dependent. At lower index values ($\no \lesssim 0.5$), all three luminosity bins are clearly separated. However, at higher index values ($\no > 0.5$), a minor separation emerges. This magnitude of separation applies primarily where the low- and intermediate-luminosity bins ($38 \lesssim \log \Lx < 42$ and $42 \lesssim \log \Lx < 44$) remain tightly collapsed together, but clearly separate and diverge upward from the high-luminosity bin ($44 \lesssim \log \Lx \lesssim 48$). Similar to the $\otnt$ index, in this high-metallicity regime, more luminous AGNs yield systematically higher estimated metallicities compared to their lower-luminosity counterparts. Overall, this separation is significantly less pronounced in $\no$ than the severe dependencies found in the $\ntwo$ and $\otnt$ diagnostics.

For the other secondary parameters ($\aox$, \Ne, and $T_{\rm BB}$), the diagnostic exhibits a distinct “fanning out” or curve separation specifically at the high-metallicity regime ($Z/Z_{\odot} \gtrsim 1.0$ or $\no \gtrsim 0.5$). For the ionizing continuum, $\aox$, the stratified curves for the harder spectral indices ($\aox = -1.1$ and $-0.8$) are nearly completely collapsed onto each other, but visibly separate and deviate upward from the softer $\aox = -1.4$ curve. Also, at high metallicities, the curve representing the lowest density bin ($100 \lesssim N_{\rm e} < 500 \text{ cm}^{-3}$) visibly detaches and diverges steeply upward from the intermediate and high-density curves. Finally, for the Big Blue Bump Temperature ($T_{\rm BB}$), while the stratified curves remain tightly collapsed onto the global fit at low to moderate metallicities (demonstrating exceptional stability in this regime), a similar divergence occurs at the extreme high-metallicity end. Here, models with lower peak temperatures ($T_{\rm BB} \lesssim 3.0 \times 10^4 \text{ K}$) diverge steeply upward from the global fit, while the highest temperature models trend slightly below it.

Because of these high-abundance divergences, if the specific electron density or spectral shape of an extremely metal-rich target is known, future studies may benefit from using the parameter-specific coefficients from Table~\ref{table_1} rather than the global fit.

\subsection{Calibration of the \texorpdfstring{\boldmath $\ns$}{N2S2}  index}
\label{n2s2_index}
The right panel of Figure~\ref{fig_5} presents the calibration for the $\ns = \log(\nii\lambda6584/\sii\lambda\lambda6716,6731)$ index. 
The global relation derived from all model points is expressed by the following relation:
 \begin{eqnarray}
       \begin{array}{l@{}l@{}l}
(Z/{\rm Z_{\odot}})\, & = \,  &  c_1^x + c_0,   \\
       \end{array}
\label{cal_2}
\end{eqnarray}
where $ x = \ns $.
The calibration shows a clear exponential relation, consistent with the complex interplay between the increasing $\rm N/S$  abundance ratio and the temperature effects at different metallicities. Similarly, this derived calibration yields an RMS dispersion of $\sim 0.121$ dex (see Equation~\ref{eqn_rms} and Figure~\ref{fig_5}) over the index range $-0.76\lesssim \ns \lesssim 0.99$. The slightly higher scatter compared to $\no$ reflects the complex dependence of the $\sii$ lines on secondary parameters, though the calibration remains robust for the majority of the AGN population.

A systematic assessment of Figure~\ref{fig_4} (right column) outlines the model parameter dependencies of the $\ns$ diagnostic.
Regarding the X-ray luminosity ($\Lx$) for the $\ns$ index, the vertical separation between calibration curves exhibits a contrasting regime-dependent behaviour. At lower index values ($\ns \lesssim 0.0$), the luminosity curves are effectively inseparable and the systematic offset is negligible, with the different luminosity bins well-mixed and following a single global trend. However, the separation becomes more apparent at higher metallicity regimes ($\ns > 0.0$), where all three luminosity bins are distinctly separated from one another. Similar to the $\otnt$ and $\no$ indices, in this high-metallicity regime, more luminous AGNs ($44 \lesssim \log \Lx \lesssim 48$) yield systematically distinct higher estimated metallicities compared to their lower-luminosity counterparts ($38 \lesssim \log \Lx < 42$ and $42 \lesssim \log \Lx < 44$). While this separation is noticeably more apparent than the divergence observed in the $\no$ diagnostic, it remains significantly less pronounced than the severe $\Lx$ dependencies characterizing the $\ntwo$ and $\otnt$ diagnostics.
  Similar to the $\no$ index but slightly more pronounced, the $\aox = -1.1$ and $-0.8$ curves clearly separate and diverge downward from the softer $\aox = -1.4$ curve at high metallicities ($Z/Z_{\odot} \gtrsim 1.5$). As clearly shown in the second row of Figure~\ref{fig_4}, the fitted curves stratified by $N_e$ exhibit a much greater separation compared to the $\no$ diagnostic. As metallicity increases ($Z/Z_{\odot} \gtrsim 1.0$), the curve representing the highest density bin $\left({\rm i.e.}~ 3\times10^3\lesssim N_{\rm e}\left[{\rm cm}^{-3}\right]\lesssim 10^4\right)$ visibly detaches and deviates downward from the intermediate- and low-density curves. This indicates a pronounced, albeit secondary, dependence on \Ne\ for the $\ns$ index, driven by the low critical densities of the $\sii$ doublet. While relatively stable, the stratified curves for different $T_{\rm BB}$ values show a minor but visible “fanning out” or separation at the extreme high-metallicity end ($Z/Z_{\odot} \gtrsim 1.5$), contrasting with the tighter $T_{\rm BB}$ curves seen in the $\no$ diagnostic at those specific metallicities.

Consequently, if the electron density (or specific spectral shape) of a target is known, it is beneficial to use the parameter-specific coefficients from Table~\ref{table_1} rather than the global fit to minimize these residual separations.

\section{Discussion}
\label{discussion}
\subsection{The influence of \texorpdfstring{\boldmath $\aox$}{aox} and extreme observational regimes}
The shape of the AGN SED is classically parameterized by the spectral index $\aox$, defined as the two-point power-law index between monochromatic luminosities at 2500 Å and 2 keV \citep{1979ApJ...234L...9T, 1981ApJ...245..357Z}. This index governs the hardness of the ionizing radiation field, exhibiting an indirect relation with the local ionization parameter \citep{2025A&A...696A.229P}. While the applicability of our semi-empirical calibrations is explicitly constrained to models featuring harder ionizing continua ($\aox \gtrsim -1.4$), direct estimations from X-ray to UV observations frequently yield significantly softer values. The typical observed range for AGN ionizing spectral index spans a broad range of $-2.0 \lesssim \aox \lesssim -0.8$ 
across cosmic time ($z \sim 0 - 7.5$), reflecting the immense intrinsic diversity of the extreme ultraviolet (EUV) continuum \citep[e.g.][]{1981ApJ...245..357Z, 1994ApJS...92...53W, 2005AJ....130..387S, 2006AJ....131.2826S, 2007ApJ...665.1004J, 2011ApJ...726...20M, 2011ApJS..196....2S, 2019A&A...630A.118V, 2021RNAAS...5..101T, 2021RAA....21....4Z}. Moreover, it is well-established that the shape of the AGN continuum depends strongly on the intrinsic luminosity of the source; AGNs with higher ultraviolet/optical luminosities systematically exhibit softer (more negative $\aox$) X-ray-to-optical spectral slopes \citep{1979ApJ...234L...9T, 1981ApJ...245..357Z, 2006AJ....131.2826S, 2010A&A...512A..34L}. 
This highlights a fundamental and well-documented discrepancy in AGN astrophysics: the $\aox$ values derived directly from continuum observations are systematically softer than the $\aox$ values strictly required by photoionization models to reproduce the optical emission-line spectra of the NLRs. 

Detailed photoionization model efforts, where the spectral index is treated as a free fitting parameter, have demonstrated that it is generally not possible to reproduce observational emission-line data using soft continua with $\aox < -1.2$ \citep[e.g.][]{2004ApJS..153....9G, 2004ApJS..153...75G, 2006MNRAS.371.1559G, 2006A&A...458..405G, 2016MNRAS.456.3354F, 2017MNRAS.468L.113D, 2020MNRAS.492.5675C, 2025A&A...696A.229P}. As shown in our diagnostic grids (Figures~\ref{fig_2} and \ref{fig_3}), models adopting soft SEDs ($\aox \lesssim -1.7$) undergo a severe structural collapse. The iso-metallicity and iso-luminosity tracks become highly degenerate, and the models entirely fail to span the locus of the BASS DR2 observational data. This persistent discrepancy between required model parameters and direct continuum observations remains a long-standing open problem in AGN astrophysics, which may be linked to the “ionizing photon deficit” or the energy budget problem in AGNs \citep[e.g.][]{1981ApJ...251..465M,1985ApJ...289..451N,1993PASP..105.1150B}, wherein the observed continuum is seemingly insufficient to power the observed nebular recombination lines. Moreover, variations in $\aox$ significantly alter the local radiation pressure distribution and heating balance, as steeper (softer) power laws reduce the high-energy photons needed to maintain the high-ionization species observed in the NLRs of Seyferts \citep[e.g.][]{2004ApJS..153....9G, 2014MNRAS.438..901S}. Consequently, for highly luminous AGN hosts with explicitly measured soft ionizing continua ($\aox \lesssim -1.7$), strong-line indices are intrinsically incapable of disentangling chemical abundance from the ionization state due to this structural collapse. For such targets, we strongly caution against the use of any strong-line metallicity calibrations (whether empirical, semi-empirical, or theoretical). Instead, metallicities in these extreme regimes should be constrained via direct \Te-based derivations or through comprehensive Bayesian forward-modeling of the full optical spectrum.

This fundamental limitation extends to other extreme or observationally constrained regimes. For instance, utilizing $\Lx$ to directly couple the photoionization grids to the physical properties of the NLR, our framework provides a more robust path to metallicity than traditional strong-line methods when $\Lx$ is known. Additionally, when the intrinsic $\Lx$ is entirely unavailable, traditional strong-line diagnostics are inherently degenerate. As detailed in \S~\ref{interp}, this degeneracy arises because traditional indices conflate genuine chemical abundance variations with the $\Lx$-driven volumetric expansion of the partially ionized zone. In such cases, the atomic symmetry of the $\no$ and $\ns$ indices provides a highly robust alternative, as their reduced sensitivity to the underlying ionization state inherently mitigates the absence of direct $\Lx$ parameterization. Furthermore, the application of strong-line calibrations derived from local ($z \sim 0$) Seyfert samples to high-redshift targets requires extreme caution. As demonstrated by studies of high-redshift galaxy evolution \citep[e.g.][]{2016Ap&SS.361...61D,2020MNRAS.491.1427S,2023MNRAS.526.3504H,2024ApJ...962...24S}, the ionization conditions and interstellar medium properties inherently evolve across cosmic time. Applying local calibrations to sources at Cosmic Noon or the Epoch of Reionization can drive significant systematic biases.  When rest-frame optical transitions such as the $\oii$ doublet shift entirely out of accessible observational bands at these high redshifts, alternative diagnostic frameworks must be adopted. In these distant regimes, utilizing UV emission-line ratios becomes highly effective. Specifically, deploying UV emission line diagnostic diagrams, such as C$\iv\lambda1549/$C\iii]$\lambda1909$ versus C$\iv\lambda1549/$He$\ii\lambda$1640 \citep[e.g.][]{2019MNRAS.486.5853D,2025MNRAS.540.1608D} and C$\iv\lambda1549/$He$\ii\lambda$1640 versus C\iii]$\lambda1909/$C$\iv\lambda1549$ \citep[e.g.][]{2006A&A...447..863N, 2009A&A...503..721M} offer a powerful mechanism to break structural degeneracies and isolate gas-phase metallicities for distant AGNs.

\subsection{Physical interpretation of the calibrations}
\label{interp}
\begin{figure*}
    \centering
\includegraphics[width=0.45\textwidth]{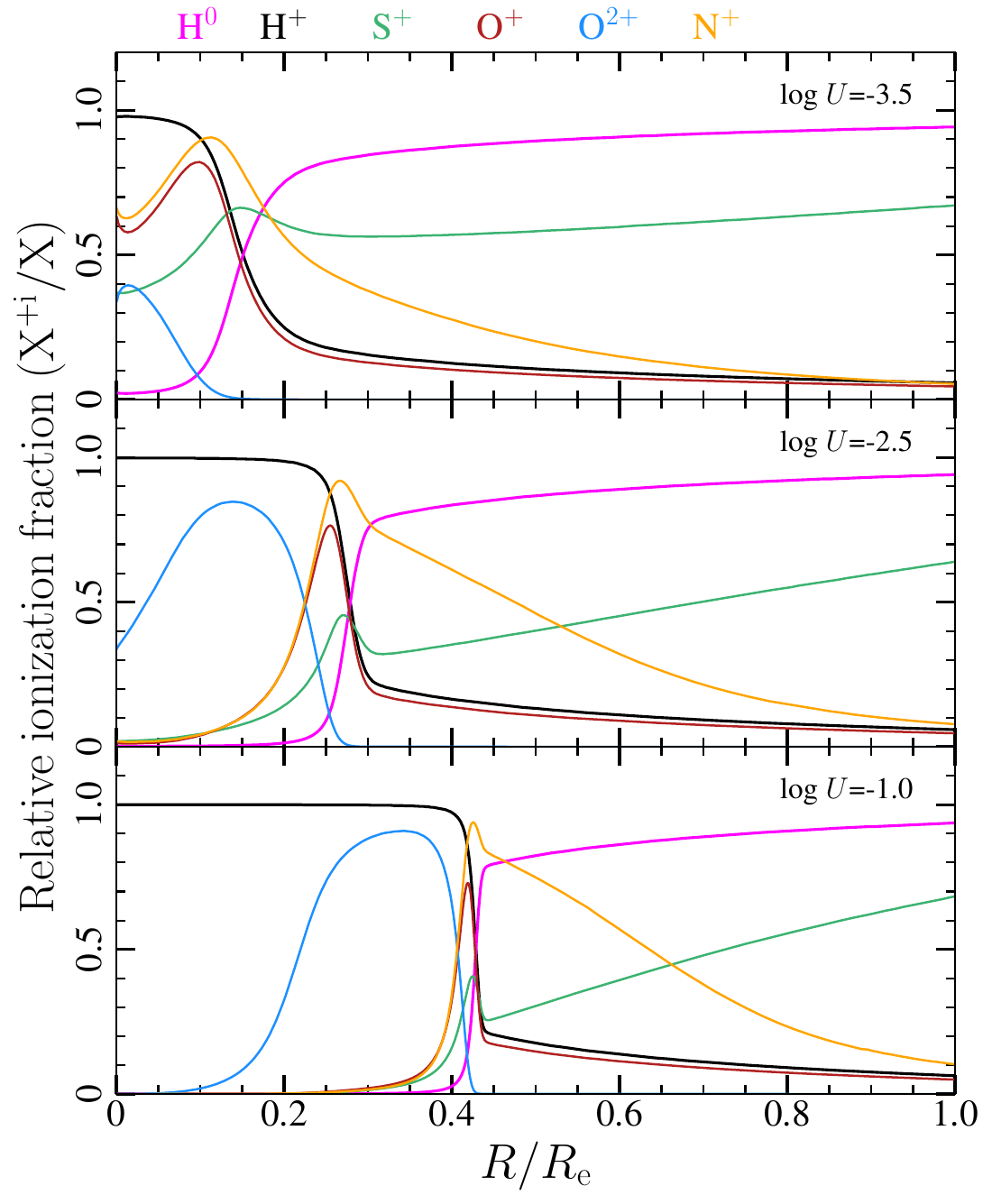}
\includegraphics[width=0.45\textwidth]{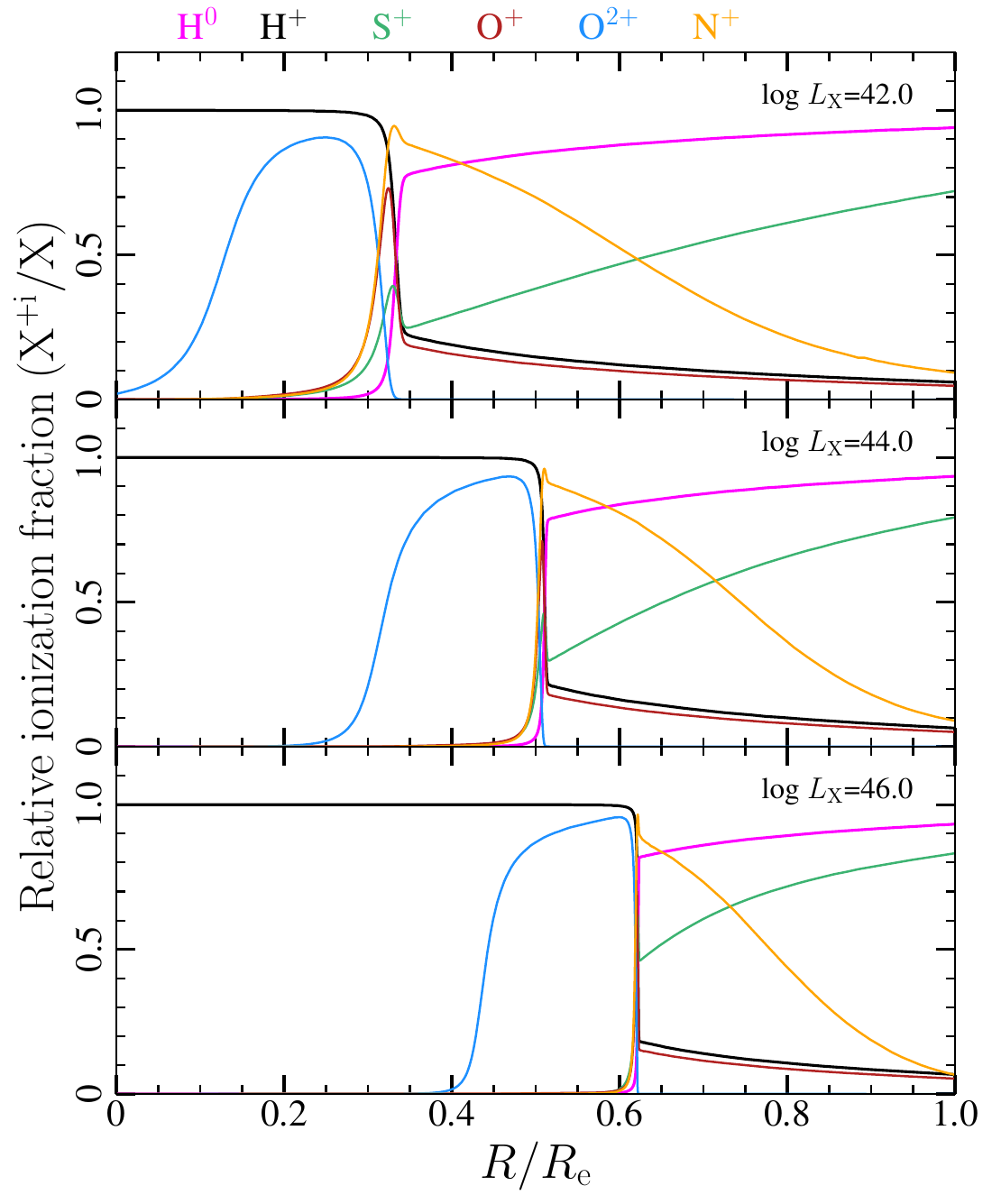}
\caption{Relative ionic abundance fraction for distinct ions ($\rm X^{+i} /X$) versus the nebular radius ($R$) from the innermost gas region normalized by the
outermost radius ($R_{\rm e}$) predicted by the photoionization models simulating NLRs of AGNs. The nebular parameters assumed in the models are spherical geometry, an inner radius of 2 pc, solar metallicity, $\aox = -1.1$, and the assumed power-law electron density profile. The left panels show models parameterized by the ionization parameter, with three values for $\log U = -3.5$, $-2.5$, and $-1.0$ (top, middle, and bottom rows, respectively). The right panels show models parameterized by the 2-10 keV X-ray luminosity, with three values for $\log \Lx = 42.0$, $44.0$, and $46.0$ (top, middle, and bottom rows, respectively).}
\label{fig_6}
\end{figure*}

To justify the use of the observable 2-10 keV X-ray luminosity, $\Lx$, as a primary calibration parameter in place of the unobservable $U$, we investigated the predicted ionization structure of key nebular ions, specifically H$^+$, H$^0$, O$^{2+}$, O$^+$, N$^+$, and S$^+$, within the NLRs. We constructed detailed photoionization models adopting the baseline nebular parameters utilized in our diagnostic grids: a spherical geometry with an inner radius of 2 pc, solar metallicity, and a SED characterized by a representative mean slope of $\aox = -1.1$ for the harder ionizing continuum of Seyfert 2 galaxies in the local universe ($z < 0.1$). To accurately simulate the stratification of the NLR gas, we assumed a density profile normalized to initial electron density, $N_0 = 10^4 \text{ cm}^{-3}$, at the inner radius, following the power law\footnote{For a detailed description of the density profile, see \citet{2018ApJ...856...46R}.} $N_{\rm e}(r) = N_0 \left(r^{-0.5}\right)$, where $r$ is the radial distance from the AGN. The resulting ionization structures are presented in Figure~\ref{fig_6}, which contrasts models parameterized by the traditional ionization parameter \citep[$\log U = -3.5, -2.5, -1.0$; see][and references therein]{2020MNRAS.492.5675C, 2025A&A...696A.229P} against those parameterized by intrinsic X-ray luminosity (i.e. $\log \Lx = 42.0, 44.0, 46.0$).

A comparison of the ionization structures reveals a distinct difference in how the fully ionized inner zone responds to changes in $U$ versus $\Lx$. In the models parameterized by $U$ (Figure~\ref{fig_6}, left panels), increasing the ionization parameter significantly shifts the hydrogen ionization front deeper into the cloud. At low ionization parameters ($\log U = -3.5$), low-ionization species such as O$^+$ peak very early, near a normalized radius of 0.1, whereas at high $U$ ($\log U = -1.0$) the higher flux of ionizing photons per atom sustains a fully ionized H$^+$ and O$^{2+}$ zone out to a normalized radius of $\sim 0.4$. In contrast, the luminosity-parameterized models (Figure~\ref{fig_6}, right panels) exhibit a remarkable self-similarity in the fully ionized zone; the relative position of the hydrogen ionization front remains consistent across three orders of magnitude in luminosity ($38 \lesssim \log \Lx \lesssim 46$). This stability arises because scaling the intrinsic luminosity while fixing the inner radius effectively scales the incident flux and the physical size of the cloud proportionally; thus, when normalized by the total cloud depth ($r/r_{\rm max}$), the inner transition zone dominated by H$^+$ and O$^{2+}$ appears structurally uniform. 

However, the most significant physical divergence occurs in the extended PIZ, or the “outer tail” of the cloud, which has profound implications for metallicity diagnostics. As the X-ray luminosity increases, the harder X-ray photons, which have mean free paths significantly longer than those of EUV photons penetrate far beyond the hydrogen ionization front into the neutral gas \citep[e.g.][]{1980ApJ...242.1041H, 1981ApJ...250..478K}. This energetic radiation creates an extensive, X-ray heated zone that naturally drives the strong emission of collisionally excited low-ionization lines, independent of shock heating \citep[][]{1983ApJ...264..105F}. Accounting for this deep X-ray continuum absorption is strictly required to successfully reproduce the complex, multi-component low-ionization spectra observed in Seyfert galaxies \citep[][]{1984A&A...131..159P}. Within this extended PIZ, low-ionization species such as $\mathrm{S^+}$ and $\mathrm{N^+}$ coexist with neutral hydrogen \citep[$\mathrm{H^0}$; e.g.][]{2004ApJS..153....9G,2019ARA&A..57..511K}. Furthermore, modern physical constraints demonstrate that the specific column densities and structural depths where this X-ray heated tail dominates $\nii$ and $\sii$ emissions are intrinsically coupled to the incident AGN radiation pressure \citep[e.g][]{2014MNRAS.438..901S}. In the high-luminosity models (e.g. $\log \Lx = 46.0$), this PIZ is substantially more voluminous compared to the $U$-parameterized models, resulting in extended tails of S$^+$ and N$^+$ emission that persist deep into the cloud. While the relative shape of these tails remains roughly self-similar across different luminosities, their absolute contribution to the integrated flux of low-ionization lines increases significantly with the hardness and intensity of the incident field.

This luminosity-driven extension of the PIZ creates a degeneracy for which standard metallicity calibrations fail to account. Because collisionally excited lines emitted from this outer zone, specifically $\nii\lambda6584$ and $\sii\lambda\lambda6716, 6731$, are naturally enhanced by the larger volume of the X-ray heated PIZ in highly luminous AGNs, an observed increase in these line ratios can be misinterpreted as an increase in chemical abundance. Standard diagnostics that scale strictly with U cannot distinguish between a true metallicity increase and this luminosity-driven enhancement. Consequently, to break this degeneracy, calibrations utilizing standard diagnostic ratios (such as $\ntwo$ and $\otnt$, as identified in \citetalias{2026MNRAS.548ag560A}) must explicitly parameterize the abundance derivation against $\Lx$.

The principal advantage of the $\no$ and $\ns$ diagnostics arises from their shared atomic symmetry, which inherently circumvents the luminosity-driven expansion of the PIZ identified in \citetalias{2026MNRAS.548ag560A}. Highly luminous AGNs are capable of propagating hard X-ray photons deep into the predominantly neutral gas, artificially boosting collisionally excited emission lines originating from this extended region. Conventional metallicity diagnostics are unable to disentangle this geometrical (volume) enhancement from a genuine increase in chemical abundance. 
In contrast, $\text{N}^+$ and $\text{O}^+$ are co-spatial and occupy essentially the same nebular volume. Any $\Lx$-driven expansion of the PIZ therefore amplifies the emissivity of both ions in a nearly identical manner. Taking their ratio effectively removes the dependence on the absolute physical scale of the emitting cloud.

The distinct performance of the $\no$ and $\ns$ indices is regulated by the combined effects of ionization stratification and collisional de-excitation. Both diagnostics employ line ratios with closely matched ionization potentials, thereby reducing degeneracies related to the shape and intensity of the incident ionizing radiation field ($\Lx$). However, the $\no$ ratio exhibits greater robustness against variations in nebular density. This robustness is primarily a consequence of the close correspondence between the ionization potentials of $\nii$ (14.53 eV) and $\oii$ (13.62 eV), which ensures a high degree of spatial co-existence within the NLR. By contrast, the $\sii$ doublet (10.36 eV) preferentially traces more extended, partially ionized regions located deeper within the gas cloud.

Fundamentally, because these transitions originate from ions with closely matched ionization potentials, they trace co-spatial emitting volumes. Consequently, any $\Lx$-driven volumetric expansion of the X-ray heated tail induces a symmetric enhancement in the integrated emissivities of both the numerator ($\nii$) and the denominator (e.g. $\oii$ or $\sii$), allowing the line ratio to cleanly cancel out this geometrical dependence. However, the critical densities ($N_{\rm c}$) of the relevant transitions differ substantially \citep[e.g.][]{2012MNRAS.427.1266V}. The $\nii$ lines remain in their low-density, linear emissivity regime up to $N_{\rm c} \approx 8.7 \times 10^{4}$ cm$^{-3}$, whereas the $\sii$ and $\oii$ doublets exhibit significantly lower critical densities, $N_{\rm c} \approx 1.5 - 3.9 \times 10^{3}$ cm$^{-3}$ and $4.5 \times 10^{3}$ cm$^{-3}$, respectively \citep[][]{1986ApJ...301..727D}. As a result, the $\no$ index is largely insensitive to variations in \Ne, since the $\nii$ and $\oii$ lines undergo collisional de-excitation in a closely synchronized fashion at high densities. In contrast, the $\ns$ index is affected by differential collisional de-excitation: the $\sii$ denominator is quenched at densities substantially below those at which the $\nii$ numerator is significantly suppressed. This leads to an enhanced sensitivity to \Ne\ and manifests as the larger residual scatter ($\sim 0.121$ dex) observed in our $\ns$ calibration.

Despite this secondary density sensitivity, both $\no$ and $\ns$ provide a fundamental advantage over standard optical diagnostics involving high-ionization lines (e.g. $\ntwo$). They are inherently robust against variations in $\Lx$, eliminating the primary source of systematic error associated with the diverse ionizing radiation fields of AGN host galaxies and making them reliable metallicity tracers.

\begin{table}
\centering
\setlength{\tabcolsep}{1.5pt} 
\caption{The values of the $c_1$ and $c_0$ coefficients and their $1\sigma$ uncertainties resulting from the exponential curve fits (Equation~\ref{exp_calibration}) obtained from the estimations in Figure~\ref{fig_4}. The fits are provided for the $\no$ and $\ns$ diagnostics, stratified by different model parameter subsets. The final row, Global fit, lists the coefficients derived from the entire model grid without parameter discrimination.}
\label{table_1}
\begin{tabular}{@{}l | r@{$\pm$}l r@{$\pm$}l | r@{$\pm$}l r@{$\pm$}l@{}}
\toprule
\multicolumn{1}{c}{} & \multicolumn{4}{c}{\textbf{$\no$}} & \multicolumn{4}{c}{\textbf{$\ns$}} \\
\hline
Parameter & \multicolumn{2}{c}{$c_1$} & \multicolumn{2}{c|}{$c_0$} & \multicolumn{2}{c}{$c_1$} & \multicolumn{2}{c}{$c_0$} \\
\hline
\multicolumn{1}{l}{$\alpha_{\rm ox}$} & \multicolumn{2}{c}{} & \multicolumn{2}{c}{} & \multicolumn{2}{c}{} & \multicolumn{2}{}{}  \\
$-1.40$ & 1.948 & 0.010 & -0.132 & 0.010 & 2.185 & 0.040 & -0.269 & 0.010 \\
$-1.10$ & 2.028 & 0.020 & -0.098 & 0.010 & 2.397 & 0.040 & -0.258 & 0.010 \\
$-0.80$ & 2.041 & 0.010 & -0.116 & 0.010 & 2.391 & 0.030 & -0.263 & 0.010 \\
\\
\multicolumn{1}{l}{$N_{\rm e}\, ({\rm cm^{-3}})$} & \multicolumn{2}{c}{} & \multicolumn{2}{c}{} & \multicolumn{2}{c}{} & \multicolumn{2}{c}{}  \\
 $\in [100, 500)$ & 2.082 & 0.010 & -0.005 & 0.010 & 2.988 & 0.020 & -0.164 & 0.010 \\
$\in [500, 3\,000)$ & 2.123 & 0.010 & -0.068 & 0.010 & 2.648 & 0.020 & -0.238 & 0.010 \\
$\in [3\,000, 10\,000]$ & 2.066 & 0.010 & -0.138 & 0.010 & 2.285 & 0.020 & -0.301 & 0.010 \\
\\
\multicolumn{1}{l}{$\log L_{\mathrm{X}}$} & \multicolumn{2}{c}{} & \multicolumn{2}{c}{} & \multicolumn{2}{c}{} & \multicolumn{2}{c}{} \\
$\in [38, 42)$  & 2.067 & 0.010 & -0.151 & 0.010 & 2.318 & 0.050 & -0.292 & 0.010 \\
$\in [42, 44)$ & 2.056 & 0.020 & -0.106 & 0.010 & 2.410 & 0.040 & -0.270 & 0.010 \\
$\in [44, 48]$ & 2.099 & 0.030 & -0.046 & 0.010 & 2.480 & 0.030 & -0.239 & 0.010 \\
\\
\multicolumn{1}{l}{$T_{\rm BB}$ [K]} & \multicolumn{2}{c}{} & \multicolumn{2}{c}{} & \multicolumn{2}{c}{} & \multicolumn{2}{c}{} \\
$1.0 \times 10^{4}$ & 2.030 & 0.010 & -0.116 & 0.010 & 2.369 & 0.030 & -0.264 & 0.010 \\
$3.0 \times 10^{4}$ & 2.006 & 0.010 & -0.118 & 0.010 & 2.340 & 0.030 & -0.263 & 0.010 \\
$5.0 \times 10^{4}$ & 1.993 & 0.010 & -0.125 & 0.010 & 2.428 & 0.040 & -0.255 & 0.010 \\
$1.0 \times 10^{5}$ & 1.978 & 0.010 & -0.123 & 0.010 & 2.369 & 0.040 & -0.254 & 0.010 \\
$1.5 \times 10^{5}$ & 1.975 & 0.010 & -0.113 & 0.010 & 2.324 & 0.040 & -0.259 & 0.010 \\
$3.0 \times 10^{5}$  & 1.990 & 0.010 & -0.093 & 0.010 & 2.215 & 0.040 & -0.251 & 0.010 \\
\hline
Global fit & 1.968 & 0.014 & -0.132 & 0.005 & 2.216 & 0.037 & -0.273 & 0.008 \\
\bottomrule
\end{tabular}
\end{table}

\subsection{Inter-comparison of the metallicity diagnostics}
\begin{figure*}
    \centering
    \includegraphics[width=1.0\textwidth]{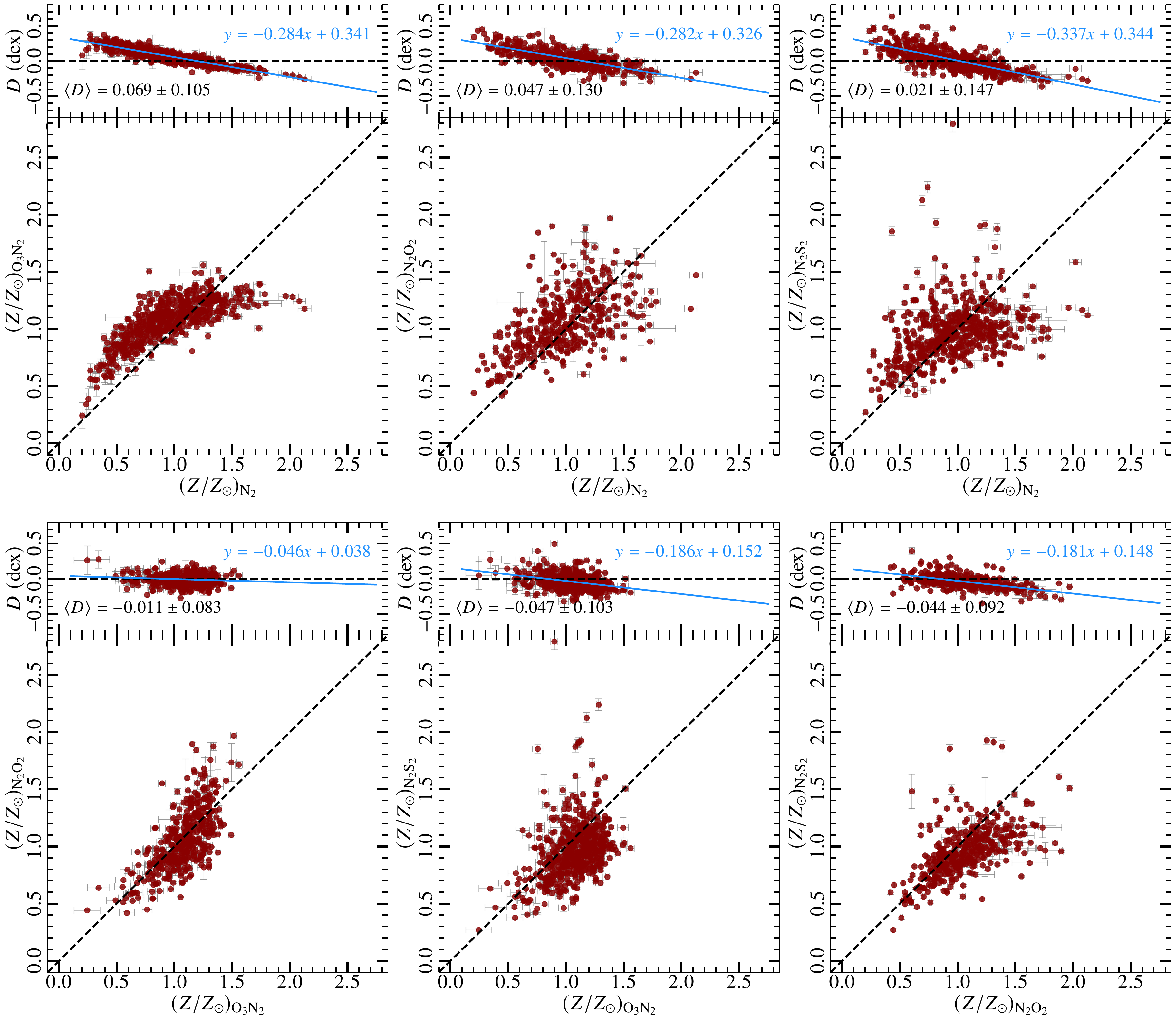}
    \vspace{-12pt} 
    \caption{Comparison of metallicity estimates derived from our four strong-line semi-empirical calibrations. The top row compares the $\ntwo$ diagnostic (from \citetalias{2026MNRAS.548ag560A}) against the $\otnt$ (\citetalias{2026MNRAS.548ag560A}), $\no$ (this work), and $\ns$ (this work) indices. The $\ntwo$ diagnostic shows the largest scatter and systematic deviations, characterized by steep slopes in the linear regression fits to the residuals (solid blue lines), confirming its residual sensitivity to ionization parameter variations. The bottom row compares the three most robust diagnostics against each other. The comparison between $\otnt$ and $\no$ (bottom-left) reveals the highest consistency, with a near-zero mean offset ($\langle D \rangle = -0.011 \pm 0.083$ dex) and a remarkably flat residual trend ($y = -0.046x + 0.038$), indicating they trace a nearly identical metallicity scale. The $\ns$ diagnostic (bottom-middle and right) shows a minor systematic offset relative to the others, explicitly attributed to the secondary density dependence of the [S II] lines. In each column, the bottom panel shows the one-to-one relation (dashed black line), while the top panel shows the logarithmic residuals in dex $\left[D = \log(Z/Z_{\odot})_{\text{ordinate}} - \log(Z/Z_{\odot})_{\text{abscissa}}\right]$. The mean difference ($\langle D \rangle$), its standard deviation, and the equation of the linear regression fit are indicated in each residual panel.}
   \label{fig_7}
\end{figure*}

{
Figure~\ref{fig_7} and Table~\ref{diag_tab} evaluate the internal consistency between the four strong-line metallicity calibrations established in our framework: the $\Lx$--dependent $\ntwo$ and $\otnt$ indices (from \citetalias{2026MNRAS.548ag560A}) and the $\Lx$--insensitive $\no$ and $\ns$ indices in this work. By applying all four calibrations to the same sample of galaxies ($N \approx 393-558$ depending on the available emission line), we can assess their agreements and identify potential systematic biases. To rigorously quantify these differences across our entire framework (for both internal and external literature comparisons), we evaluate the logarithmic residuals $D = \log(Z/Z_{\odot})_{\mathrm{ordinate}} - \log(Z/Z_{\odot})_{\mathrm{abscissa}}$ for each pair of diagnostics. The systematic shifts are estimated by calculating the mean offset, $\langle D \rangle$, which corresponds to the logarithm of the geometric mean ratio of the derivations:  
\begin{equation}
\langle D \rangle = \log \left[ \left( \prod_{i=1}^{N} \frac{(Z/Z_\odot)_{\mathrm{Literature}, i}}{(Z/Z_\odot)_{\mathrm{This~work}, i}} \right)^{\frac{1}{N}} \right]. 
\label{eqn_diff}
\end{equation} Additionally, we compute the standard deviation ($\sigma$) of these residuals to estimate the overall scatter.

The top row of Figure~\ref{fig_7}  compares the $\ntwo$  diagnostic against the other three. While correlations are evident, the $\ntwo$ index exhibits the largest scatter and systematic deviations across all comparisons. First, in the comparison between $\ntwo$ and $\otnt$ (top-left panel), we observe a clear systematic deviation ($y = -0.284x + 0.341$; $\langle D \rangle = 0.069 \pm 0.105$ dex). For the comparison between $\ntwo$ and the robust $\no$ diagnostic (top-middle panel), the residuals reveal a strong systematic trend, quantified by the linear fit $y = -0.282x + 0.326$ and a mean offset of $\langle D \rangle = 0.047 \pm 0.130$ dex. This steep slope indicates that $\ntwo$ systematically underpredicts metallicity at the low-$Z$ end and overpredicts at the high-$Z$ end relative to $\no$. This behaviour is similarly pronounced in the comparison for $\ntwo$ versus $\ns$ (top-right panel), where the systematic trend reaches a slope of $y = -0.337x + 0.344$ $\left(\langle D \rangle = 0.021 \pm 0.147\, \text{dex} \right)$. These trends collectively confirm that the $\ntwo$ index retains a significant residual sensitivity to the ionization parameter.  

In contrast, the bottom row demonstrates the remarkable stability of the diagnostics that control for ionization effects. The comparison between the $\otnt$ and $\no$ calibrations (bottom-left panel) reveals the highest degree of consistency in our framework. The analysis of the residuals shows a nearly flat linear trend ($y = -0.046x + 0.038$) and a negligible mean offset of $\langle D \rangle = -0.011 \pm 0.083$ dex. This near-zero slope and offset indicate that $\otnt$ (when explicitly corrected for $\Lx$ as prescribed in \citetalias{2026MNRAS.548ag560A}) and $\no$ (inherently robust) are essentially tracing the exact same metallicity scale across the entire AGN population. The stability of the $\otnt$ index in non-star-forming regimes is firmly corroborated by empirical studies \citep[e.g.][]{2017MNRAS.466.3217Z, 2019MNRAS.485..367K,2019MNRAS.489.4721V}. Because low-ionization emission regions (LIERs) and diffuse ionized gas (DIG) share the extended partially ionized zones characteristic of AGN environments, its demonstrated resilience against contamination in these regions further supports its application to Seyfert galaxies. Finally, we assess the $\ns$ diagnostic. When compared to $\no$ (bottom-right panel), the $\ns$ index introduces a residual trend ($y = -0.181x + 0.148$) and a minor systematic offset where $\ns$ yields values lower by $-0.044 \pm 0.092$ dex. A similar deviation and offset are found relative to $\otnt$ in the bottom-middle panel ($y = -0.186x + 0.152$ and mean offset $-0.047 \pm 0.103$ dex). We explicitly attribute these persistent, $\ns$-driven deviations to the secondary density dependence of the $\sii$ lines discussed in \S~\ref{interp}. Despite these shifts, the monotonic agreement remains strong, confirming that $\ns$ is a reliable alternative when data from the blue spectroscopic coverage is unavailable, though $\no$ remains the most robust and stable calibration. 
}

\begin{table}
\centering
\setlength{\tabcolsep}{1.5pt}
\caption{Statistical parameters for the metallicity comparisons shown in Figures~\ref{fig_7} and \ref{fig_8}. The linear fits represent the residual trends $\left[D = \log(Z/Z_{\odot})_{\text{ordinate}} - \log(Z/Z_{\odot})_{\text{abscissa}}\right]$, expressed as $y = {\rm m}x + {\rm c}$.}
\label{diag_tab}
\resizebox{\columnwidth}{!}{%
\begin{tabular}{lcccccc}
\toprule
\multirow{2}{*}{Comparison ($y$ vs. $x$)} & \multicolumn{2}{c}{Linear Fit} & \multicolumn{2}{c}{Pearson} & \multicolumn{2}{c}{Spearman} \\
\cmidrule(lr){2-3} \cmidrule(lr){4-5} \cmidrule(lr){6-7}
 & m & c & $r$ & $p$-value & $\rho$ & $p$-value \\
\midrule
\multicolumn{7}{c}{\textit{Internal Comparisons}} \\
\midrule
$\ntwo$ vs. $\otnt$   & $-0.284$ & $0.341$ & $0.755$ & $5.20 \times 10^{-104}$ & $0.771$ & $8.55 \times 10^{-111}$ \\
$\ntwo$ vs. $\no$   & $-0.282$ & $0.326$ & $0.547$ & $3.77 \times 10^{-32}$  & $0.573$ & $9.44 \times 10^{-36}$  \\
$\ntwo$ vs. $\ns$   & $-0.337$ & $0.344$ & $0.362$ & $2.47 \times 10^{-18}$  & $0.427$ & $1.19 \times 10^{-25}$  \\
$\otnt$ vs. $\no$ & $-0.046$ & $0.038$ & $0.677$ & $2.02 \times 10^{-54}$  & $0.672$ & $1.85 \times 10^{-53}$  \\
$\otnt$ vs. $\ns$ & $-0.186$ & $0.152$ & $0.384$ & $1.18 \times 10^{-20}$  & $0.408$ & $2.10 \times 10^{-23}$  \\
$\no$ vs. $\ns$ & $-0.181$ & $0.148$ & $0.583$ & $1.23 \times 10^{-37}$  & $0.637$ & $3.87 \times 10^{-46}$  \\
\midrule
\multicolumn{7}{c}{\textit{Literature Comparisons}} \\
\midrule
\Te\ vs. $\no$ & $-0.564$ & $0.516$ & $-0.176$ & $5.42 \times 10^{-03}$ & $-0.167$ & $8.12 \times 10^{-03}$ \\
\Te\ vs. $\ns$ & $-0.635$ & $0.425$ & $-0.088$ & $1.26 \times 10^{-01}$ & $-0.107$ & $6.28 \times 10^{-02}$ \\
\SBcalib{F1} vs. $\no$ & $-0.275$ & $-0.115$ & $0.501$ & $2.57 \times 10^{-26}$ & $0.498$ & $5.58 \times 10^{-26}$ \\
\SBcalib{F1} vs. $\ns$ & $-0.318$ & $-0.065$ & $0.369$ & $5.81 \times 10^{-19}$ & $0.406$ & $4.86 \times 10^{-23}$ \\
\SBcalib{F2} vs. $\no$ & $0.080$ & $0.057$ & $0.741$ & $7.80 \times 10^{-62}$ & $0.772$ & $4.88 \times 10^{-70}$ \\
\SBcalib{F2} vs. $\ns$ & $-0.027$ & $0.191$ & $0.560$ & $6.45 \times 10^{-30}$ & $0.622$ & $1.81 \times 10^{-38}$ \\
\citetalias{2020MNRAS.492.5675C} vs. $\no$ & $0.031$ & $-0.118$ & $0.540$ & $3.08 \times 10^{-31}$ & $0.573$ & $9.44 \times 10^{-36}$ \\
\citetalias{2020MNRAS.492.5675C} vs. $\ns$ & $-0.080$ & $0.016$ & $0.355$ & $1.07 \times 10^{-17}$ & $0.427$ & $1.19 \times 10^{-25}$ \\
\citetalias{2021MNRAS.507..466D} vs. $\no$ & $-0.574$ & $0.374$ & $-0.164$ & $1.21 \times 10^{-03}$ & $-0.229$ & $5.44 \times 10^{-06}$ \\
\citetalias{2021MNRAS.507..466D} vs. $\ns$ & $-0.547$ & $0.255$ & $-0.080$ & $6.53 \times 10^{-02}$ & $-0.121$ & $4.99 \times 10^{-03}$ \\
\bottomrule
\end{tabular}%
}
\end{table}

\subsection{Comparison with literature calibrations}
We assess the statistical precision of our new calibrations by calculating the residual dispersion of the model grid points around the best-fit polynomial functions. We derive RMS dispersions of $\sim 0.081$ dex for the $\no$ index and $\sim 0.121$ dex for the $\ns$ index. Notably, these dispersions are significantly lower than those obtained for the standard optical diagnostics discussed in \citetalias{2026MNRAS.548ag560A}, where the $\ntwo$ and $\otnt$ indices yielded comparable $1\sigma$ dispersions of $\sim 0.22$ dex and $\sim 0.20$ dex, respectively.  This improvement highlights the robustness of the $\no$ and $\ns$ diagnostics against secondary parameters in the AGN environment. The $\no$ index, in particular, benefits from the similar ionization potentials of the $\nii$ and $\oii$ lines, which makes the ratio less sensitive to variations in the $U$ and the hardness of the ionizing radiation field \citep[e.g.][]{2002ApJS..142...35K, 2019MNRAS.489.2652P}. Similarly, while the $\ns$ index shows a marginally higher scatter due to the density dependence of the $\sii$ doublet, it remains a highly stable tracer compared to indices involving high-ionization lines like $\oiii$. These results confirm that, when available, $\no$ and $\ns$ provide the most precise metallicity constraints for AGNs.

\begin{figure*}
    \centering
    \includegraphics[width=1.0\textwidth]{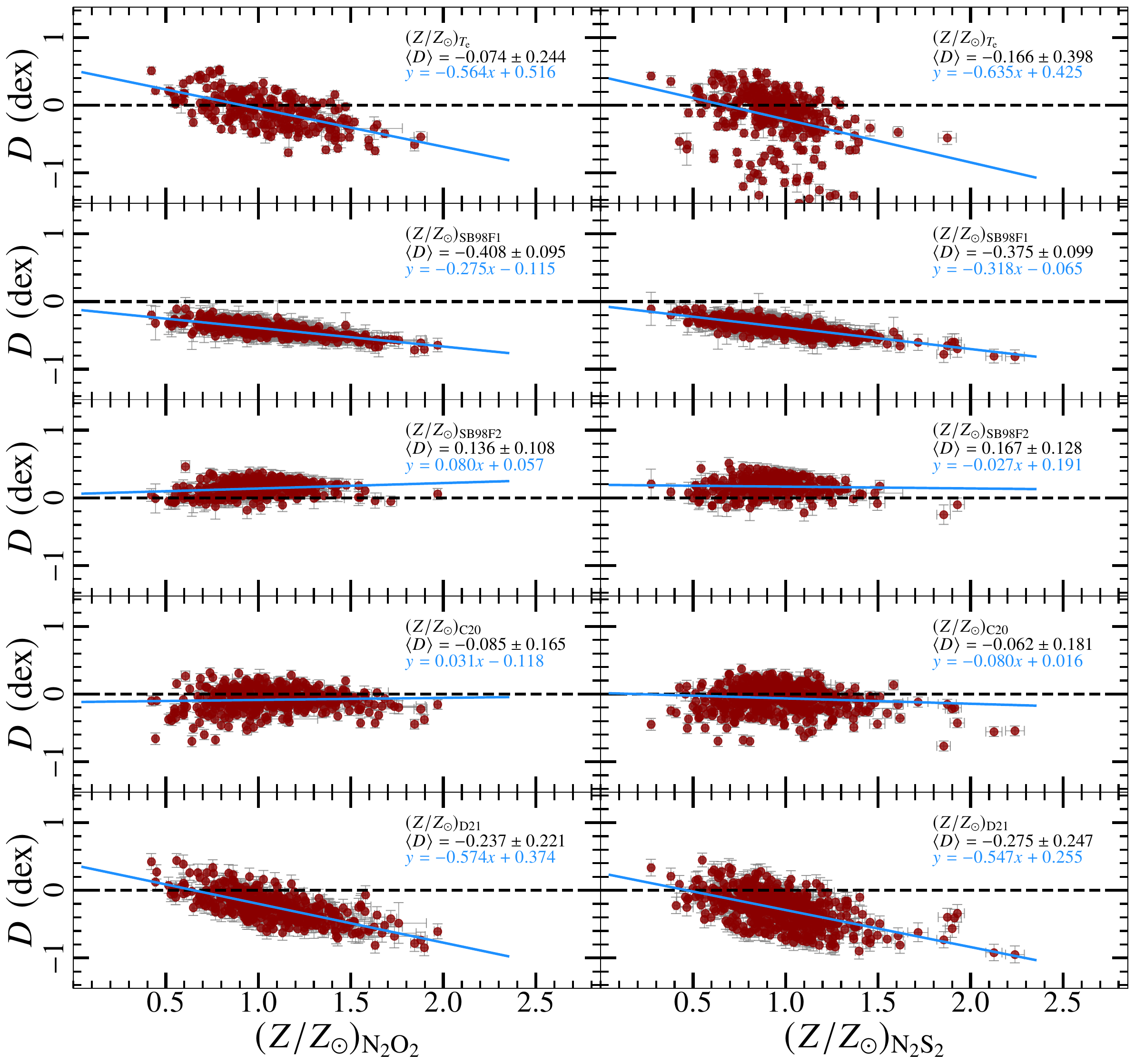}
    \vspace{-12pt} 
    \caption{Comparison of metallicities derived from our new $\no$ and $\ns$ calibrations (x-axes) against direct \Te~derivations and several literature methods (y-axes). The y-axes in the top sub-panels represent the logarithmic metallicity difference in dex ($D = \log(Z/Z_{\odot})_{\text{Literature}} - \log(Z/Z_{\odot})_{\text{This work}}$) between the external methods and our new calibrations. {\it Top row:} comparison with the \Te-based derivations (showing mean systematic offsets of $\langle D \rangle \approx -0.166$ to $-0.074$ dex). {\it Second row:} comparison with the \SBcalib{F1} calibration, where the exceedingly small dispersion at a fixed index occurs because the soft-continuum models saturate and fail to dynamically resolve metallicity variations. {\it Third row:} comparison with the \SBcalib{F2} calibration. {\it Fourth row:} comparison with the semi-empirical \citet[][i.e. \citetalias{2020MNRAS.492.5675C}]{2020MNRAS.492.5675C} calibration, which exhibits the best agreement ($\langle D \rangle \approx -0.062$ dex for $\ns$). {\it Bottom row:} comparison with the empirical, \Te-based \citet[][i.e. \citetalias{2021MNRAS.507..466D}]{2021MNRAS.507..466D} calibration. The solid blue lines represent linear regression fits to the residuals. The steep negative slopes observed in these fits highlight the severe dynamic range compression inherent to older literature calibrations, which produce a significantly narrower range of metallicities than our models. The mean difference ($\langle D \rangle$), its standard deviation, and the linear fit equation are indicated in each residual panel.
    }
    \label{fig_8}
\end{figure*}

To evaluate the performance and reliability of our new $\Lx$-based diagnostics, we compared the metallicities derived from our $\no$, and $\ns$ calibrations with those obtained using widely adopted literature calibrations (i.e. \citetalias{1998AJ....115..909S}, \citetalias{2020MNRAS.492.5675C}, and \citetalias{2021MNRAS.507..466D}), as well as the $T_{\rm e}$-method. By applying these various calibrations to the same sample of AGNs ($N \approx 233 - 547$, depending on the available emission lines required for each specific diagnostic), we can assess their overall agreement and identify any potential systematic biases between the empirical, photoionization, and $\Lx$-dependent approaches. To quantify these differences, we analyze the logarithmic residuals $D = \log(Z/Z_{\odot})_{\rm Literature} - \log(Z/Z_{\odot})_{\rm This~ work}$ for each pair of diagnostics. We estimate systematic shifts by calculating the mean offset, $\langle D \rangle$, utilizing the geometric mean ratio relation (see Equation~\ref{eqn_diff}), alongside the standard deviation ($\sigma$) to estimate the overall scatter. Figure~\ref{fig_8} shows these comparisons. 
Across all evaluated methods, we note that while the mean systematic offsets provide a baseline comparison, the residual distributions exhibit substantial point-to-point scatter, with extreme outliers reaching an absolute magnitude of $\sim 1.4$ dex, highlighting the severe variance inherent in comparing disparate calibration scales.

We explicitly compare our derived metallicities against direct \Te-based derivations (top row of Figure~\ref{fig_8}). An analysis of the residuals reveals modest systematic offsets between our semi-empirical calibrations and the direct \Te~estimates. Specifically, for the $\no$ diagnostic, we find a mean difference of $\langle D \rangle = -0.074 \pm 0.244$ dex. The $\ns$ index similarly exhibits a systematic deviation, yielding a mean offset of $\langle D \rangle = -0.166 \pm 0.398$ dex.  The large systematic offsets seen in \citetalias{2026MNRAS.548ag560A} when comparing with the theoretical calibrations of \citetalias{1998AJ....115..909S} persist for the $\no$ and $\ns$ diagnostics. This is demonstrated by the comparison with the \SBcalib{F1} calibration, which systematically underestimates metallicities by $\sim -0.408$ to $\sim -0.375$ dex across both diagnostics, and the comparison with the \SBcalib{F2} calibration, which systematically overestimates metallicities by $\sim 0.136$ to $0.167$ dex.  In contrast, our framework shows good agreement with the semi-empirical calibration of \citetalias{2020MNRAS.492.5675C}. The $\ns$ diagnostic shows a mean offset of only $\langle D \rangle = -0.062 \pm 0.181$ dex, and the comparison with $\no$ shows a similarly small systematic offset ($\langle D \rangle = -0.085 \pm 0.165$ dex). Furthermore, our calibrations yield systematically higher metallicities than the empirical, \Te-based strong-line calibration of \citetalias{2021MNRAS.507..466D} (bottom row of Figure~\ref{fig_8}), with mean differences of $\langle D \rangle \approx -0.275$ dex and $-0.237$  dex for $\ns$ and $\no$, respectively. This result is fully consistent with the well-documented ADF (also known as the \Te~problem) observed in nebular astrophysics, where theoretical calibrations are known to produce higher metallicity estimates than empirical calibrations based on the \Te-method. This fundamental offset is a known systematic effect, highlighting the importance of using a consistent calibration scale when comparing different samples.

As highlighted by the solid blue regression lines in Figure~\ref{fig_8}, the residual distributions exhibit a remarkably similar, strong anti-correlation (steep negative slopes) across the direct \Te~derivations, the empirical \citetalias{2021MNRAS.507..466D} calibration, and the theoretical \SBcalib{F1} calibration. This shared trend fundamentally arises because all three estimates are deeply linked to specific \Te~and \Ne~dependencies that ultimately compress their dynamic range. Because the \citetalias{2021MNRAS.507..466D} calibration is empirically tied to the \Te-method, it intrinsically inherits the exact same temperature scale and density-derived emissivity biases as direct \Te~estimates. Meanwhile, although \citetalias{1998AJ....115..909S} is a theoretical calibration, it was constructed using a significantly softer ionizing continuum (resulting in a cooler equilibrium gas state) and a baseline model grid fixed at a density of $N_{\rm e} = 300 \text{ cm}^{-3}$. To account for density variations across different objects, \SBcalib{F1} and \SBcalib{F2} employ a simple linear logarithmic density correction. However, this classical, decoupled treatment of \Ne~fail to fully capture the complex collisional de-excitation gradients and the extreme volumetric expansion of the partially ionized zone driven by hard X-ray luminosities in modern AGN samples. Consequently, in all three methods (i.e. \Te-method, \SBcalib{F1}, and \citetalias{2021MNRAS.507..466D}), the diagnostic line ratios saturate prematurely, severely restricting the overall range of metallicities they can resolve. When these tightly clustered, narrow-band literature values are compared against our high-dynamic-range, luminosity-coupled models, the resulting logarithmic residual analytically manifests as the steep negative slopes observed in Figure~\ref{fig_8}. In contrast, the \SBcalib{F2} calibration, which utilizes the $\otot$ ratio to specifically resolve this high-metallicity degeneracy, exhibits a much flatter residual distribution.

The agreement between these comparisons and the results from \citetalias{2026MNRAS.548ag560A} demonstrates that our new $\no$ and $\ns$ calibrations are consistent with established metallicity scales. Finally, they provide the distinct advantage of being robust against variations in $\Lx$, thereby facilitating their broader observational application.

\section{Summary and Conclusions}
\label{conclusions}
In this study, we expand our semi-empirical metallicity framework for Seyfert galaxies by introducing new, robust calibrations for the $\no$ and $\ns$ strong-line indices.  At the core of our methodology is the use of {\sc cloudy} photoionization simulations parameterized by the observable 2--10 keV X-ray luminosity ($\Lx$), effectively bypassing the unobservable ionization parameter in favour of a directly measurable physical quantity. The principal conclusions of this study are summarized as follows:

\begin{enumerate}
    \item We have developed new semi-empirical calibrations for estimating gas-phase metallicities from the NLRs using the $\no$ and $\ns$ indices. We find that both are powerful tracers of metallicity in the high-metallicity environments typical of AGN hosts.

\item  We systematically investigated the influence of secondary parameters on our models. While both diagnostics show remarkable robustness against variations in $\aox$, $\Lx$, and $T_{\rm BB}$, the $\ns$ index exhibits a visibly greater dependence on the gas density (\Ne) compared to the $\no$ index, as demonstrated by the larger stratification of the fitted curves in the right column of Figure~\ref{fig_4}.

\item  Our most significant finding, which resolves the contradictions found in \citetalias{2026MNRAS.548ag560A}, is that both new diagnostic indices ($\no$ and $\ns$) exhibit a high degree of robustness against variations in the X-ray luminosity, $\Lx$. Unlike the $\ntwo$ and $\otnt$ indices, which required explicit $\Lx$-dependent corrections to account for the ionization parameter, $\no$ and $\ns$ are naturally insensitive to $\Lx$. This is a direct result of their physical construction, which compares ions of similar ionization potentials.

\item The $\no$ index serves as our most precise metallicity tracer, exhibiting an RMS residual dispersion of $\sim 0.081$ dex, while the $\ns$ index offers a comparable precision of an RMS dispersion of $\approx 0.121$ dex. These values represent a factor of $\sim 2$ improvement in precision compared to the standard optical diagnostics ($1\sigma \approx 0.20$ dex from  $\ntwo$ and $\otnt$ indices).

\item A direct inter-comparison of the four strong-line calibrations established in this series confirms the robustness of the symmetric diagnostics. While the $\ntwo$ index exhibits steep systematic deviations driven by residual ionization sensitivity, the $\no$ and $\Lx$-corrected $\otnt$ indices trace a nearly identical, flat metallicity scale (mean offset $\langle D \rangle = -0.011 \pm 0.083$ dex). Conversely, introducing the $\ns$ index drives systematic deviations relative to both $\no$ and $\otnt$. The $\ns$ index shows a consistent systematic offset of $\sim -0.04$ dex relative to $\no$ and $\otnt$, likely driven by its density sensitivity.

\item Our comparison with calibrations from the literature validates this new framework. When compared against direct \Te-based derivations, we observe modest systematic offsets ($\langle D \rangle \approx -0.17$ to $-0.08$ dex). Furthermore, we find good agreement with the semi-empirical \citetalias{2020MNRAS.492.5675C} calibration ($\langle D \rangle \approx -0.09$ dex for $\no$ and $\langle D \rangle \approx -0.06$ dex for $\ns$). The calibrations show expected systematic offsets when compared to the \citetalias{1998AJ....115..909S} methods (i.e. $\approx -0.4$ dex for \SBcalib{F1} and $\approx +0.14 - +0.17$ dex for \SBcalib{F2}), driven by fundamental differences in the underlying assumptions regarding the ionizing continuum. Finally, our derivations yield systematically higher metallicities than empirical, \Te-based strong-line methods (\citetalias{2021MNRAS.507..466D}, offset $\sim -0.28$ dex to $\sim -0.24$ dex), consistent with the well-documented ADF.

\item For AGN hosts with explicitly measured soft ionizing continua ($\aox \lesssim -1.7$), strong-line indices are intrinsically incapable of disentangling chemical abundance from the ionization state due to the structural collapse of the emission region. The use of these strong-line calibrations is therefore discouraged for such targets. Instead, metallicities in these extreme regimes should be constrained via direct \Te-based derivations or through comprehensive Bayesian forward-modeling of the full optical spectrum.
\end{enumerate} 

This work completes our set of self-consistent, luminosity-dependent calibrations for four key strong-line diagnostics. By replacing the unobservable ionization parameter $U$ with the observable $\Lx$, we have created a more direct and intuitive link between theoretical models and the properties of real AGNs. This framework provides a robust and comprehensive set of tools for future studies aimed at accurately estimating the chemical evolution of AGN host galaxies. Consequently, the $\no$ index, by virtue of its atomic symmetry and insensitivity to both $\Lx$ and \Ne\, represents an optimal, unbiased optical metallicity tracer for AGNs. Where blue spectroscopic coverage is unavailable, the $\ns$ index serves as a highly reliable substitute, provided its mild density dependence is accounted for in high-density environments.

\section*{Acknowledgements}
We thank the anonymous referee for their insightful comments, which significantly improved the quality of this paper. MA gratefully acknowledges support from Fundação de Amparo à Pesquisa do Estado de São Paulo
(FAPESP, Processo: 2024/03727-3). OLD is grateful to Fundação de Amparo à Pesquisa do Estado de
São Paulo (FAPESP) and Conselho Nacional de Desenvolvimento Científico e Tecnológico (CNPq). RAR acknowledges support from the Conselho Nacional de Desenvolvimento Cient\'ifico e Tecnol\'ogico (CNPq; Proj. 303450/2022-3, 403398/2023-1, \& 441722/2023-7) and Coordena\c c\~ao de Aperfei\c coamento de Pessoal de N\'ivel Superior (CAPES;  Proj. 88887.894973/2023-00). 
RR acknowledges support from  Conselho Nacional de Desenvolvimento Cient\'{i}fico e Tecnol\'ogico  (CNPq, Proj. CNPq-445231/2024-6,311223/2020-6, 404238/2021-1, and 310413/2025-7), Funda\c{c}\~ao de amparo \`{a} pesquisa do Rio Grande do Sul (FAPERGS, Proj. 19/1750-2 and 24/2551-0001282-6) and Coordena\c{c}\~ao de Aperfei\c{c}oamento de Pessoal de N\'{i}vel Superior (CAPES, 88881.109987/2025-01). JVM acknowledges support from the Spanish grants: PID2022-136598NB-C32 and Severo Ochoa CEX2021-001131-S funded by MICIU/AEI/ 10.13039/501100011033.

\section*{Data Availability}
The observational data used in this study are publicly available from the \textit{Swift} Burst Alert Telescope (BAT) AGN Spectroscopic Survey (BASS) Data Release 2, accessible at \url{https://www.bass-survey.com/dr2.html}. The data products and analysis results generated specifically for this paper are available from the corresponding author upon reasonable request.


\bibliographystyle{mnras}
\bibliography{ref} 

@ARTICLE{2023MNRAS.524.5640R,
       author = {{Riffel}, Rog{\'e}rio and {Mallmann}, Nicolas D. and {Rembold}, Sandro B. and {Ilha}, Gabriele S. and {Riffel}, Rogemar A. and {Storchi-Bergmann}, Thaisa and {Ruschel-Dutra}, Daniel and {Vazdekis}, Alexandre and {Mart{\'\i}n-Navarro}, Ignacio and {Schimoia}, Jaderson S. and {Ramos Almeida}, Cristina and {da Costa}, Luiz N. and {Vila-Verde}, Glauber C. and {Gatto}, Lara},
        title = "{Mapping the stellar population and gas excitation of MaNGA galaxies with MEGACUBES. Results for AGN versus control sample}",
      journal = {\mnras},
         year = 2023,
        month = oct,
       volume = {524},
       number = {4},
        pages = {5640-5657},
          doi = {10.1093/mnras/stad2234},
archivePrefix = {arXiv},
       eprint = {2307.11474},
 primaryClass = {astro-ph.GA},
       adsurl = {https://ui.adsabs.harvard.edu/abs/2023MNRAS.524.5640R}
}

@ARTICLE{1999ARA&A..37..487H,
       author = {{Hamann}, Fred and {Ferland}, Gary},
        title = "{Elemental Abundances in Quasistellar Objects: Star Formation and Galactic Nuclear Evolution at High Redshifts}",
      journal = {\araa},
         year = 1999,
        month = jan,
       volume = {37},
        pages = {487-531},
          doi = {10.1146/annurev.astro.37.1.487},
archivePrefix = {arXiv},
       eprint = {astro-ph/9904223},
 primaryClass = {astro-ph},
       adsurl = {https://ui.adsabs.harvard.edu/abs/1999ARA&A..37..487H}
}

@ARTICLE{2019A&ARv..27....3M,
       author = {{Maiolino}, R. and {Mannucci}, F.},
        title = "{De re metallica: the cosmic chemical evolution of galaxies}",
      journal = {\aapr},
         year = 2019,
        month = feb,
       volume = {27},
       number = {1},
          eid = {3},
        pages = {3},
          doi = {10.1007/s00159-018-0112-2},
archivePrefix = {arXiv},
       eprint = {1811.09642},
 primaryClass = {astro-ph.GA},
       adsurl = {https://ui.adsabs.harvard.edu/abs/2019A&ARv..27....3M}
}

@ARTICLE{2023MNRAS.520.1687A,
       author = {{Armah}, Mark and {Riffel}, Rog{\'e}rio and {Dors}, O.~L. and {Oh}, Kyuseok and {Koss}, Michael J. and {Ricci}, Claudio and {Trakhtenbrot}, Benny and {Valerdi}, Mabel and {Riffel}, Rogemar A. and {Krabbe}, Angela C.},
        title = "{Oxygen abundances in the narrow line regions of Seyfert galaxies and the metallicity-luminosity relation}",
      journal = {\mnras},
         year = 2023,
        month = apr,
       volume = {520},
       number = {2},
        pages = {1687-1703},
          doi = {10.1093/mnras/stad217},
archivePrefix = {arXiv},
       eprint = {2301.07596},
 primaryClass = {astro-ph.GA},
       adsurl = {https://ui.adsabs.harvard.edu/abs/2023MNRAS.520.1687A}
}

@book{2006agna.book.....O,
       author = {{Osterbrock}, Donald E. and {Ferland}, Gary J.},
        title     = {{Astrophysics of Gaseous Nebulae and Active Galactic Nuclei}},
  edition   = {2nd},
  publisher = {University Science Books},
  address   = {Sausalito, California, USA},
         year = 2006,
       adsurl = {https://ui.adsabs.harvard.edu/abs/2006agna.book.....O}
}

@ARTICLE{2019ARA&A..57..511K,
       author = {{Kewley}, Lisa J. and {Nicholls}, David C. and {Sutherland}, Ralph S.},
        title = "{Understanding Galaxy Evolution Through Emission Lines}",
      journal = {\araa},
         year = 2019,
        month = aug,
       volume = {57},
        pages = {511-570},
          doi = {10.1146/annurev-astro-081817-051832},
archivePrefix = {arXiv},
       eprint = {1910.09730},
 primaryClass = {astro-ph.GA},
       adsurl = {https://ui.adsabs.harvard.edu/abs/2019ARA&A..57..511K}
}

@ARTICLE{2015ApJ...798...99B,
       author = {{Blanc}, Guillermo A. and {Kewley}, Lisa and {Vogt}, Fr{\'e}d{\'e}ric P.~A. and {Dopita}, Michael A.},
        title = "{IZI: Inferring the Gas Phase Metallicity (Z) and Ionization Parameter (q) of Ionized Nebulae Using Bayesian Statistics}",
      journal = {\apj},
         year = 2015,
        month = jan,
       volume = {798},
       number = {2},
          eid = {99},
        pages = {99},
          doi = {10.1088/0004-637X/798/2/99},
archivePrefix = {arXiv},
       eprint = {1410.8146},
 primaryClass = {astro-ph.GA},
       adsurl = {https://ui.adsabs.harvard.edu/abs/2015ApJ...798...99B}
}

@ARTICLE{2020MNRAS.492..468D,
       author = {{Dors}, O.~L. and {Freitas-Lemes}, P. and {Am{\^o}res}, E.~B. and {P{\'e}rez-Montero}, E. and {Cardaci}, M.~V. and {H{\"a}gele}, G.~F. and {Armah}, M. and {Krabbe}, A.~C. and {Fa{\'u}ndez-Abans}, M.},
        title = "{Chemical abundances of Seyfert 2 AGNs - I. Comparing oxygen abundances from distinct methods using SDSS}",
      journal = {\mnras},
         year = 2020,
        month = feb,
       volume = {492},
       number = {1},
        pages = {468-479},
          doi = {10.1093/mnras/stz3492},
archivePrefix = {arXiv},
       eprint = {1912.04236},
 primaryClass = {astro-ph.GA},
       adsurl = {https://ui.adsabs.harvard.edu/abs/2020MNRAS.492..468D}
}

@ARTICLE{2020MNRAS.496.3209D,
       author = {{Dors}, O.~L. and {Maiolino}, R. and {Cardaci}, M.~V. and {H{\"a}gele}, G.~F. and {Krabbe}, A.~C. and {P{\'e}rez-Montero}, E. and {Armah}, M.},
        title = "{Chemical abundances of Seyfert 2 AGNs - III. Reducing the oxygen abundance discrepancy}",
      journal = {\mnras},
         year = 2020,
        month = aug,
       volume = {496},
       number = {3},
        pages = {3209-3221},
          doi = {10.1093/mnras/staa1781},
archivePrefix = {arXiv},
       eprint = {2006.09152},
 primaryClass = {astro-ph.GA},
       adsurl = {https://ui.adsabs.harvard.edu/abs/2020MNRAS.496.3209D}
}

@ARTICLE{2004MNRAS.355..229E,
       author = {{Esteban}, C. and {Peimbert}, M. and {Garc{\'\i}a-Rojas}, J. and {Ruiz}, M.~T. and {Peimbert}, A. and {Rodr{\'\i}guez}, M.},
        title = "{A reappraisal of the chemical composition of the Orion nebula based on Very Large Telescope echelle spectrophotometry}",
      journal = {\mnras},
         year = 2004,
        month = nov,
       volume = {355},
       number = {1},
        pages = {229-247},
          doi = {10.1111/j.1365-2966.2004.08313.x},
archivePrefix = {arXiv},
       eprint = {astro-ph/0408249},
 primaryClass = {astro-ph},
       adsurl = {https://ui.adsabs.harvard.edu/abs/2004MNRAS.355..229E}
}

@ARTICLE{2007ApJ...670..457G,
       author = {{Garc{\'\i}a-Rojas}, Jorge and {Esteban}, C{\'e}sar},
        title = "{On the Abundance Discrepancy Problem in H II Regions}",
      journal = {\apj},
         year = 2007,
        month = nov,
       volume = {670},
       number = {1},
        pages = {457-470},
          doi = {10.1086/521871},
archivePrefix = {arXiv},
       eprint = {0707.3518},
 primaryClass = {astro-ph},
       adsurl = {https://ui.adsabs.harvard.edu/abs/2007ApJ...670..457G}
}

@ARTICLE{2017PASP..129h2001P,
       author = {{Peimbert}, Manuel and {Peimbert}, Antonio and {Delgado-Inglada}, Gloria},
        title = "{Nebular Spectroscopy: A Guide on Hii Regions and Planetary Nebulae}",
      journal = {\pasp},
         year = 2017,
        month = aug,
       volume = {129},
       number = {978},
        pages = {082001},
          doi = {10.1088/1538-3873/aa72c3},
archivePrefix = {arXiv},
       eprint = {1705.06323},
 primaryClass = {astro-ph.GA},
       adsurl = {https://ui.adsabs.harvard.edu/abs/2017PASP..129h2001P}
}

@ARTICLE{1967ApJ...150..825P,
       author = {{Peimbert}, Manuel},
        title = "{Temperature Determinations of H II Regions}",
      journal = {\apj},
         year = "1967",
        month = "Dec",
       volume = {150},
        pages = {825},
          doi = {10.1086/149385},
       adsurl = {https://ui.adsabs.harvard.edu/abs/1967ApJ...150..825P}
}

@ARTICLE{2021MNRAS.506L..11R,
       author = {{Riffel}, Rogemar A. and {Dors}, Oli L. and {Krabbe}, Angela C. and {Esteban}, C{\'e}sar},
        title = "{Electron temperature fluctuations in Seyfert galaxies}",
      journal = {\mnras},
         year = 2021,
        month = sep,
       volume = {506},
       number = {1},
        pages = {L11-L15},
          doi = {10.1093/mnrasl/slab064},
archivePrefix = {arXiv},
       eprint = {2106.03623},
 primaryClass = {astro-ph.GA},
       adsurl = {https://ui.adsabs.harvard.edu/abs/2021MNRAS.506L..11R}
}

@ARTICLE{2002ApJ...572..753D,
       author = {{Dopita}, Michael A. and {Groves}, Brent A. and {Sutherland}, Ralph S. and {Binette}, Luc and {Cecil}, Gerald},
        title = "{Are the Narrow-Line Regions in Active Galaxies Dusty and Radiation Pressure Dominated?}",
      journal = {\apj},
         year = 2002,
        month = jun,
       volume = {572},
       number = {2},
        pages = {753-761},
          doi = {10.1086/340429},
archivePrefix = {arXiv},
       eprint = {astro-ph/0203360},
 primaryClass = {astro-ph},
       adsurl = {https://ui.adsabs.harvard.edu/abs/2002ApJ...572..753D}
}

@ARTICLE{2013ApJ...774..100K,
       author = {{Kewley}, Lisa J. and {Dopita}, Michael A. and {Leitherer}, Claus and {Dav{\'e}}, Romeel and {Yuan}, Tiantian and {Allen}, Mark and {Groves}, Brent and {Sutherland}, Ralph},
        title = "{Theoretical Evolution of Optical Strong Lines across Cosmic Time}",
      journal = {\apj},
         year = 2013,
        month = sep,
       volume = {774},
       number = {2},
          eid = {100},
        pages = {100},
          doi = {10.1088/0004-637X/774/2/100},
archivePrefix = {arXiv},
       eprint = {1307.0508},
 primaryClass = {astro-ph.CO},
       adsurl = {https://ui.adsabs.harvard.edu/abs/2013ApJ...774..100K}
}

@ARTICLE{2021MNRAS.501.1370D,
       author = {{Dors}, O.~L. and {Contini}, M. and {Riffel}, R.~A. and {P{\'e}rez-Montero}, E. and {Krabbe}, A.~C. and {Cardaci}, M.~V. and {H{\"a}gele}, G.~F.},
        title = "{Chemical abundances of Seyfert 2 AGNs - IV. Composite models calculated by photoionization + shocks}",
      journal = {\mnras},
         year = 2021,
        month = feb,
       volume = {501},
       number = {1},
        pages = {1370-1383},
          doi = {10.1093/mnras/staa3707},
archivePrefix = {arXiv},
       eprint = {2011.12103},
 primaryClass = {astro-ph.GA},
       adsurl = {https://ui.adsabs.harvard.edu/abs/2021MNRAS.501.1370D}
}

@ARTICLE{2006A&A...448..955I,
       author = {{Izotov}, Y.~I. and {Stasi{\'n}ska}, G. and {Meynet}, G. and {Guseva}, N.~G. and {Thuan}, T.~X.},
        title = "{The chemical composition of metal-poor emission-line galaxies in the Data Release 3 of the Sloan Digital Sky Survey}",
      journal = {\aap},
         year = 2006,
        month = mar,
       volume = {448},
       number = {3},
        pages = {955-970},
          doi = {10.1051/0004-6361:20053763},
archivePrefix = {arXiv},
       eprint = {astro-ph/0511644},
 primaryClass = {astro-ph},
       adsurl = {https://ui.adsabs.harvard.edu/abs/2006A&A...448..955I}
}

@ARTICLE{2019MNRAS.489.2652P,
       author = {{P{\'e}rez-Montero}, E. and {Dors}, O.~L. and {V{\'\i}lchez}, J.~M. and {Garc{\'\i}a-Benito}, R. and {Cardaci}, M.~V. and {H{\"a}gele}, G.~F.},
        title = "{A bayesian-like approach to derive chemical abundances in type-2 active galactic nuclei based on photoionization models}",
      journal = {\mnras},
         year = 2019,
        month = oct,
       volume = {489},
       number = {2},
        pages = {2652-2668},
          doi = {10.1093/mnras/stz2278},
archivePrefix = {arXiv},
       eprint = {1908.04827},
 primaryClass = {astro-ph.GA},
       adsurl = {https://ui.adsabs.harvard.edu/abs/2019MNRAS.489.2652P}
}

@ARTICLE{1998AJ....116.2805V,
       author = {{van Zee}, Liese and {Salzer}, John J. and {Haynes}, Martha P. and {O'Donoghue}, Aileen A. and {Balonek}, Thomas J.},
        title = "{Spectroscopy of Outlying H II Regions in Spiral Galaxies: Abundances and Radial Gradients}",
      journal = {\aj},
         year = 1998,
        month = dec,
       volume = {116},
       number = {6},
        pages = {2805-2833},
          doi = {10.1086/300647},
archivePrefix = {arXiv},
       eprint = {astro-ph/9808315},
 primaryClass = {astro-ph},
       adsurl = {https://ui.adsabs.harvard.edu/abs/1998AJ....116.2805V}
}

@ARTICLE{1979A&A....78..200A,
       author = {{Alloin}, D. and {Collin-Souffrin}, S. and {Joly}, M. and {Vigroux}, L.},
        title = "{Nitrogen and oxygen abundances in galaxies.}",
      journal = {\aap},
         year = 1979,
        month = sep,
       volume = {78},
        pages = {200-216},
       adsurl = {https://ui.adsabs.harvard.edu/abs/1979A&A....78..200A}
}

@ARTICLE{1979MNRAS.189...95P,
       author = {{Pagel}, B.~E.~J. and {Edmunds}, M.~G. and {Blackwell}, D.~E. and {Chun}, M.~S. and {Smith}, G.},
        title = "{On the composition of H II regions in southern galaxies - I. NGC 300 and 1365.}",
      journal = {\mnras},
         year = 1979,
        month = oct,
       volume = {189},
        pages = {95-113},
          doi = {10.1093/mnras/189.1.95},
       adsurl = {https://ui.adsabs.harvard.edu/abs/1979MNRAS.189...95P}
}

@ARTICLE{1998AJ....115..909S,
       author = {{Storchi-Bergmann}, Thaisa and {Schmitt}, Henrique R. and
         {Calzetti}, Daniela and {Kinney}, Anne L.},
        title = "{Chemical Abundance Calibrations for the Narrow-Line Region of Active Galaxies}",
      journal = {\aj},
         year = "1998",
        month = "Mar",
       volume = {115},
       number = {3},
        pages = {909 (SB98)-914},
          doi = {10.1086/300242},
archivePrefix = {arXiv},
       eprint = {astro-ph/9711302},
 primaryClass = {astro-ph},
       adsurl = {https://ui.adsabs.harvard.edu/abs/1998AJ....115..909S},
}

@ARTICLE{2016Ap&SS.361...61D,
       author = {{Dopita}, Michael A. and {Kewley}, Lisa J. and {Sutherland}, Ralph S. and {Nicholls}, David C.},
        title = "{Chemical abundances in high-redshift galaxies: a powerful new emission line diagnostic}",
      journal = {\apss},
         year = 2016,
        month = feb,
       volume = {361},
          eid = {61},
        pages = {61},
          doi = {10.1007/s10509-016-2657-8},
archivePrefix = {arXiv},
       eprint = {1601.01337},
 primaryClass = {astro-ph.GA},
       adsurl = {https://ui.adsabs.harvard.edu/abs/2016Ap&SS.361...61D}
}

@ARTICLE{2020MNRAS.491.1427S,
       author = {{Sanders}, Ryan L. and {Shapley}, Alice E. and {Reddy}, Naveen A. and {Kriek}, Mariska and {Siana}, Brian and {Coil}, Alison L. and {Mobasher}, Bahram and {Shivaei}, Irene and {Freeman}, William R. and {Azadi}, Mojegan and {Price}, Sedona H. and {Leung}, Gene and {Fetherolf}, Tara and {de Groot}, Laura and {Zick}, Tom and {Fornasini}, Francesca M. and {Barro}, Guillermo},
        title = "{The MOSDEF survey: direct-method metallicities and ISM conditions at z {\ensuremath{\sim}} 1.5-3.5}",
      journal = {\mnras},
         year = 2020,
        month = jan,
       volume = {491},
       number = {1},
        pages = {1427-1455},
          doi = {10.1093/mnras/stz3032},
archivePrefix = {arXiv},
       eprint = {1907.00013},
 primaryClass = {astro-ph.GA},
       adsurl = {https://ui.adsabs.harvard.edu/abs/2020MNRAS.491.1427S}
}

@ARTICLE{2023MNRAS.526.3504H,
       author = {{Hirschmann}, Michaela and {Charlot}, Stephane and {Somerville}, Rachel S.},
        title = "{High-redshift metallicity calibrations for JWST spectra: insights from line emission in cosmological simulations}",
      journal = {\mnras},
         year = 2023,
        month = dec,
       volume = {526},
       number = {3},
        pages = {3504-3518},
          doi = {10.1093/mnras/stad2745},
archivePrefix = {arXiv},
       eprint = {2305.03753},
 primaryClass = {astro-ph.GA},
       adsurl = {https://ui.adsabs.harvard.edu/abs/2023MNRAS.526.3504H}
}

@ARTICLE{2024ApJ...962...24S,
       author = {{Sanders}, Ryan L. and {Shapley}, Alice E. and {Topping}, Michael W. and {Reddy}, Naveen A. and {Brammer}, Gabriel B.},
        title = "{Direct T $_{e}$-based Metallicities of z = 2─9 Galaxies with JWST/NIRSpec: Empirical Metallicity Calibrations Applicable from Reionization to Cosmic Noon}",
      journal = {\apj},
         year = 2024,
        month = feb,
       volume = {962},
       number = {1},
          eid = {24},
        pages = {24},
          doi = {10.3847/1538-4357/ad15fc},
archivePrefix = {arXiv},
       eprint = {2303.08149},
 primaryClass = {astro-ph.GA},
       adsurl = {https://ui.adsabs.harvard.edu/abs/2024ApJ...962...24S}
}

@ARTICLE{1979ApJ...234L...9T,
       author = {{Tananbaum}, H. and {Avni}, Y. and {Branduardi}, G. and {Elvis}, M. and {Fabbiano}, G. and {Feigelson}, E. and {Giacconi}, R. and {Henry}, J.~P. and {Pye}, J.~P. and {Soltan}, A. and {Zamorani}, G.},
        title = "{X-ray studies of quasars with the Einstein Observatory.}",
      journal = {\apjl},
         year = 1979,
        month = nov,
       volume = {234},
        pages = {L9-L13},
          doi = {10.1086/183100},
       adsurl = {https://ui.adsabs.harvard.edu/abs/1979ApJ...234L...9T}
}

@ARTICLE{1981ApJ...245..357Z,
       author = {{Zamorani}, G. and {Henry}, J.~P. and {Maccacaro}, T. and {Tananbaum}, H. and {Soltan}, A. and {Avni}, Y. and {Liebert}, J. and {Stocke}, J. and {Strittmatter}, P.~A. and {Weymann}, R.~J. and {Smith}, M.~G. and {Condon}, J.~J.},
        title = "{X-ray studies of quasars with the Einstein Observatory II.}",
      journal = {\apj},
         year = 1981,
        month = apr,
       volume = {245},
        pages = {357-374},
          doi = {10.1086/158815},
       adsurl = {https://ui.adsabs.harvard.edu/abs/1981ApJ...245..357Z}
}

@ARTICLE{1994ApJS...92...53W,
       author = {{Wilkes}, Belinda J. and {Tananbaum}, Harvey and {Worrall}, D.~M. and {Avni}, Yoram and {Oey}, M.~S. and {Flanagan}, Joan},
        title = "{The Einstein Database of IPC X-Ray Observations of Optically Selected and Radio-selected Quasars. I.}",
      journal = {\apjs},
         year = 1994,
        month = may,
       volume = {92},
        pages = {53},
          doi = {10.1086/191959},
       adsurl = {https://ui.adsabs.harvard.edu/abs/1994ApJS...92...53W}
}

@ARTICLE{2005AJ....130..387S,
       author = {{Strateva}, Iskra V. and {Brandt}, W.~N. and {Schneider}, Donald P. and {Vanden Berk}, Daniel G. and {Vignali}, Cristian},
        title = "{Soft X-Ray and Ultraviolet Emission Relations in Optically Selected AGN Samples}",
      journal = {\aj},
         year = 2005,
        month = aug,
       volume = {130},
       number = {2},
        pages = {387-405},
          doi = {10.1086/431247},
archivePrefix = {arXiv},
       eprint = {astro-ph/0503009},
 primaryClass = {astro-ph},
       adsurl = {https://ui.adsabs.harvard.edu/abs/2005AJ....130..387S}
}

@ARTICLE{2006AJ....131.2826S,
       author = {{Steffen}, A.~T. and {Strateva}, I. and {Brandt}, W.~N. and {Alexander}, D.~M. and {Koekemoer}, A.~M. and {Lehmer}, B.~D. and {Schneider}, D.~P. and {Vignali}, C.},
        title = "{The X-Ray-to-Optical Properties of Optically Selected Active Galaxies over Wide Luminosity and Redshift Ranges}",
      journal = {\aj},
         year = 2006,
        month = jun,
       volume = {131},
       number = {6},
        pages = {2826-2842},
          doi = {10.1086/503627},
archivePrefix = {arXiv},
       eprint = {astro-ph/0602407},
 primaryClass = {astro-ph},
       adsurl = {https://ui.adsabs.harvard.edu/abs/2006AJ....131.2826S}
}

@ARTICLE{2007ApJ...665.1004J,
       author = {{Just}, D.~W. and {Brandt}, W.~N. and {Shemmer}, O. and {Steffen}, A.~T. and {Schneider}, D.~P. and {Chartas}, G. and {Garmire}, G.~P.},
        title = "{The X-Ray Properties of the Most Luminous Quasars from the Sloan Digital Sky Survey}",
      journal = {\apj},
         year = 2007,
        month = aug,
       volume = {665},
       number = {2},
        pages = {1004-1022},
          doi = {10.1086/519990},
archivePrefix = {arXiv},
       eprint = {0705.3059},
 primaryClass = {astro-ph},
       adsurl = {https://ui.adsabs.harvard.edu/abs/2007ApJ...665.1004J}
}

@ARTICLE{2011ApJS..196....2S,
       author = {{Shang}, Zhaohui and {Brotherton}, Michael S. and {Wills}, Beverley J. and {Wills}, D. and {Cales}, Sabrina L. and {Dale}, Daniel A. and {Green}, Richard F. and {Runnoe}, Jessie C. and {Nemmen}, Rodrigo S. and {Gallagher}, Sarah C. and {Ganguly}, Rajib and {Hines}, Dean C. and {Kelly}, Benjamin J. and {Kriss}, Gerard A. and {Li}, Jun and {Tang}, Baitian and {Xie}, Yanxia},
        title = "{The Next Generation Atlas of Quasar Spectral Energy Distributions from Radio to X-Rays}",
      journal = {\apjs},
         year = 2011,
        month = sep,
       volume = {196},
       number = {1},
          eid = {2},
        pages = {2},
          doi = {10.1088/0067-0049/196/1/2},
archivePrefix = {arXiv},
       eprint = {1107.1855},
 primaryClass = {astro-ph.CO},
       adsurl = {https://ui.adsabs.harvard.edu/abs/2011ApJS..196....2S}
}

@ARTICLE{2011ApJ...726...20M,
       author = {{Miller}, B.~P. and {Brandt}, W.~N. and {Schneider}, D.~P. and {Gibson}, R.~R. and {Steffen}, A.~T. and {Wu}, Jianfeng},
        title = "{X-ray Emission from Optically Selected Radio-intermediate and Radio-loud Quasars}",
      journal = {\apj},
         year = 2011,
        month = jan,
       volume = {726},
       number = {1},
          eid = {20},
        pages = {20},
          doi = {10.1088/0004-637X/726/1/20},
archivePrefix = {arXiv},
       eprint = {1010.4804},
 primaryClass = {astro-ph.CO},
       adsurl = {https://ui.adsabs.harvard.edu/abs/2011ApJ...726...20M}
}

@ARTICLE{2019A&A...630A.118V,
       author = {{Vito}, F. and {Brandt}, W.~N. and {Bauer}, F.~E. and {Calura}, F. and {Gilli}, R. and {Luo}, B. and {Shemmer}, O. and {Vignali}, C. and {Zamorani}, G. and {Brusa}, M. and {Civano}, F. and {Comastri}, A. and {Nanni}, R.},
        title = "{The X-ray properties of z > 6 quasars: no evident evolution of accretion physics in the first Gyr of the Universe}",
      journal = {\aap},
         year = 2019,
        month = oct,
       volume = {630},
          eid = {A118},
        pages = {A118},
          doi = {10.1051/0004-6361/201936217},
archivePrefix = {arXiv},
       eprint = {1908.09849},
 primaryClass = {astro-ph.GA},
       adsurl = {https://ui.adsabs.harvard.edu/abs/2019A&A...630A.118V}
}

@ARTICLE{2021RNAAS...5..101T,
       author = {{Timlin}, John and {Zhu}, Shifu and {Brandt}, W.~N. and {Laor}, Ari},
        title = "{The {\ensuremath{\alpha}}$_{ox}$-He II EW Connection in Radio-loud Quasars}",
      journal = {Research Notes of the American Astronomical Society},
         year = 2021,
        month = apr,
       volume = {5},
       number = {4},
          eid = {101},
        pages = {101},
          doi = {10.3847/2515-5172/abfbe5},
archivePrefix = {arXiv},
       eprint = {2104.14407},
 primaryClass = {astro-ph.HE},
       adsurl = {https://ui.adsabs.harvard.edu/abs/2021RNAAS...5..101T}
}

@ARTICLE{2021RAA....21....4Z,
       author = {{Zhou}, Min-Hua and {Gu}, Min-Feng},
        title = "{The composite X-ray spectra of radio-loud and radio-quiet SDSS quasars}",
      journal = {Research in Astronomy and Astrophysics},
         year = 2021,
        month = jan,
       volume = {21},
       number = {1},
          eid = {004},
        pages = {004},
          doi = {10.1088/1674-4527/21/1/4},
archivePrefix = {arXiv},
       eprint = {2007.01049},
 primaryClass = {astro-ph.HE},
       adsurl = {https://ui.adsabs.harvard.edu/abs/2021RAA....21....4Z}
}

@ARTICLE{2010A&A...512A..34L,
       author = {{Lusso}, E. and {Comastri}, A. and {Vignali}, C. and {Zamorani}, G. and {Brusa}, M. and {Gilli}, R. and {Iwasawa}, K. and {Salvato}, M. and {Civano}, F. and {Elvis}, M. and {Merloni}, A. and {Bongiorno}, A. and {Trump}, J.~R. and {Koekemoer}, A.~M. and {Schinnerer}, E. and {Le Floc'h}, E. and {Cappelluti}, N. and {Jahnke}, K. and {Sargent}, M. and {Silverman}, J. and {Mainieri}, V. and {Fiore}, F. and {Bolzonella}, M. and {Le F{\`e}vre}, O. and {Garilli}, B. and {Iovino}, A. and {Kneib}, J.~P. and {Lamareille}, F. and {Lilly}, S. and {Mignoli}, M. and {Scodeggio}, M. and {Vergani}, D.},
        title = "{The X-ray to optical-UV luminosity ratio of X-ray selected type 1 AGN in XMM-COSMOS}",
      journal = {\aap},
         year = 2010,
        month = mar,
       volume = {512},
          eid = {A34},
        pages = {A34},
          doi = {10.1051/0004-6361/200913298},
archivePrefix = {arXiv},
       eprint = {0912.4166},
 primaryClass = {astro-ph.CO},
       adsurl = {https://ui.adsabs.harvard.edu/abs/2010A&A...512A..34L}
}

@ARTICLE{1981ApJ...251..465M,
       author = {{MacAlpine}, G.~M.},
        title = "{On He II lambda 4686 emission and the question of reddening in quasars and Seyfert galaxies}",
      journal = {\apj},
         year = 1981,
        month = dec,
       volume = {251},
        pages = {465-470},
          doi = {10.1086/159484},
       adsurl = {https://ui.adsabs.harvard.edu/abs/1981ApJ...251..465M}
}

@ARTICLE{1985ApJ...289..451N,
       author = {{Netzer}, H.},
        title = "{The far-ultraviolet continuum of quasars and the universe at Z greater than 4}",
      journal = {\apj},
         year = 1985,
        month = feb,
       volume = {289},
        pages = {451-456},
          doi = {10.1086/162906},
       adsurl = {https://ui.adsabs.harvard.edu/abs/1985ApJ...289..451N}
}

@ARTICLE{1993PASP..105.1150B,
       author = {{Binette}, Luc and {Fosbury}, Robert A. and {Parker}, Dylan},
        title = "{Interpretation of the Ionizing Photon Deficit of AGN}",
      journal = {\pasp},
         year = 1993,
        month = oct,
       volume = {105},
        pages = {1150},
          doi = {10.1086/133293},
       adsurl = {https://ui.adsabs.harvard.edu/abs/1993PASP..105.1150B}
}

@ARTICLE{2014MNRAS.438..901S,
       author = {{Stern}, Jonathan and {Laor}, Ari and {Baskin}, Alexei},
        title = "{Radiation pressure confinement - I. Ionized gas in the ISM of AGN hosts}",
      journal = {\mnras},
         year = 2014,
        month = feb,
       volume = {438},
       number = {2},
        pages = {901-921},
          doi = {10.1093/mnras/stt1843},
archivePrefix = {arXiv},
       eprint = {1309.7825},
 primaryClass = {astro-ph.CO},
       adsurl = {https://ui.adsabs.harvard.edu/abs/2014MNRAS.438..901S}
}

@ARTICLE{2017MNRAS.466.3217Z,
       author = {{Zhang}, Kai and {Yan}, Renbin and {Bundy}, Kevin and {Bershady}, Matthew and {Haffner}, L. Matthew and {Walterbos}, Ren{\'e} and {Maiolino}, Roberto and {Tremonti}, Christy and {Thomas}, Daniel and {Drory}, Niv and {Jones}, Amy and {Belfiore}, Francesco and {S{\'a}nchez}, Sebastian F. and {Diamond-Stanic}, Aleksandar M. and {Bizyaev}, Dmitry and {Nitschelm}, Christian and {Andrews}, Brett and {Brinkmann}, Jon and {Brownstein}, Joel R. and {Cheung}, Edmond and {Li}, Cheng and {Law}, David R. and {Roman Lopes}, Alexandre and {Oravetz}, Daniel and {Pan}, Kaike and {Storchi Bergmann}, Thaisa and {Simmons}, Audrey},
        title = "{SDSS-IV MaNGA: the impact of diffuse ionized gas on emission-line ratios, interpretation of diagnostic diagrams and gas metallicity measurements}",
      journal = {\mnras},
         year = 2017,
        month = apr,
       volume = {466},
       number = {3},
        pages = {3217-3243},
          doi = {10.1093/mnras/stw3308},
archivePrefix = {arXiv},
       eprint = {1612.02000},
 primaryClass = {astro-ph.GA},
       adsurl = {https://ui.adsabs.harvard.edu/abs/2017MNRAS.466.3217Z}
}

@ARTICLE{2019MNRAS.485..367K,
       author = {{Kumari}, Nimisha and {Maiolino}, Roberto and {Belfiore}, Francesco and {Curti}, Mirko},
        title = "{Metallicity calibrations for diffuse ionized gas and low-ionization emission regions}",
      journal = {\mnras},
         year = 2019,
        month = may,
       volume = {485},
       number = {1},
        pages = {367-381},
          doi = {10.1093/mnras/stz366},
archivePrefix = {arXiv},
       eprint = {1902.01408},
 primaryClass = {astro-ph.GA},
       adsurl = {https://ui.adsabs.harvard.edu/abs/2019MNRAS.485..367K}
}

@ARTICLE{2019MNRAS.489.4721V,
       author = {{Vale Asari}, N. and {Couto}, G.~S. and {Cid Fernandes}, R. and {Stasi{\'n}ska}, G. and {de Amorim}, A.~L. and {Ruschel-Dutra}, D. and {Werle}, A. and {Florido}, T.~Z.},
        title = "{Diffuse ionized gas and its effects on nebular metallicity estimates of star-forming galaxies}",
      journal = {\mnras},
         year = 2019,
        month = nov,
       volume = {489},
       number = {4},
        pages = {4721-4733},
          doi = {10.1093/mnras/stz2470},
archivePrefix = {arXiv},
       eprint = {1907.08635},
 primaryClass = {astro-ph.GA},
       adsurl = {https://ui.adsabs.harvard.edu/abs/2019MNRAS.489.4721V}
}

@ARTICLE{2012MNRAS.427.1266V,
       author = {{Vaona}, L. and {Ciroi}, S. and {Di Mille}, F. and {Cracco}, V. and {La Mura}, G. and {Rafanelli}, P.},
        title = "{Spectral properties of the narrow-line region in Seyfert galaxies selected from the SDSS-DR7}",
      journal = {\mnras},
         year = 2012,
        month = dec,
       volume = {427},
       number = {2},
        pages = {1266-1283},
          doi = {10.1111/j.1365-2966.2012.22060.x},
archivePrefix = {arXiv},
       eprint = {1210.5201},
 primaryClass = {astro-ph.CO},
       adsurl = {https://ui.adsabs.harvard.edu/abs/2012MNRAS.427.1266V}
}

@ARTICLE{1986ApJ...301..727D,
       author = {{De Robertis}, M.~M. and {Osterbrock}, D.~E.},
        title = "{An Analysis of the Narrow-Line Profiles in Seyfert 2 Galaxies}",
      journal = {\apj},
         year = 1986,
        month = feb,
       volume = {301},
        pages = {727},
          doi = {10.1086/163939},
       adsurl = {https://ui.adsabs.harvard.edu/abs/1986ApJ...301..727D}
}

@ARTICLE{2020MNRAS.492.5675C,
       author = {{Carvalho}, S.~P. and {Dors}, O.~L. and {Cardaci}, M.~V. and {H{\"a}gele}, G.~F. and {Krabbe}, A.~C. and {P{\'e}rez-Montero}, E. and {Monteiro}, A.~F. and {Armah}, M. and {Freitas-Lemes}, P.},
        title = "{Chemical abundances of Seyfert 2 AGNs - II. N2 metallicity calibration based on SDSS}",
      journal = {\mnras},
         year = 2020,
        month = mar,
       volume = {492},
       number = {4},
        pages = {5675 (C20)-5683},
          doi = {10.1093/mnras/staa193},
archivePrefix = {arXiv},
       eprint = {2001.07126},
 primaryClass = {astro-ph.GA},
       adsurl = {https://ui.adsabs.harvard.edu/abs/2020MNRAS.492.5675C},
     note = {(C20)}
}

@ARTICLE{2021MNRAS.507..466D,
       author = {{Dors}, Oli L.},
        title = "{Chemical abundances in Seyfert galaxies - VI. Empirical abundance calibration}",
      journal = {\mnras},
         year = 2021,
        month = oct,
       volume = {507},
       number = {1},
        pages = {466 (D21)-474},
          doi = {10.1093/mnras/stab2166},
       adsurl = {https://ui.adsabs.harvard.edu/abs/2021MNRAS.507..466D}
}

@ARTICLE{2000AJ....120.1579Y,
       author = {{York}, Donald G. and {Adelman}, J. and {Anderson}, Jr., John E. and {Anderson}, Scott F. and {Annis}, James and {Bahcall}, Neta A. and {Bakken}, J.~A. and {Barkhouser}, Robert and {Bastian}, Steven and {Berman}, Eileen and {Boroski}, William N. and {Bracker}, Steve and {Briegel}, Charlie and {Briggs}, John W. and {Brinkmann}, J. and {Brunner}, Robert and {Burles}, Scott and {Carey}, Larry and {Carr}, Michael A. and {Castander}, Francisco J. and {Chen}, Bing and {Colestock}, Patrick L. and {Connolly}, A.~J. and {Crocker}, J.~H. and {Csabai}, Istv{\'a}n and {Czarapata}, Paul C. and {Davis}, John Eric and {Doi}, Mamoru and {Dombeck}, Tom and {Eisenstein}, Daniel and {Ellman}, Nancy and {Elms}, Brian R. and {Evans}, Michael L. and {Fan}, Xiaohui and {Federwitz}, Glenn R. and {Fiscelli}, Larry and {Friedman}, Scott and {Frieman}, Joshua A. and {Fukugita}, Masataka and {Gillespie}, Bruce and {Gunn}, James E. and {Gurbani}, Vijay K. and {de Haas}, Ernst and {Haldeman}, Merle and {Harris}, Frederick H. and {Hayes}, J. and {Heckman}, Timothy M. and {Hennessy}, G.~S. and {Hindsley}, Robert B. and {Holm}, Scott and {Holmgren}, Donald J. and {Huang}, Chi-hao and {Hull}, Charles and {Husby}, Don and {Ichikawa}, Shin-Ichi and {Ichikawa}, Takashi and {Ivezi{\'c}}, {\v{Z}}eljko and {Kent}, Stephen and {Kim}, Rita S.~J. and {Kinney}, E. and {Klaene}, Mark and {Kleinman}, A.~N. and {Kleinman}, S. and {Knapp}, G.~R. and {Korienek}, John and {Kron}, Richard G. and {Kunszt}, Peter Z. and {Lamb}, D.~Q. and {Lee}, B. and {Leger}, R. French and {Limmongkol}, Siriluk and {Lindenmeyer}, Carl and {Long}, Daniel C. and {Loomis}, Craig and {Loveday}, Jon and {Lucinio}, Rich and {Lupton}, Robert H. and {MacKinnon}, Bryan and {Mannery}, Edward J. and {Mantsch}, P.~M. and {Margon}, Bruce and {McGehee}, Peregrine and {McKay}, Timothy A. and {Meiksin}, Avery and {Merelli}, Aronne and {Monet}, David G. and {Munn}, Jeffrey A. and {Narayanan}, Vijay K. and {Nash}, Thomas and {Neilsen}, Eric and {Neswold}, Rich and {Newberg}, Heidi Jo and {Nichol}, R.~C. and {Nicinski}, Tom and {Nonino}, Mario and {Okada}, Norio and {Okamura}, Sadanori and {Ostriker}, Jeremiah P. and {Owen}, Russell and {Pauls}, A. George and {Peoples}, John and {Peterson}, R.~L. and {Petravick}, Donald and {Pier}, Jeffrey R. and {Pope}, Adrian and {Pordes}, Ruth and {Prosapio}, Angela and {Rechenmacher}, Ron and {Quinn}, Thomas R. and {Richards}, Gordon T. and {Richmond}, Michael W. and {Rivetta}, Claudio H. and {Rockosi}, Constance M. and {Ruthmansdorfer}, Kurt and {Sandford}, Dale and {Schlegel}, David J. and {Schneider}, Donald P. and {Sekiguchi}, Maki and {Sergey}, Gary and {Shimasaku}, Kazuhiro and {Siegmund}, Walter A. and {Smee}, Stephen and {Smith}, J. Allyn and {Snedden}, S. and {Stone}, R. and {Stoughton}, Chris and {Strauss}, Michael A. and {Stubbs}, Christopher and {SubbaRao}, Mark and {Szalay}, Alexander S. and {Szapudi}, Istvan and {Szokoly}, Gyula P. and {Thakar}, Anirudda R. and {Tremonti}, Christy and {Tucker}, Douglas L. and {Uomoto}, Alan and {Vanden Berk}, Dan and {Vogeley}, Michael S. and {Waddell}, Patrick and {Wang}, Shu-i. and {Watanabe}, Masaru and {Weinberg}, David H. and {Yanny}, Brian and {Yasuda}, Naoki and {SDSS Collaboration}},
        title = "{The Sloan Digital Sky Survey: Technical Summary}",
      journal = {\aj},
         year = 2000,
        month = sep,
       volume = {120},
       number = {3},
        pages = {1579-1587},
          doi = {10.1086/301513},
archivePrefix = {arXiv},
       eprint = {astro-ph/0006396},
 primaryClass = {astro-ph},
       adsurl = {https://ui.adsabs.harvard.edu/abs/2000AJ....120.1579Y}
}

@ARTICLE{2015ApJ...798....7B,
       author = {{Bundy}, Kevin and {Bershady}, Matthew A. and {Law}, David R. and {Yan}, Renbin and {Drory}, Niv and {MacDonald}, Nicholas and {Wake}, David A. and {Cherinka}, Brian and {S{\'a}nchez-Gallego}, Jos{\'e} R. and {Weijmans}, Anne-Marie and {Thomas}, Daniel and {Tremonti}, Christy and {Masters}, Karen and {Coccato}, Lodovico and {Diamond-Stanic}, Aleksandar M. and {Arag{\'o}n-Salamanca}, Alfonso and {Avila-Reese}, Vladimir and {Badenes}, Carles and {Falc{\'o}n-Barroso}, J{\'e}sus and {Belfiore}, Francesco and {Bizyaev}, Dmitry and {Blanc}, Guillermo A. and {Bland-Hawthorn}, Joss and {Blanton}, Michael R. and {Brownstein}, Joel R. and {Byler}, Nell and {Cappellari}, Michele and {Conroy}, Charlie and {Dutton}, Aaron A. and {Emsellem}, Eric and {Etherington}, James and {Frinchaboy}, Peter M. and {Fu}, Hai and {Gunn}, James E. and {Harding}, Paul and {Johnston}, Evelyn J. and {Kauffmann}, Guinevere and {Kinemuchi}, Karen and {Klaene}, Mark A. and {Knapen}, Johan H. and {Leauthaud}, Alexie and {Li}, Cheng and {Lin}, Lihwai and {Maiolino}, Roberto and {Malanushenko}, Viktor and {Malanushenko}, Elena and {Mao}, Shude and {Maraston}, Claudia and {McDermid}, Richard M. and {Merrifield}, Michael R. and {Nichol}, Robert C. and {Oravetz}, Daniel and {Pan}, Kaike and {Parejko}, John K. and {Sanchez}, Sebastian F. and {Schlegel}, David and {Simmons}, Audrey and {Steele}, Oliver and {Steinmetz}, Matthias and {Thanjavur}, Karun and {Thompson}, Benjamin A. and {Tinker}, Jeremy L. and {van den Bosch}, Remco C.~E. and {Westfall}, Kyle B. and {Wilkinson}, David and {Wright}, Shelley and {Xiao}, Ting and {Zhang}, Kai},
        title = "{Overview of the SDSS-IV MaNGA Survey: Mapping nearby Galaxies at Apache Point Observatory}",
      journal = {\apj},
         year = 2015,
        month = jan,
       volume = {798},
       number = {1},
          eid = {7},
        pages = {7},
          doi = {10.1088/0004-637X/798/1/7},
archivePrefix = {arXiv},
       eprint = {1412.1482},
 primaryClass = {astro-ph.GA},
       adsurl = {https://ui.adsabs.harvard.edu/abs/2015ApJ...798....7B}
}

@ARTICLE{2010SPIE.7735E..08B,
       author = {{Bacon}, R. and {Accardo}, M. and {Adjali}, L. and {Anwand}, H. and {Bauer}, S. and {Biswas}, I. and {Blaizot}, J. and {Boudon}, D. and {Brau-Nogue}, S. and {Brinchmann}, J. and {Caillier}, P. and {Capoani}, L. and {Carollo}, C.~M. and {Contini}, T. and {Couderc}, P. and {Daguis{\'e}}, E. and {Deiries}, S. and {Delabre}, B. and {Dreizler}, S. and {Dubois}, J. and {Dupieux}, M. and {Dupuy}, C. and {Emsellem}, E. and {Fechner}, T. and {Fleischmann}, A. and {Fran{\c{c}}ois}, M. and {Gallou}, G. and {Gharsa}, T. and {Glindemann}, A. and {Gojak}, D. and {Guiderdoni}, B. and {Hansali}, G. and {Hahn}, T. and {Jarno}, A. and {Kelz}, A. and {Koehler}, C. and {Kosmalski}, J. and {Laurent}, F. and {Le Floch}, M. and {Lilly}, S.~J. and {Lizon}, J.-L. and {Loupias}, M. and {Manescau}, A. and {Monstein}, C. and {Nicklas}, H. and {Olaya}, J.-C. and {Pares}, L. and {Pasquini}, L. and {P{\'e}contal-Rousset}, A. and {Pell{\'o}}, R. and {Petit}, C. and {Popow}, E. and {Reiss}, R. and {Remillieux}, A. and {Renault}, E. and {Roth}, M. and {Rupprecht}, G. and {Serre}, D. and {Schaye}, J. and {Soucail}, G. and {Steinmetz}, M. and {Streicher}, O. and {Stuik}, R. and {Valentin}, H. and {Vernet}, J. and {Weilbacher}, P. and {Wisotzki}, L. and {Yerle}, N.},
        title = "{The MUSE second-generation VLT instrument}",
      journal = {Proc. SPIE},
         year = 2010,
       volume = {7735},
        pages = {773508},
          doi = {10.1117/12.856027},
   archivePrefix = {arXiv},
       eprint = {2211.16795},
 primaryClass = {astro-ph.IM},
       adsurl = {https://ui.adsabs.harvard.edu/abs/2010SPIE.7735E..08B}
}

@ARTICLE{2004PASP..116..425H,
       author = {{Hook}, I.~M. and {J{\o}rgensen}, Inger and {Allington-Smith}, J.~R. and {Davies}, R.~L. and {Metcalfe}, N. and {Murowinski}, R.~G. and {Crampton}, D.},
        title = "{The Gemini-North Multi-Object Spectrograph: Performance in Imaging, Long-Slit, and Multi-Object Spectroscopic Modes}",
      journal = {\pasp},
         year = 2004,
        month = may,
       volume = {116},
       number = {819},
        pages = {425-440},
          doi = {10.1086/383624},
       adsurl = {https://ui.adsabs.harvard.edu/abs/2004PASP..116..425H}
}

@ARTICLE{2022AJ....164..207D,
       author = {{DESI Collaboration} and {Abareshi}, B. and {Aguilar}, J. and {Ahlen}, S. and {Alam}, Shadab and {Alexander}, David M. and {Alfarsy}, R. and {Allen}, L. and {Allende Prieto}, C. and {Alves}, O. and {Ameel}, J. and {Armengaud}, E. and {Asorey}, J. and {Aviles}, Alejandro and {Bailey}, S. and {Balaguera-Antol{\'\i}nez}, A. and {Ballester}, O. and {Baltay}, C. and {Bault}, A. and {Beltran}, S.~F. and {Benavides}, B. and {BenZvi}, S. and {Berti}, A. and {Besuner}, R. and {Beutler}, Florian and {Bianchi}, D. and {Blake}, C. and {Blanc}, P. and {Blum}, R. and {Bolton}, A. and {Bose}, S. and {Bramall}, D. and {Brieden}, S. and {Brodzeller}, A. and {Brooks}, D. and {Brownewell}, C. and {Buckley-Geer}, E. and {Cahn}, R.~N. and {Cai}, Z. and {Canning}, R. and {Capasso}, R. and {Carnero Rosell}, A. and {Carton}, P. and {Casas}, R. and {Castander}, F.~J. and {Cervantes-Cota}, J.~L. and {Chabanier}, S. and {Chaussidon}, E. and {Chuang}, C. and {Circosta}, C. and {Cole}, S. and {Cooper}, A.~P. and {da Costa}, L. and {Cousinou}, M.-C. and {Cuceu}, A. and {Davis}, T.~M. and {Dawson}, K. and {de la Cruz-Noriega}, R. and {de la Macorra}, A. and {de Mattia}, A. and {Della Costa}, J. and {Demmer}, P. and {Derwent}, M. and {Dey}, A. and {Dey}, B. and {Dhungana}, G. and {Ding}, Z. and {Dobson}, C. and {Doel}, P. and {Donald-McCann}, J. and {Donaldson}, J. and {Douglass}, K. and {Duan}, Y. and {Dunlop}, P. and {Edelstein}, J. and {Eftekharzadeh}, S. and {Eisenstein}, D.~J. and {Enriquez-Vargas}, M. and {Escoffier}, S. and {Evatt}, M. and {Fagrelius}, P. and {Fan}, X. and {Fanning}, K. and {Fawcett}, V.~A. and {Ferraro}, S. and {Ereza}, J. and {Flaugher}, B. and {Font-Ribera}, A. and {Forero-Romero}, J.~E. and {Frenk}, C.~S. and {Fromenteau}, S. and {G{\"a}nsicke}, B.~T. and {Garcia-Quintero}, C. and {Garrison}, L. and {Gazta{\~n}aga}, E. and {Gerardi}, F. and {Gil-Mar{\'\i}n}, H. and {Gontcho A Gontcho}, S. and {Gonzalez-Morales}, Alma X. and {Gonzalez-de-Rivera}, G. and {Gonzalez-Perez}, V. and {Gordon}, C. and {Graur}, O. and {Green}, D. and {Grove}, C. and {Gruen}, D. and {Gutierrez}, G. and {Guy}, J. and {Hahn}, C. and {Harris}, S. and {Herrera}, D. and {Herrera-Alcantar}, Hiram K. and {Honscheid}, K. and {Howlett}, C. and {Huterer}, D. and {Ir{\v{s}}i{\v{c}}}, V. and {Ishak}, M. and {Jelinsky}, P. and {Jiang}, L. and {Jimenez}, J. and {Jing}, Y.~P. and {Joyce}, R. and {Jullo}, E. and {Juneau}, S. and {Kara{\c{c}}ayl{\i}}, N.~G. and {Karamanis}, M. and {Karcher}, A. and {Karim}, T. and {Kehoe}, R. and {Kent}, S. and {Kirkby}, D. and {Kisner}, T. and {Kitaura}, F. and {Koposov}, S.~E. and {Kov{\'a}cs}, A. and {Kremin}, A. and {Krolewski}, Alex and {L'Huillier}, B. and {Lahav}, O. and {Lambert}, A. and {Lamman}, C. and {Lan}, Ting-Wen and {Landriau}, M. and {Lane}, S. and {Lang}, D. and {Lange}, J.~U. and {Lasker}, J. and {Le Guillou}, L. and {Leauthaud}, A. and {Le Van Suu}, A. and {Levi}, Michael E. and {Li}, T.~S. and {Magneville}, C. and {Manera}, M. and {Manser}, Christopher J. and {Marshall}, B. and {Martini}, Paul and {McCollam}, W. and {McDonald}, P. and {Meisner}, Aaron M. and {Mena-Fern{\'a}ndez}, J. and {Meneses-Rizo}, J. and {Mezcua}, M. and {Miller}, T. and {Miquel}, R. and {Montero-Camacho}, P. and {Moon}, J. and {Moustakas}, J. and {Mueller}, E. and {Mu{\~n}oz-Guti{\'e}rrez}, Andrea and {Myers}, Adam D. and {Nadathur}, S. and {Najita}, J. and {Napolitano}, L. and {Neilsen}, E. and {Newman}, Jeffrey A. and {Nie}, J.~D. and {Ning}, Y. and {Niz}, G. and {Norberg}, P. and {Noriega}, Hern{\'a}n E. and {O'Brien}, T. and {Obuljen}, A. and {Palanque-Delabrouille}, N. and {Palmese}, A. and {Zhiwei}, P. and {Pappalardo}, D. and {PENG}, X. and {Percival}, W.~J. and {Perruchot}, S. and {Pogge}, R. and {Poppett}, C. and {Porredon}, A. and {Prada}, F. and {Prochaska}, J. and {Pucha}, R. and {P{\'e}rez-Fern{\'a}ndez}, A. and {P{\'e}rez-R{\`a}fols}, I. and {Rabinowitz}, D. and {Raichoor}, A.},
        title = "{Overview of the Instrumentation for the Dark Energy Spectroscopic Instrument}",
      journal = {\aj},
         year = 2022,
        month = nov,
       volume = {164},
       number = {5},
          eid = {207},
        pages = {207},
          doi = {10.3847/1538-3881/ac882b},
archivePrefix = {arXiv},
       eprint = {2205.10939},
 primaryClass = {astro-ph.IM},
       adsurl = {https://ui.adsabs.harvard.edu/abs/2022AJ....164..207D}
}

@ARTICLE{2023PASP..135d8002R,
       author = {{Rigby}, Jane R. and {Lightsey}, Paul A. and {Garc{\'\i}a Mar{\'\i}n}, Macarena and {Bowers}, Charles W. and {Smith}, Erin C. and {Glasse}, Alistair and {McElwain}, Michael W. and {Rieke}, George H. and {Chary}, Ranga-Ram and {Liu}, Xiang (Cate) and {Clampin}, Mark and {Kimble}, Randy A. and {Kinzel}, Wayne and {Laidler}, Vicki and {Mehalick}, Kimberly I. and {Noriega-Crespo}, Alberto and {Shivaei}, Irene and {Skelton}, Dennis and {Stark}, Christopher and {Temim}, Tea and {Wei}, Zongying and {Willott}, Chris J.},
        title = "{How Dark the Sky: The JWST Backgrounds}",
      journal = {\pasp},
         year = 2023,
        month = apr,
       volume = {135},
       number = {1046},
          eid = {048002},
        pages = {048002},
          doi = {10.1088/1538-3873/acbcf4},
archivePrefix = {arXiv},
       eprint = {2211.09890},
 primaryClass = {astro-ph.IM},
       adsurl = {https://ui.adsabs.harvard.edu/abs/2023PASP..135d8002R}
}

@ARTICLE{2025arXiv251001033C,
       author = {{Curtis-Lake}, Emma and {Cameron}, Alex J. and {Bunker}, Andrew J. and {Scholtz}, Jan and {Carniani}, Stefano and {Parlanti}, Eleonora and {D'Eugenio}, Francesco and {Jakobsen}, Peter and {Willmer}, Christopher N.~A. and {Arribas}, Santiago and {Baker}, William M. and {Charlot}, St{\'e}phane and {Chevallard}, Jacopo and {Circosta}, Chiara and {Curti}, Mirko and {Eisenstein}, Daniel J. and {Hainline}, Kevin and {Ji}, Zhiyuan and {Johnson}, Benjamin D. and {Jones}, Gareth C. and {Maiolino}, Roberto and {Maseda}, Michael V. and {P{\'e}rez-Gonz{\'a}lez}, Pablo G. and {Rawle}, Tim and {Rieke}, Marcia and {Rinaldi}, Pierluigi and {Robertson}, Brant and {Rodr{\'\i}gez Del Pino}, Bruno and {Saxena}, Aayush and {Shivaei}, Irene and {Smit}, Renske and {Tacchella}, Sandro and {{\"U}bler}, Hannah and {Venturi}, Giacomo and {Williams}, Christina C. and {Willott}, Chris and {Duan}, Qiao},
        title = "{JADES Data Release 4 Paper I: Sample Selection, Observing Strategy and Redshifts of the complete spectroscopic sample}",
      journal = {arXiv e-prints},
         year = 2025,
        month = oct,
          eid = {arXiv:2510.01033},
        pages = {arXiv:2510.01033},
          doi = {10.48550/arXiv.2510.01033},
archivePrefix = {arXiv},
       eprint = {2510.01033},
 primaryClass = {astro-ph.GA},
       adsurl = {https://ui.adsabs.harvard.edu/abs/2025arXiv251001033C}
}

@ARTICLE{2025arXiv251001034S,
       author = {{Scholtz}, J. and {Carniani}, S. and {Parlanti}, E. and {D'Eugenio}, F. and {Curtis-Lake}, E. and {Jakobsen}, P. and {Bunker}, A.~J. and {Cameron}, A.~J. and {Arribas}, S. and {Baker}, W.~M. and {Charlot}, S. and {Chevellard}, J. and {Circosta}, C. and {Curti}, M. and {Duan}, Q. and {Eisenstein}, D.~J. and {Hainline}, K. and {Ji}, Z. and {Johnson}, B.~D. and {Jones}, G.~C. and {Kumari}, N. and {Maiolino}, R. and {Maseda}, M.~V. and {Perna}, M. and {P{\'e}rez-Gonz{\'a}lez}, P.~G. and {Rawle}, T. and {Rieke}, M. and {Rinaldi}, P. and {Robertson}, B. and {Saxena}, A. and {Shivaei}, I. and {Silcock}, M.~S. and {Sun}, Y. and {Rodr{\'\i}guez Del Pino}, B. and {Tacchella}, S. and {{\"U}bler}, H. and {Venturi}, G. and {Williams}, C.~C. and {Willmer}, C.~N.~A. and {Willott}, C. and {Witstok}, J.},
        title = "{JADES Data Release 4 -- Paper II: Data reduction, analysis and emission-line fluxes of the complete spectroscopic sample}",
      journal = {arXiv e-prints},
         year = 2025,
        month = oct,
          eid = {arXiv:2510.01034},
        pages = {arXiv:2510.01034},
          doi = {10.48550/arXiv.2510.01034},
archivePrefix = {arXiv},
       eprint = {2510.01034},
 primaryClass = {astro-ph.GA},
       adsurl = {https://ui.adsabs.harvard.edu/abs/2025arXiv251001034S}
}

@ARTICLE{1981PASP...93....5B,
   author = {{Baldwin}, J.~A. and {Phillips}, M.~M. and {Terlevich}, R.},
    title = "{Classification parameters for the emission-line spectra of extragalactic objects}",
  journal = {\pasp},
     year = 1981,
    month = feb,
   volume = 93,
    pages = {5-19},
      doi = {10.1086/130766},
   adsurl = {http://adsabs.harvard.edu/abs/1981PASP...93....5B}
}

@ARTICLE{1987ApJS...63..295V,
       author = {{Veilleux}, Sylvain and {Osterbrock}, Donald E.},
        title = "{Spectral Classification of Emission-Line Galaxies}",
      journal = {\apjs},
         year = 1987,
        month = feb,
       volume = {63},
        pages = {295},
          doi = {10.1086/191166},
       adsurl = {https://ui.adsabs.harvard.edu/abs/1987ApJS...63..295V}
}

@ARTICLE{2006MNRAS.372..961K,
  author        = {Kewley, L. J. and Groves, B. and Kauffmann, G. and Heckman, T.},
  year          = {2006},
  title         = {{The host galaxies and classification of active galactic nuclei}},
  journal       = {\mnras},
  volume        = {372},
  number        = {3},
  pages         = {961--976},
  doi           = {10.1111/j.1365-2966.2006.10859.x},
  adsurl        = {https://ui.adsabs.harvard.edu/abs/2006MNRAS.372..961K}
}

@ARTICLE{2004ApJS..153....9G,
       author = {{Groves}, Brent A. and {Dopita}, Michael A. and {Sutherland}, Ralph S.},
        title = "{Dusty, Radiation Pressure-Dominated Photoionization. I. Model Description, Structure, and Grids}",
      journal = {\apjs},
         year = 2004,
        month = jul,
       volume = {153},
       number = {1},
        pages = {9-73},
          doi = {10.1086/421113},
archivePrefix = {arXiv},
       eprint = {astro-ph/0404175},
 primaryClass = {astro-ph},
       adsurl = {https://ui.adsabs.harvard.edu/abs/2004ApJS..153....9G}
}

@ARTICLE{2004ApJS..153...75G,
       author = {{Groves}, Brent A. and {Dopita}, Michael A. and {Sutherland}, Ralph S.},
        title = "{Dust in Photoionized Nebulae. I. The Effect on Emission-Line Ratios and the DUSTY Code}",
      journal = {\apjs},
         year = 2004,
        month = jul,
       volume = {153},
       number = {1},
        pages = {75-91},
          doi = {10.1086/421113},
       adsurl = {https://ui.adsabs.harvard.edu/abs/2004ApJS..153...75G}
}

@ARTICLE{2006A&A...458..405G,
       author = {{Groves}, B. and {Dopita}, M. and {Sutherland}, R.},
        title = "{The infrared emission from the narrow line region}",
      journal = {\aap},
         year = 2006,
        month = nov,
       volume = {458},
       number = {2},
        pages = {405-416},
          doi = {10.1051/0004-6361:20065097},
archivePrefix = {arXiv},
       eprint = {astro-ph/0608057},
 primaryClass = {astro-ph},
       adsurl = {https://ui.adsabs.harvard.edu/abs/2006A&A...458..405G}
}

@ARTICLE{2017MNRAS.468L.113D,
       author = {{Dors}, Jr., O.~L. and {Arellano-C{\'o}rdova}, K.~Z. and {Cardaci}, M.~V. and {H{\"a}gele}, G.~F.},
        title = "{New quantitative nitrogen abundance estimations in a sample of Seyfert 2 active galactic nuclei}",
      journal = {\mnras},
         year = 2017,
        month = jun,
       volume = {468},
       number = {1},
        pages = {L113-L117},
          doi = {10.1093/mnrasl/slx036},
archivePrefix = {arXiv},
       eprint = {1703.03250},
 primaryClass = {astro-ph.GA},
       adsurl = {https://ui.adsabs.harvard.edu/abs/2017MNRAS.468L.113D}
}

@ARTICLE{2008ApJ...681.1183K,
       author = {{Kewley}, Lisa J. and {Ellison}, Sara L.},
        title = "{Metallicity Calibrations and the Mass-Metallicity Relation for Star-forming Galaxies}",
      journal = {\apj},
         year = 2008,
        month = jul,
       volume = {681},
       number = {2},
        pages = {1183-1204},
          doi = {10.1086/587500},
archivePrefix = {arXiv},
       eprint = {0801.1849},
 primaryClass = {astro-ph},
       adsurl = {https://ui.adsabs.harvard.edu/abs/2008ApJ...681.1183K}
}

@ARTICLE{2002ApJS..142...35K,
       author = {{Kewley}, L.~J. and {Dopita}, M.~A.},
        title = "{Using Strong Lines to Estimate Abundances in Extragalactic H II Regions and Starburst Galaxies}",
      journal = {\apjs},
         year = 2002,
        month = sep,
       volume = {142},
       number = {1},
        pages = {35-52},
          doi = {10.1086/341326},
archivePrefix = {arXiv},
       eprint = {astro-ph/0206495},
 primaryClass = {astro-ph},
       adsurl = {https://ui.adsabs.harvard.edu/abs/2002ApJS..142...35K}
}

@ARTICLE{2006A&A...459...85N,
       author = {{Nagao}, T. and {Maiolino}, R. and {Marconi}, A.},
        title = "{Gas metallicity diagnostics in star-forming galaxies}",
      journal = {\aap},
         year = 2006,
        month = nov,
       volume = {459},
       number = {1},
        pages = {85-101},
          doi = {10.1051/0004-6361:20065216},
archivePrefix = {arXiv},
       eprint = {astro-ph/0603580},
 primaryClass = {astro-ph},
       adsurl = {https://ui.adsabs.harvard.edu/abs/2006A&A...459...85N}
}

@ARTICLE{2016ApJ...816...23S,
       author = {{Sanders}, Ryan L. and {Shapley}, Alice E. and {Kriek}, Mariska and {Reddy}, Naveen A. and {Freeman}, William R. and {Coil}, Alison L. and {Siana}, Brian and {Mobasher}, Bahram and {Shivaei}, Irene and {Price}, Sedona H. and {de Groot}, Laura},
        title = "{The MOSDEF Survey: Electron Density and Ionization Parameter at z \raisebox{-0.5ex}\textasciitilde 2.3}",
      journal = {\apj},
         year = 2016,
        month = jan,
       volume = {816},
       number = {1},
          eid = {23},
        pages = {23},
          doi = {10.3847/0004-637X/816/1/23},
archivePrefix = {arXiv},
       eprint = {1509.03636},
 primaryClass = {astro-ph.GA},
       adsurl = {https://ui.adsabs.harvard.edu/abs/2016ApJ...816...23S}
}

@ARTICLE{2006A&A...447..863N,
       author = {{Nagao}, T. and {Maiolino}, R. and {Marconi}, A.},
        title = "{Gas metallicity in the narrow-line regions of high-redshift active galactic nuclei}",
      journal = {\aap},
         year = 2006,
        month = mar,
       volume = {447},
       number = {3},
        pages = {863-876},
          doi = {10.1051/0004-6361:20054127},
archivePrefix = {arXiv},
       eprint = {astro-ph/0508652},
 primaryClass = {astro-ph},
       adsurl = {https://ui.adsabs.harvard.edu/abs/2006A&A...447..863N}
}

@ARTICLE{2009A&A...503..721M,
       author = {{Matsuoka}, K. and {Nagao}, T. and {Maiolino}, R. and {Marconi}, A. and {Taniguchi}, Y.},
        title = "{Chemical evolution of high-redshift radio galaxies}",
      journal = {\aap},
         year = 2009,
        month = sep,
       volume = {503},
       number = {3},
        pages = {721-730},
          doi = {10.1051/0004-6361/200811478},
archivePrefix = {arXiv},
       eprint = {0905.1581},
 primaryClass = {astro-ph.CO},
       adsurl = {https://ui.adsabs.harvard.edu/abs/2009A&A...503..721M}
}

@ARTICLE{2014MNRAS.443.1291D,
       author = {{Dors}, Oli L. and {Cardaci}, M{\'o}nica V. and {H{\"a}gele}, Guillermo F. and {Krabbe}, {\^A}ngela C.},
        title = "{Metallicity evolution of AGNs from UV emission lines based on a new index}",
      journal = {\mnras},
         year = 2014,
        month = sep,
       volume = {443},
       number = {2},
        pages = {1291-1300},
          doi = {10.1093/mnras/stu1218},
archivePrefix = {arXiv},
       eprint = {1406.4832},
 primaryClass = {astro-ph.GA},
       adsurl = {https://ui.adsabs.harvard.edu/abs/2014MNRAS.443.1291D}
}

@ARTICLE{2026MNRAS.548ag560A,
       author = {{Armah}, Mark and {Dors}, O.~L. and {Riffel}, Rog{\'e}rio and {Cardaci}, M.~V. and {H{\"a}gele}, G.~F. and {Riffel}, Rogemar A. and {V{\'\i}lchez}, J.~M.},
        title = "{Identifying AGNs from X-ray detections ─ I: Metallicity calibrations in AGNs with X-ray luminosity as the primary input parameter}",
      journal = {\mnras},
         year = 2026,
        month = may,
       volume = {548},
       number = {1},
          eid = {stag560},
        pages = {stag560},
          doi = {10.1093/mnras/stag560},
archivePrefix = {arXiv},
       eprint = {2603.19181},
 primaryClass = {astro-ph.GA},
       adsurl = {https://ui.adsabs.harvard.edu/abs/2026MNRAS.548ag560A}
}

@ARTICLE{2017MNRAS.467.1507C,
       author = {{Castro}, C.~S. and {Dors}, O.~L. and {Cardaci}, M.~V. and {H{\"a}gele}, G.~F.},
        title = "{New metallicity calibration for Seyfert 2 galaxies based on the N2O2 index}",
      journal = {\mnras},
         year = 2017,
        month = may,
       volume = {467},
       number = {2},
        pages = {1507 (C17)-1514},
          doi = {10.1093/mnras/stx150},
archivePrefix = {arXiv},
       eprint = {1701.04997},
 primaryClass = {astro-ph.GA},
       adsurl = {https://ui.adsabs.harvard.edu/abs/2017MNRAS.467.1507C}
}

@ARTICLE{2021AJ....161...52L,
       author = {{Law}, David R. and {Westfall}, Kyle B. and {Bershady}, Matthew A. and {Cappellari}, Michele and {Yan}, Renbin and {Belfiore}, Francesco and {Bizyaev}, Dmitry and {Brownstein}, Joel R. and {Chen}, Yanping and {Cherinka}, Brian and {Drory}, Niv and {Lazarz}, Daniel and {Shetty}, Shravan},
        title = "{SDSS-IV MaNGA: Modeling the Spectral Line-spread Function to Subpercent Accuracy}",
      journal = {\aj},
         year = 2021,
        month = feb,
       volume = {161},
       number = {2},
          eid = {52},
        pages = {52},
          doi = {10.3847/1538-3881/abcaa2},
archivePrefix = {arXiv},
       eprint = {2011.04675},
 primaryClass = {astro-ph.IM},
       adsurl = {https://ui.adsabs.harvard.edu/abs/2021AJ....161...52L}
}

@ARTICLE{2013ApJS..208....5N,
       author = {{Newman}, Jeffrey A. and {Cooper}, Michael C. and {Davis}, Marc and {Faber}, S.~M. and {Coil}, Alison L. and {Guhathakurta}, Puragra and {Koo}, David C. and {Phillips}, Andrew C. and {Conroy}, Charlie and {Dutton}, Aaron A. and {Finkbeiner}, Douglas P. and {Gerke}, Brian F. and {Rosario}, David J. and {Weiner}, Benjamin J. and {Willmer}, C.~N.~A. and {Yan}, Renbin and {Harker}, Justin J. and {Kassin}, Susan A. and {Konidaris}, N.~P. and {Lai}, Kamson and {Madgwick}, Darren S. and {Noeske}, K.~G. and {Wirth}, Gregory D. and {Connolly}, A.~J. and {Kaiser}, N. and {Kirby}, Evan N. and {Lemaux}, Brian C. and {Lin}, Lihwai and {Lotz}, Jennifer M. and {Luppino}, G.~A. and {Marinoni}, C. and {Matthews}, Daniel J. and {Metevier}, Anne and {Schiavon}, Ricardo P.},
        title = "{The DEEP2 Galaxy Redshift Survey: Design, Observations, Data Reduction, and Redshifts}",
      journal = {\apjs},
         year = 2013,
        month = sep,
       volume = {208},
       number = {1},
          eid = {5},
        pages = {5},
          doi = {10.1088/0067-0049/208/1/5},
archivePrefix = {arXiv},
       eprint = {1203.3192},
 primaryClass = {astro-ph.CO},
       adsurl = {https://ui.adsabs.harvard.edu/abs/2013ApJS..208....5N}
}

@ARTICLE{2009MNRAS.398..949P,
       author = {{P{\'e}rez-Montero}, Enrique and {Contini}, Thierry},
        title = "{The impact of the nitrogen-to-oxygen ratio on ionized nebula diagnostics based on [NII] emission lines}",
      journal = {\mnras},
         year = 2009,
        month = sep,
       volume = {398},
       number = {2},
        pages = {949-960},
          doi = {10.1111/j.1365-2966.2009.15145.x},
archivePrefix = {arXiv},
       eprint = {0905.4621},
 primaryClass = {astro-ph.CO},
       adsurl = {https://ui.adsabs.harvard.edu/abs/2009MNRAS.398..949P}
}

@ARTICLE{2013ApJS..207...21N,
       author = {{Nicholls}, David C. and {Dopita}, Michael A. and {Sutherland}, Ralph S. and {Kewley}, Lisa J. and {Palay}, Ethan},
        title = "{Measuring Nebular Temperatures: The Effect of New Collision Strengths with Equilibrium and {\ensuremath{\kappa}}-distributed Electron Energies}",
      journal = {\apjs},
         year = 2013,
        month = aug,
       volume = {207},
       number = {2},
          eid = {21},
        pages = {21},
          doi = {10.1088/0067-0049/207/2/21},
archivePrefix = {arXiv},
       eprint = {1306.2023},
 primaryClass = {astro-ph.GA},
       adsurl = {https://ui.adsabs.harvard.edu/abs/2013ApJS..207...21N}
}

@ARTICLE{2017ApJ...850...74K,
       author = {{Koss}, Michael and {Trakhtenbrot}, Benny and {Ricci}, Claudio and {Lamperti}, Isabella and {Oh}, Kyuseok and {Berney}, Simon and {Schawinski}, Kevin and {Balokovi{\'c}}, Mislav and {Baronchelli}, Linda and {Crenshaw}, D. Michael and {Fischer}, Travis and {Gehrels}, Neil and {Harrison}, Fiona and {Hashimoto}, Yasuhiro and {Hogg}, Drew and {Ichikawa}, Kohei and {Masetti}, Nicola and {Mushotzky}, Richard and {Sartori}, Lia and {Stern}, Daniel and {Treister}, Ezequiel and {Ueda}, Yoshihiro and {Veilleux}, Sylvain and {Winter}, Lisa},
        title = "{BAT AGN Spectroscopic Survey. I. Spectral Measurements, Derived Quantities, and AGN Demographics}",
      journal = {\apj},
         year = 2017,
        month = nov,
       volume = {850},
       number = {1},
          eid = {74},
        pages = {74},
          doi = {10.3847/1538-4357/aa8ec9},
archivePrefix = {arXiv},
       eprint = {1707.08123},
 primaryClass = {astro-ph.HE},
       adsurl = {https://ui.adsabs.harvard.edu/abs/2017ApJ...850...74K}
}

@ARTICLE{2017MNRAS.464.1466O,
       author = {{Oh}, Kyuseok and {Schawinski}, Kevin and {Koss}, Michael and {Trakhtenbrot}, Benny and {Lamperti}, Isabella and {Ricci}, Claudio and {Mushotzky}, Richard and {Veilleux}, Sylvain and {Berney}, Simon and {Crenshaw}, D. Michael and {Gehrels}, Neil and {Harrison}, Fiona and {Masetti}, Nicola and {Soto}, Kurt T. and {Stern}, Daniel and {Treister}, Ezequiel and {Ueda}, Yoshihiro},
        title = "{BAT AGN Spectroscopic Survey - III. An observed link between AGN Eddington ratio and narrow-emission-line ratios}",
      journal = {\mnras},
         year = 2017,
        month = jan,
       volume = {464},
       number = {2},
        pages = {1466-1473},
          doi = {10.1093/mnras/stw2467},
archivePrefix = {arXiv},
       eprint = {1609.08625},
 primaryClass = {astro-ph.GA},
       adsurl = {https://ui.adsabs.harvard.edu/abs/2017MNRAS.464.1466O}
}

@ARTICLE{2022ApJS..261....1K,
       author = {{Koss}, Michael J. and {Trakhtenbrot}, Benny and {Ricci}, Claudio and {Bauer}, Franz E. and {Treister}, Ezequiel and {Mushotzky}, Richard and {Urry}, C. Megan and {Ananna}, Tonima T. and {Balokovi{\'c}}, Mislav and {den Brok}, Jakob S. and {Cenko}, S. Bradley and {Harrison}, Fiona and {Ichikawa}, Kohei and {Lamperti}, Isabella and {Lein}, Amy and {Mej{\'\i}a-Restrepo}, Julian E. and {Oh}, Kyuseok and {Pacucci}, Fabio and {Pfeifle}, Ryan W. and {Powell}, Meredith C. and {Privon}, George C. and {Ricci}, Federica and {Salvato}, Mara and {Schawinski}, Kevin and {Shimizu}, Taro and {Smith}, Krista L. and {Stern}, Daniel},
        title = "{BASS. XXI. The Data Release 2 Overview}",
      journal = {\apjs},
         year = 2022,
        month = jul,
       volume = {261},
       number = {1},
          eid = {1},
        pages = {1},
          doi = {10.3847/1538-4365/ac6c8f},
archivePrefix = {arXiv},
       eprint = {2207.12428},
 primaryClass = {astro-ph.GA},
       adsurl = {https://ui.adsabs.harvard.edu/abs/2022ApJS..261....1K}
}

@ARTICLE{2021A&A...653A.141A,
       author = {{Asplund}, M. and {Amarsi}, A.~M. and {Grevesse}, N.},
        title = "{The chemical make-up of the Sun: A 2020 vision}",
      journal = {\aap},
         year = 2021,
        month = sep,
       volume = {653},
          eid = {A141},
        pages = {A141},
          doi = {10.1051/0004-6361/202140445},
archivePrefix = {arXiv},
       eprint = {2105.01661},
 primaryClass = {astro-ph.SR},
       adsurl = {https://ui.adsabs.harvard.edu/abs/2021A&A...653A.141A}
}

@ARTICLE{2021A&A...652C...4P,
       author = {{Planck Collaboration} and {Aghanim}, N. and {Akrami}, Y. and {Ashdown}, M. and {Aumont}, J. and {Baccigalupi}, C. and {Ballardini}, M. and {Banday}, A.~J. and {Barreiro}, R.~B. and {Bartolo}, N. and {Basak}, S. and {Battye}, R. and {Benabed}, K. and {Bernard}, J. -P. and {Bersanelli}, M. and {Bielewicz}, P. and {Bock}, J.~J. and {Bond}, J.~R. and {Borrill}, J. and {Bouchet}, F.~R. and {Boulanger}, F. and {Bucher}, M. and {Burigana}, C. and {Butler}, R.~C. and {Calabrese}, E. and {Cardoso}, J. -F. and {Carron}, J. and {Challinor}, A. and {Chiang}, H.~C. and {Chluba}, J. and {Colombo}, L.~P.~L. and {Combet}, C. and {Contreras}, D. and {Crill}, B.~P. and {Cuttaia}, F. and {de Bernardis}, P. and {de Zotti}, G. and {Delabrouille}, J. and {Delouis}, J. -M. and {Di Valentino}, E. and {Diego}, J.~M. and {Dor{\'e}}, O. and {Douspis}, M. and {Ducout}, A. and {Dupac}, X. and {Dusini}, S. and {Efstathiou}, G. and {Elsner}, F. and {En{\ss}lin}, T.~A. and {Eriksen}, H.~K. and {Fantaye}, Y. and {Farhang}, M. and {Fergusson}, J. and {Fernandez-Cobos}, R. and {Finelli}, F. and {Forastieri}, F. and {Frailis}, M. and {Fraisse}, A.~A. and {Franceschi}, E. and {Frolov}, A. and {Galeotta}, S. and {Galli}, S. and {Ganga}, K. and {G{\'e}nova-Santos}, R.~T. and {Gerbino}, M. and {Ghosh}, T. and {Gonz{\'a}lez-Nuevo}, J. and {G{\'o}rski}, K.~M. and {Gratton}, S. and {Gruppuso}, A. and {Gudmundsson}, J.~E. and {Hamann}, J. and {Handley}, W. and {Hansen}, F.~K. and {Herranz}, D. and {Hildebrandt}, S.~R. and {Hivon}, E. and {Huang}, Z. and {Jaffe}, A.~H. and {Jones}, W.~C. and {Karakci}, A. and {Keih{\"a}nen}, E. and {Keskitalo}, R. and {Kiiveri}, K. and {Kim}, J. and {Kisner}, T.~S. and {Knox}, L. and {Krachmalnicoff}, N. and {Kunz}, M. and {Kurki-Suonio}, H. and {Lagache}, G. and {Lamarre}, J. -M. and {Lasenby}, A. and {Lattanzi}, M. and {Lawrence}, C.~R. and {Le Jeune}, M. and {Lemos}, P. and {Lesgourgues}, J. and {Levrier}, F. and {Lewis}, A. and {Liguori}, M. and {Lilje}, P.~B. and {Lilley}, M. and {Lindholm}, V. and {L{\'o}pez-Caniego}, M. and {Lubin}, P.~M. and {Ma}, Y. -Z. and {Mac{\'\i}as-P{\'e}rez}, J.~F. and {Maggio}, G. and {Maino}, D. and {Mandolesi}, N. and {Mangilli}, A. and {Marcos-Caballero}, A. and {Maris}, M. and {Martin}, P.~G. and {Martinelli}, M. and {Mart{\'\i}nez-Gonz{\'a}lez}, E. and {Matarrese}, S. and {Mauri}, N. and {McEwen}, J.~D. and {Meinhold}, P.~R. and {Melchiorri}, A. and {Mennella}, A. and {Migliaccio}, M. and {Millea}, M. and {Mitra}, S. and {Miville-Desch{\^e}nes}, M. -A. and {Molinari}, D. and {Montier}, L. and {Morgante}, G. and {Moss}, A. and {Natoli}, P. and {N{\o}rgaard-Nielsen}, H.~U. and {Pagano}, L. and {Paoletti}, D. and {Partridge}, B. and {Patanchon}, G. and {Peiris}, H.~V. and {Perrotta}, F. and {Pettorino}, V. and {Piacentini}, F. and {Polastri}, L. and {Polenta}, G. and {Puget}, J. -L. and {Rachen}, J.~P. and {Reinecke}, M. and {Remazeilles}, M. and {Renzi}, A. and {Rocha}, G. and {Rosset}, C. and {Roudier}, G. and {Rubi{\~n}o-Mart{\'\i}n}, J.~A. and {Ruiz-Granados}, B. and {Salvati}, L. and {Sandri}, M. and {Savelainen}, M. and {Scott}, D. and {Shellard}, E.~P.~S. and {Sirignano}, C. and {Sirri}, G. and {Spencer}, L.~D. and {Sunyaev}, R. and {Suur-Uski}, A. -S. and {Tauber}, J.~A. and {Tavagnacco}, D. and {Tenti}, M. and {Toffolatti}, L. and {Tomasi}, M. and {Trombetti}, T. and {Valenziano}, L. and {Valiviita}, J. and {Van Tent}, B. and {Vibert}, L. and {Vielva}, P. and {Villa}, F. and {Vittorio}, N. and {Wandelt}, B.~D. and {Wehus}, I.~K. and {White}, M. and {White}, S.~D.~M. and {Zacchei}, A. and {Zonca}, A.},
        title = "{Planck 2018 results. VI. Cosmological parameters (Corrigendum)}",
      journal = {\aap},
         year = 2021,
        month = aug,
       volume = {652},
          eid = {C4},
        pages = {C4},
          doi = {10.1051/0004-6361/201833910e},
       adsurl = {https://ui.adsabs.harvard.edu/abs/2021A&A...652C...4P}
}

@ARTICLE{2013RMxAA..49..137F,
       author = {{Ferland}, G.~J. and {Porter}, R.~L. and {van Hoof}, P.~A.~M. and {Williams}, R.~J.~R. and {Abel}, N.~P. and {Lykins}, M.~L. and {Shaw}, G. and {Henney}, W.~J. and {Stancil}, P.~C.},
        title = "{The 2013 Release of Cloudy}",
      journal = {\rmxaa},
         year = 2013,
        month = apr,
       volume = {49},
        pages = {137-163},
          doi = {10.48550/arXiv.1302.4485},
archivePrefix = {arXiv},
       eprint = {1302.4485},
 primaryClass = {astro-ph.GA},
       adsurl = {https://ui.adsabs.harvard.edu/abs/2013RMxAA..49..137F}
}

@ARTICLE{2017RMxAA..53..385F,
       author = {{Ferland}, G.~J. and {Chatzikos}, M. and {Guzm{\'a}n}, F. and {Lykins}, M.~L. and {van Hoof}, P.~A.~M. and {Williams}, R.~J.~R. and {Abel}, N.~P. and {Badnell}, N.~R. and {Keenan}, F.~P. and {Porter}, R.~L. and {Stancil}, P.~C.},
        title = "{The 2017 Release Cloudy}",
      journal = {\rmxaa},
         year = 2017,
        month = oct,
       volume = {53},
        pages = {385-438},
          doi = {10.48550/arXiv.1705.10877},
archivePrefix = {arXiv},
       eprint = {1705.10877},
 primaryClass = {astro-ph.GA},
       adsurl = {https://ui.adsabs.harvard.edu/abs/2017RMxAA..53..385F}
}

@ARTICLE{2023RNAAS...7..246G,
       author = {{Gunasekera}, Chamani M. and {van Hoof}, Peter A.~M. and {Chatzikos}, Marios and {Ferland}, Gary J.},
        title = "{The 23.01 Release of Cloudy}",
      journal = {Research Notes of the American Astronomical Society},
         year = 2023,
        month = nov,
       volume = {7},
       number = {11},
          eid = {246},
        pages = {246},
          doi = {10.3847/2515-5172/ad0e75},
archivePrefix = {arXiv},
       eprint = {2311.10163},
 primaryClass = {astro-ph.GA},
       adsurl = {https://ui.adsabs.harvard.edu/abs/2023RNAAS...7..246G}
}

@ARTICLE{2022MNRAS.514.5506D,
       author = {{Dors}, O.~L. and {Valerdi}, M. and {Freitas-Lemes}, P. and {Krabbe}, A.~C. and {Riffel}, R.~A. and {Am{\^o}res}, E.~B. and {Riffel}, R. and {Armah}, M. and {Monteiro}, A.~F. and {Oliveira}, C.~B.},
        title = "{Chemical abundances in Seyfert galaxies - IX. Helium abundance estimates}",
      journal = {\mnras},
         year = 2022,
        month = aug,
       volume = {514},
       number = {4},
        pages = {5506-5527},
          doi = {10.1093/mnras/stac1722},
archivePrefix = {arXiv},
       eprint = {2206.09836},
 primaryClass = {astro-ph.GA},
       adsurl = {https://ui.adsabs.harvard.edu/abs/2022MNRAS.514.5506D}
}

@ARTICLE{2024MNRAS.534.3040D,
       author = {{Dors}, O.~L. and {Cardaci}, M.~V. and {H{\"a}gele}, G.~F. and {Valerdi}, M. and {Ilha}, G.~S. and {Oliveira}, C.~B. and {Riffel}, R.~A. and {Flury}, S.~R. and {Arellano-C{\'o}rdova}, K.~Z. and {Storchi-Bergmann}, T. and {Riffel}, R. and {Almeida}, G.~C. and {Morais}, I.~N.},
        title = "{Direct estimates of nitrogen abundance for Seyfert 2 nuclei}",
      journal = {\mnras},
         year = 2024,
        month = nov,
       volume = {534},
       number = {4},
        pages = {3040-3054},
          doi = {10.1093/mnras/stae2253},
archivePrefix = {arXiv},
       eprint = {2405.13906},
 primaryClass = {astro-ph.GA},
       adsurl = {https://ui.adsabs.harvard.edu/abs/2024MNRAS.534.3040D}
}

@ARTICLE{2024A&A...690A..76G,
       author = {{Gravity Collaboration} and {Amorim}, A. and {Bourdarot}, G. and {Brandner}, W. and {Cao}, Y. and {Cl{\'e}net}, Y. and {Davies}, R. and {de Zeeuw}, P.~T. and {Dexter}, J. and {Drescher}, A. and {Eckart}, A. and {Eisenhauer}, F. and {Fabricius}, M. and {Feuchtgruber}, H. and {F{\"o}rster Schreiber}, N.~M. and {Garcia}, P.~J.~V. and {Genzel}, R. and {Gillessen}, S. and {Gratadour}, D. and {H{\"o}nig}, S. and {Kishimoto}, M. and {Lacour}, S. and {Lutz}, D. and {Millour}, F. and {Netzer}, H. and {Ott}, T. and {Perraut}, K. and {Perrin}, G. and {Peterson}, B.~M. and {Petrucci}, P.~O. and {Pfuhl}, O. and {Prieto}, A. and {Rabien}, S. and {Rouan}, D. and {Santos}, D.~J.~D. and {Shangguan}, J. and {Shimizu}, T. and {Sternberg}, A. and {Straubmeier}, C. and {Sturm}, E. and {Tacconi}, L.~J. and {Tristram}, K.~R.~W. and {Widmann}, F. and {Woillez}, J.},
        title = "{VLTI/GRAVITY interferometric measurements of the innermost dust structure sizes around active galactic nuclei}",
      journal = {\aap},
         year = 2024,
        month = oct,
       volume = {690},
          eid = {A76},
        pages = {A76},
          doi = {10.1051/0004-6361/202450746},
archivePrefix = {arXiv},
       eprint = {2407.13458},
 primaryClass = {astro-ph.GA},
       adsurl = {https://ui.adsabs.harvard.edu/abs/2024A&A...690A..76G}
}

@ARTICLE{2022ApJS..261....4O,
       author = {{Oh}, Kyuseok and {Koss}, Michael J. and {Ueda}, Yoshihiro and {Stern}, Daniel and {Ricci}, Claudio and {Trakhtenbrot}, Benny and {Powell}, Meredith C. and {den Brok}, Jakob S. and {Lamperti}, Isabella and {Mushotzky}, Richard and {Ricci}, Federica and {B{\"a}r}, Rudolf E. and {Rojas}, Alejandra F. and {Ichikawa}, Kohei and {Riffel}, Rog{\'e}rio and {Treister}, Ezequiel and {Harrison}, Fiona and {Urry}, C. Megan and {Bauer}, Franz E. and {Schawinski}, Kevin},
        title = "{BASS. XXIV. The BASS DR2 Spectroscopic Line Measurements and AGN Demographics}",
      journal = {\apjs},
         year = 2022,
        month = jul,
       volume = {261},
       number = {1},
          eid = {4},
        pages = {4},
          doi = {10.3847/1538-4365/ac5b68},
archivePrefix = {arXiv},
       eprint = {2203.00017},
 primaryClass = {astro-ph.GA},
       adsurl = {https://ui.adsabs.harvard.edu/abs/2022ApJS..261....4O}
}

@ARTICLE{1978ApJ...223...56K,
       author = {{Koski}, A.~T.},
        title = "{Spectrophotometry of Seyfert 2 galaxies and narrow-line radio galaxies.}",
      journal = {\apj},
         year = 1978,
        month = jul,
       volume = {223},
        pages = {56-73},
          doi = {10.1086/156235},
       adsurl = {https://ui.adsabs.harvard.edu/abs/1978ApJ...223...56K}
}

@ARTICLE{1981ApJ...249..462O,
       author = {{Osterbrock}, D.~E.},
        title = "{Seyfert galaxies with weak broad H alpha emission lines}",
      journal = {\apj},
         year = 1981,
        month = oct,
       volume = {249},
        pages = {462-470},
          doi = {10.1086/159306},
       adsurl = {https://ui.adsabs.harvard.edu/abs/1981ApJ...249..462O}
}

@ARTICLE{2001ApJ...556..121K,
       author = {{Kewley}, L.~J. and {Dopita}, M.~A. and {Sutherland}, R.~S. and {Heisler}, C.~A. and {Trevena}, J.},
        title = "{Theoretical Modeling of Starburst Galaxies}",
      journal = {\apj},
         year = 2001,
        month = jul,
       volume = {556},
       number = {1},
        pages = {121-140},
          doi = {10.1086/321545},
archivePrefix = {arXiv},
       eprint = {astro-ph/0106324},
 primaryClass = {astro-ph},
       adsurl = {https://ui.adsabs.harvard.edu/abs/2001ApJ...556..121K}
}

@ARTICLE{2011MNRAS.413.1687C,
       author = {{Cid Fernandes}, R. and {Stasi{\'n}ska}, G. and {Mateus}, A. and {Vale Asari}, N.},
        title = "{A comprehensive classification of galaxies in the Sloan Digital Sky Survey: how to tell true from fake AGN?}",
      journal = {\mnras},
         year = 2011,
        month = may,
       volume = {413},
       number = {3},
        pages = {1687-1699},
          doi = {10.1111/j.1365-2966.2011.18244.x},
archivePrefix = {arXiv},
       eprint = {1012.4426},
 primaryClass = {astro-ph.CO},
       adsurl = {https://ui.adsabs.harvard.edu/abs/2011MNRAS.413.1687C}
}

@ARTICLE{1989ApJ...345..245C,
       author = {{Cardelli}, Jason A. and {Clayton}, Geoffrey C. and {Mathis}, John S.},
        title = "{The Relationship between Infrared, Optical, and Ultraviolet Extinction}",
      journal = {\apj},
         year = "1989",
        month = "Oct",
       volume = {345},
        pages = {245},
          doi = {10.1086/167900},
       adsurl = {https://ui.adsabs.harvard.edu/abs/1989ApJ...345..245C}
}

@ARTICLE{2000ApJ...542..224D,
       author = {{Dopita}, M.~A. and {Kewley}, L.~J. and {Heisler}, C.~A. and {Sutherland}, R.~S.},
        title = "{A Theoretical Recalibration of the Extragalactic H II Region Sequence}",
      journal = {\apj},
         year = 2000,
        month = oct,
       volume = {542},
       number = {1},
        pages = {224-234},
          doi = {10.1086/309538},
       adsurl = {https://ui.adsabs.harvard.edu/abs/2000ApJ...542..224D}
}

@ARTICLE{2018ApJ...856...46R,
       author = {{Revalski}, M. and {Crenshaw}, D.~M. and {Kraemer}, S.~B. and {Fischer}, T.~C. and {Schmitt}, H.~R. and {Machuca}, C.},
        title = "{Quantifying Feedback from Narrow Line Region Outflows in Nearby Active Galaxies. I. Spatially Resolved Mass Outflow Rates for the Seyfert 2 Galaxy Markarian 573}",
      journal = {\apj},
         year = 2018,
        month = mar,
       volume = {856},
       number = {1},
          eid = {46},
        pages = {46},
          doi = {10.3847/1538-4357/aab107},
archivePrefix = {arXiv},
       eprint = {1802.07734},
 primaryClass = {astro-ph.GA},
       adsurl = {https://ui.adsabs.harvard.edu/abs/2018ApJ...856...46R}
}

@ARTICLE{2025A&A...696A.229P,
       author = {{P{\'e}rez-Montero}, E. and {Fern{\'a}ndez-Ontiveros}, J.~A. and {P{\'e}rez-D{\'\i}az}, B. and {V{\'\i}lchez}, J.~M. and {Amor{\'\i}n}, R.},
        title = "{Exploring the hardness of the ionizing radiation with the infrared softness diagram: II. Bimodal distributions in both the ionizing continuum slope and the excitation in active galactic nuclei}",
      journal = {\aap},
         year = 2025,
        month = apr,
       volume = {696},
          eid = {A229},
        pages = {A229},
          doi = {10.1051/0004-6361/202453276},
archivePrefix = {arXiv},
       eprint = {2503.09267},
 primaryClass = {astro-ph.GA},
       adsurl = {https://ui.adsabs.harvard.edu/abs/2025A&A...696A.229P}
}

@ARTICLE{1980ApJ...242.1041H,
       author = {{Halpern}, J.~P. and {Grindlay}, J.~E.},
        title = "{X-ray photoionized nebulae}",
      journal = {\apj},
         year = 1980,
        month = dec,
       volume = {242},
        pages = {1041-1055},
          doi = {10.1086/158535},
       adsurl = {https://ui.adsabs.harvard.edu/abs/1980ApJ...242.1041H}
}

@ARTICLE{1981ApJ...250..478K,
       author = {{Kwan}, J. and {Krolik}, J.~H.},
        title = "{The formation of emission lines in quasars and Seyfert nuclei}",
      journal = {\apj},
         year = 1981,
        month = nov,
       volume = {250},
        pages = {478-507},
          doi = {10.1086/159395},
       adsurl = {https://ui.adsabs.harvard.edu/abs/1981ApJ...250..478K}
}

@PHDTHESIS{1982PhDT.........4H,
       author = {{Halpern}, J.~P.},
        title = "{X-ray spectra of active galactic nuclei}",
       school = {Harvard University, Massachusetts},
         year = 1982,
        month = mar,
       adsurl = {https://ui.adsabs.harvard.edu/abs/1982PhDT.........4H}
}

@ARTICLE{1983ApJ...264..105F,
       author = {{Ferland}, G.~J. and {Netzer}, H.},
        title = "{Are there any shock-heated galaxies ?}",
      journal = {\apj},
         year = 1983,
        month = jan,
       volume = {264},
        pages = {105-113},
          doi = {10.1086/160577},
       adsurl = {https://ui.adsabs.harvard.edu/abs/1983ApJ...264..105F}
}

@ARTICLE{1983ApJ...269L..37H,
       author = {{Halpern}, J.~P. and {Steiner}, J.~E.},
        title = "{Low ionization active galactic nuclei : X-ray or shock heated ?}",
      journal = {\apjl},
         year = 1983,
        month = jun,
       volume = {269},
        pages = {L37-L41},
          doi = {10.1086/184051},
       adsurl = {https://ui.adsabs.harvard.edu/abs/1983ApJ...269L..37H}
}

@ARTICLE{1984ApL....24...43G,
       author = {{Gaskell}, C.~M.},
        title = "{Reddening of the Narrow-Line Regions of Active Galactic Nuclei and the Intrinsic Balmer Decrement II}",
      journal = {\aplett},
         year = 1984,
        month = jan,
       volume = {24},
        pages = {43},
       adsurl = {https://ui.adsabs.harvard.edu/abs/1984ApL....24...43G}
}

@ARTICLE{1984PASP...96..393G,
       author = {{Gaskell}, C.~M. and {Ferland}, G.~J.},
        title = "{Theoretical hydrogen-line ratios for the narrow-line regions of active galactic nuclei}",
      journal = {\pasp},
         year = 1984,
        month = jun,
       volume = {96},
        pages = {393-397},
          doi = {10.1086/131352},
       adsurl = {https://ui.adsabs.harvard.edu/abs/1984PASP...96..393G}
}

@ARTICLE{1984A&A...131..159P,
       author = {{Pequignot}, D.},
        title = "{Photoionization models of low-ionization active nuclei of galaxies : the case of NGC 1052.}",
      journal = {\aap},
         year = 1984,
        month = feb,
       volume = {131},
        pages = {159-168},
       adsurl = {https://ui.adsabs.harvard.edu/abs/1984A&A...131..159P}
}

@ARTICLE{2025MNRAS.542.3181D,
       author = {{Dors}, O.~L. and {Oliveira}, C.~B. and {Cardaci}, M.~V. and {H{\"a}gele}, G.~F. and {Armah}, Mark and {Riffel}, R.~A. and {Ramos Vieira}, L. and {Almeida}, G.~C. and {Morais}, I.~N. and {Santos}, P.~C.},
        title = "{Metallicity of active galactic nuclei from ultraviolet and optical emission lines ─ II. Revisiting the C43 metallicity calibration and its implications}",
      journal = {\mnras},
         year = 2025,
        month = oct,
       volume = {542},
       number = {4},
        pages = {3181-3197},
          doi = {10.1093/mnras/staf1407},
archivePrefix = {arXiv},
       eprint = {2508.05397},
 primaryClass = {astro-ph.GA},
       adsurl = {https://ui.adsabs.harvard.edu/abs/2025MNRAS.542.3181D}
}

@ARTICLE{1936Natur.138..503P,
       author = {{Page}, T.~L.},
        title = "{Chemical Composition of the Planetary Nebul{\ae}}",
      journal = {\nat},
         year = 1936,
        month = sep,
       volume = {138},
       number = {3490},
        pages = {503-504},
          doi = {10.1038/138503a0},
       adsurl = {https://ui.adsabs.harvard.edu/abs/1936Natur.138..503P}
}

@ARTICLE{1939LicOB..19....1B,
       author = {{Bowen}, Ira Sprague and {Wyse}, Arthur Bambridge},
        title = "{The spectra and chemical composition of the gaseous nebulae, NGC 6572, 7027, 7662}",
      journal = {Lick Observatory Bulletin},
         year = 1939,
        month = jan,
       volume = {495},
        pages = {1},
          doi = {10.5479/ADS/bib/1939LicOB.19.1B},
       adsurl = {https://ui.adsabs.harvard.edu/abs/1939LicOB..19....1B}
}

@ARTICLE{1942ApJ....95..356W,
       author = {{Wyse}, A.~B.},
        title = "{The Spectra of Ten Gaseous Nebulae.}",
      journal = {\apj},
         year = 1942,
        month = may,
       volume = {95},
        pages = {356},
          doi = {10.1086/144409},
       adsurl = {https://ui.adsabs.harvard.edu/abs/1942ApJ....95..356W}
}

@ARTICLE{1954ApJ...120..401A,
       author = {{Aller}, Lawrence H.},
        title = "{The Composition of the Planetary Nebula NGC 7027.}",
      journal = {\apj},
         year = 1954,
        month = nov,
       volume = {120},
        pages = {401},
          doi = {10.1086/145931},
       adsurl = {https://ui.adsabs.harvard.edu/abs/1954ApJ...120..401A}
}

@ARTICLE{1959ApJ...130...45A,
       author = {{Aller}, Lawrence H. and {Liller}, William},
        title = "{Photoelectric Spectrophotometry of Gaseous Nebulae. I. The Orion Nebulae.}",
      journal = {\apj},
         year = 1959,
        month = jul,
       volume = {130},
        pages = {45},
          doi = {10.1086/146695},
       adsurl = {https://ui.adsabs.harvard.edu/abs/1959ApJ...130...45A}
}

@ARTICLE{1975ApJ...197..535O,
       author = {{Osterbrock}, Donald E. and {Miller}, Joseph S.},
        title = "{The Optical Emission-Line Spectrum of Cygnus a}",
      journal = {\apj},
         year = 1975,
        month = may,
       volume = {197},
        pages = {535-544},
          doi = {10.1086/153541},
       adsurl = {https://ui.adsabs.harvard.edu/abs/1975ApJ...197..535O}
}

@ARTICLE{1992A&A...266..117A,
       author = {{Alloin}, D. and {Bica}, E. and {Bonatto}, C. and {Prugniel}, P.},
        title = "{ESO 138 G1 : a high excitation Seyfert 2 nucleus in a low luminosity early-type galaxy.}",
      journal = {\aap},
         year = 1992,
        month = dec,
       volume = {266},
        pages = {117-126},
       adsurl = {https://ui.adsabs.harvard.edu/abs/1992A&A...266..117A}
}

@ARTICLE{2008ApJ...687..133I,
       author = {{Izotov}, Yuri I. and {Thuan}, Trinh X.},
        title = "{Active Galactic Nuclei in Four Metal-poor Dwarf Emission-Line Galaxies}",
      journal = {\apj},
         year = 2008,
        month = nov,
       volume = {687},
       number = {1},
        pages = {133-140},
          doi = {10.1086/591660},
archivePrefix = {arXiv},
       eprint = {0807.2029},
 primaryClass = {astro-ph},
       adsurl = {https://ui.adsabs.harvard.edu/abs/2008ApJ...687..133I}
}

@ARTICLE{2015MNRAS.453.4102D,
       author = {{Dors}, O.~L. and {Cardaci}, M.~V. and {H{\"a}gele}, G.~F. and {Rodrigues}, I. and {Grebel}, E.~K. and {Pilyugin}, L.~S. and {Freitas-Lemes}, P. and {Krabbe}, A.~C.},
        title = "{On the central abundances of active galactic nuclei and star-forming galaxies}",
      journal = {\mnras},
         year = 2015,
        month = nov,
       volume = {453},
       number = {4},
        pages = {4102-4111},
          doi = {10.1093/mnras/stv1916},
archivePrefix = {arXiv},
       eprint = {1508.07802},
 primaryClass = {astro-ph.GA},
       adsurl = {https://ui.adsabs.harvard.edu/abs/2015MNRAS.453.4102D}
}

@ARTICLE{1969BOTT....5....3P,
       author = {{Peimbert}, M. and {Costero}, R.},
        title = "{Chemical Abundances in Galactic HII Regions}",
      journal = {Boletin de los Observatorios Tonantzintla y Tacubaya},
         year = 1969,
        month = may,
       volume = {5},
        pages = {3-22},
       adsurl = {https://ui.adsabs.harvard.edu/abs/1969BOTT....5....3P}
}

@ARTICLE{2026ApJ...998..175M,
       author = {{Morishita}, Takahiro and {Stiavelli}, Massimo and {Mason}, Charlotte A. and {Tripodi}, Roberta and {Chiaberge}, Marco and {Schuldt}, Stefan and {Willott}, Chris J. and {Zhang}, Yechi},
        title = "{A Nitrogen-rich AGN Powering a Large Ionizing Bubble at z = 8.63}",
      journal = {\apj},
         year = 2026,
        month = feb,
       volume = {998},
       number = {1},
          eid = {175},
        pages = {175},
          doi = {10.3847/1538-4357/ae3a7c},
archivePrefix = {arXiv},
       eprint = {2508.01372},
 primaryClass = {astro-ph.GA},
       adsurl = {https://ui.adsabs.harvard.edu/abs/2026ApJ...998..175M}
}

@ARTICLE{2026ApJ...998....5Z,
       author = {{Zhu}, Peixin and {Trussler}, James and {Kewley}, Lisa J.},
        title = "{The Nature of Nitrogen-enhanced High-redshift Galaxies}",
      journal = {\apj},
         year = 2026,
        month = feb,
       volume = {998},
       number = {1},
          eid = {5},
        pages = {5},
          doi = {10.3847/1538-4357/ae28d4},
archivePrefix = {arXiv},
       eprint = {2512.04043},
 primaryClass = {astro-ph.GA},
       adsurl = {https://ui.adsabs.harvard.edu/abs/2026ApJ...998....5Z}
}

@ARTICLE{2024MNRAS.535..881J,
       author = {{Ji}, Xihan and {{\"U}bler}, Hannah and {Maiolino}, Roberto and {D'Eugenio}, Francesco and {Arribas}, Santiago and {Bunker}, Andrew J. and {Charlot}, St{\'e}phane and {Perna}, Michele and {Rodr{\'\i}guez Del Pino}, Bruno and {B{\"o}ker}, Torsten and {Cresci}, Giovanni and {Curti}, Mirko and {Kumari}, Nimisha and {Lamperti}, Isabella},
        title = "{GA-NIFS: an extremely nitrogen-loud and chemically stratified galaxy at z   5.55}",
      journal = {\mnras},
         year = 2024,
        month = nov,
       volume = {535},
       number = {1},
        pages = {881-908},
          doi = {10.1093/mnras/stae2375},
archivePrefix = {arXiv},
       eprint = {2404.04148},
 primaryClass = {astro-ph.GA},
       adsurl = {https://ui.adsabs.harvard.edu/abs/2024MNRAS.535..881J}
}

@ARTICLE{2026ApJ...998..141Z,
       author = {{Zhang}, Yechi and {Morishita}, Takahiro and {Stiavelli}, Massimo},
        title = "{Potential Nitrogen Enrichment via Direct-collapse Wolf─Rayet Stars in a z = 4.7 Star-forming Galaxy}",
      journal = {\apj},
         year = 2026,
        month = feb,
       volume = {998},
       number = {1},
          eid = {141},
        pages = {141},
          doi = {10.3847/1538-4357/ae3825},
archivePrefix = {arXiv},
       eprint = {2502.04817},
 primaryClass = {astro-ph.GA},
       adsurl = {https://ui.adsabs.harvard.edu/abs/2026ApJ...998..141Z}
}

@ARTICLE{2026MNRAS.545f2107S,
       author = {{Scholtz}, J. and {Silcock}, M.~S. and {Curtis-Lake}, E. and {Maiolino}, R. and {Carniani}, S. and {D'Eugenio}, F. and {Ji}, X. and {Jakobsen}, P. and {Hainline}, K. and {Arribas}, S. and {Baker}, W.~M. and {Bhatawdekar}, R. and {Bunker}, A.~J. and {Charlot}, S. and {Chevallard}, J. and {Curti}, M. and {Eisenstein}, Daniel J. and {Isobe}, Y. and {Jones}, G.~C. and {Parlanti}, E. and {P{\'e}rez-Gonz{\'a}lez}, P.~G. and {Rinaldi}, P. and {Robertson}, B. and {Tacchella}, S. and {{\"U}bler}, H. and {Williams}, C.~C. and {Willott}, C. and {Witstok}, J.},
        title = "{JADES: carbon-enhanced, nitrogen-normal compact galaxy at z = 11.2}",
      journal = {\mnras},
         year = 2026,
        month = jan,
       volume = {545},
       number = {3},
          eid = {staf2107},
        pages = {staf2107},
          doi = {10.1093/mnras/staf2107},
archivePrefix = {arXiv},
       eprint = {2507.17809},
 primaryClass = {astro-ph.GA},
       adsurl = {https://ui.adsabs.harvard.edu/abs/2026MNRAS.545f2107S}
}

@ARTICLE{2025ApJ...994L..29Z,
       author = {{Zhu}, Peixin and {Kewley}, Lisa J. and {Hsiao}, Tiger Yu-Yang and {Trussler}, James},
        title = "{Only Nitrogen-enhanced Galaxies Have Detectable Ultraviolet Nitrogen Emission Lines at High Redshift}",
      journal = {\apjl},
         year = 2025,
        month = nov,
       volume = {994},
       number = {1},
          eid = {L29},
        pages = {L29},
          doi = {10.3847/2041-8213/ae1c43},
archivePrefix = {arXiv},
       eprint = {2511.03681},
 primaryClass = {astro-ph.GA},
       adsurl = {https://ui.adsabs.harvard.edu/abs/2025ApJ...994L..29Z}
}

@ARTICLE{2020MNRAS.496.2191F,
       author = {{Flury}, Sophia R. and {Moran}, Edward C.},
        title = "{Chemical abundances in active galaxies}",
      journal = {\mnras},
         year = 2020,
        month = aug,
       volume = {496},
       number = {2},
        pages = {2191-2203},
          doi = {10.1093/mnras/staa1563},
archivePrefix = {arXiv},
       eprint = {2006.01113},
 primaryClass = {astro-ph.GA},
       adsurl = {https://ui.adsabs.harvard.edu/abs/2020MNRAS.496.2191F}
}

@ARTICLE{2019MNRAS.486.5853D,
       author = {{Dors}, O.~L. and {Monteiro}, A.~F. and {Cardaci}, M.~V. and {H{\"a}gele}, G.~F. and {Krabbe}, A.~C.},
        title = "{Semi-empirical metallicity calibrations based on ultraviolet emission lines of type-2 AGNs}",
      journal = {\mnras},
         year = 2019,
        month = jul,
       volume = {486},
       number = {4},
        pages = {5853-5866},
          doi = {10.1093/mnras/stz1242},
archivePrefix = {arXiv},
       eprint = {1905.00691},
 primaryClass = {astro-ph.GA},
       adsurl = {https://ui.adsabs.harvard.edu/abs/2019MNRAS.486.5853D}
}

@ARTICLE{2025MNRAS.540.1608D,
       author = {{Dors}, O.~L. and {Oliveira}, C.~B. and {Cardaci}, M.~V. and {H{\"a}gele}, G.~F. and {Morais}, I.~N. and {Ji}, X. and {Riffel}, R.~A. and {Riffel}, R. and {Mezcua}, M. and {Almeida}, G.~C. and {Santos}, P.~C. and {de Mellos}, M.~S.~Z.},
        title = "{Metallicity of active galactic nuclei from ultraviolet and optical emission lines ─ I. Carbon abundance dependence}",
      journal = {\mnras},
         year = 2025,
        month = jun,
       volume = {540},
       number = {2},
        pages = {1608-1625},
          doi = {10.1093/mnras/staf727},
archivePrefix = {arXiv},
       eprint = {2505.00095},
 primaryClass = {astro-ph.GA},
       adsurl = {https://ui.adsabs.harvard.edu/abs/2025MNRAS.540.1608D}
}

@ARTICLE{2024ApJ...977..187Z,
       author = {{Zhu}, Peixin and {Kewley}, Lisa J. and {Sutherland}, Ralph S.},
        title = "{Theoretical Diagnostics for Narrow-line Regions of Active Galactic Nuclei}",
      journal = {\apj},
         year = 2024,
        month = dec,
       volume = {977},
       number = {2},
          eid = {187},
        pages = {187},
          doi = {10.3847/1538-4357/ad8f37},
archivePrefix = {arXiv},
       eprint = {2411.04103},
 primaryClass = {astro-ph.GA},
       adsurl = {https://ui.adsabs.harvard.edu/abs/2024ApJ...977..187Z}
}

@ARTICLE{1974ApJ...191..309S,
       author = {{Shields}, G.~A.},
        title = "{X-ray ionization and the helium abundance in 3C 120.}",
      journal = {\apj},
         year = 1974,
        month = jul,
       volume = {191},
        pages = {309-316},
          doi = {10.1086/152969},
       adsurl = {https://ui.adsabs.harvard.edu/abs/1974ApJ...191..309S}
}

@misc{Sutherland_2018,
       author = {{Sutherland}, Ralph and {Dopita}, Mike and {Binette}, Luc and {Groves}, Brent},
        title = "{MAPPINGS V: Astrophysical plasma modeling code}",
 howpublished = {Astrophysics Source Code Library, record ascl:1807.005},
         year = 2018,
        month = jul,
          eid = {ascl:1807.005},
archivePrefix = {ascl},
       eprint = {\href{https://ui.adsabs.harvard.edu/abs/2018ascl.soft07005S}{1807.005}},
       adsurl = {https://ui.adsabs.harvard.edu/abs/2018ascl.soft07005S}
}


\bsp	
\label{lastpage}






\end{document}